\documentclass[aps,prb,nofootinbib,twocolumn]{revtex4-2}
\usepackage{latexsym}
\usepackage{amssymb}
\usepackage[pdftex]{graphicx,color}
\usepackage{bm}
\usepackage[english]{babel}
\usepackage[latin1]{inputenc}
\usepackage{amsmath}
\usepackage{amsfonts}
\usepackage{setspace}
\usepackage{physics}
\usepackage[title]{appendix}
\usepackage[center]{titlesec}
\usepackage{hyperref}

\begin{document}

\title{Many-body constraints and non-thermal behavior in 1D open systems with Haldane exclusion statistics}

\author{Bo Xiong$^1$}

\author{Fiona Burnell$^{1,2}$}
\affiliation{$^1$ School of Physics and Astronomy, University of Minnesota, Minneapolis, Minnesota 55455, USA}
\affiliation{$^2$ Institute for Advanced Study, Princeton, New Jersey 08540, USA}

\begin{abstract}
We study the impact of the  inter-level energy constraints imposed by Haldane exclusion statistics on energy relaxation processes in one-dimensional systems coupled to a bosonic bath.  By formulating a second-quantized description of the relevant Fock space, we identify certain universal features of this relaxation dynamics, and show that it is generically slower than that of spinless fermions.  Our study focuses on the Calogero-Sutherland model, which realizes Haldane Exclusion statistics exactly in one dimension; however our results apply to any system that has the associated pattern of inter-level occupancy constraints in Fock space.  
\end{abstract}

\maketitle

\section{Introduction}

The impact of constraints on the dynamics of quantum mechanical systems has received a resurgence of interest in recent years\cite{PXP-Scar,Prem17,Constrained-MBL,FibonacciETH,timpanaro,Ho2019,KhemaniShatter,IadecolaSchecter,MoudgalyaConstrained,Gromov20,Magnifico2020}.  Much of this interest has focused on whether, and when, constraints facilitate or even necessitate violations of the entanglement thermalization hypothesis (ETH)\cite{Deutsch,Srednicki}.   Notably, in low-dimensions constraints can effectively isolate certain sectors of the Hilbert space\cite{PXP-Scar,MoudgalyaConstrained,KhemaniShatter,IadecolaSchecter} thereby preventing at least some eigenstates from being thermal.  

In the study of constrained dynamics, attention to date has focused primarily on systems with spatially local constraints, or constraints arising from symmetry.  (See, however, Refs. \cite{Mesterhazy2015,Wald2021}.)  Here, we focus on another interesting possibility: constraints due to unconventional exclusion statistics.  
The possibility of exclusion statistics that are neither fermionic nor bosonic was first raised by Haldane\cite{Haldane}, following the discovery that quasiparticles exhibiting  fractional (or anyonic) {\it exchange} statistics\cite{Leinaas, Wilczek1} are likely realized in the fractional quantum Hall effect\cite{Laughlin,Camino}.  
Inspired by counting arguments relevant to Laughlin states\cite{Laughlin}, Haldane proposed a growth of the many-body Hilbert space with particle number that is intermediate between fermions and bosons.  This counting, and the corresponding exclusion statistics, subsequently became known as Haldane exclusion statistics (HES).  
Though in the FQHE HES is related to anyonic exchange statistics\cite{Murthy2}, unlike the latter HES is not specific to two spatial dimensions.  In fact, it occurs generically in one-dimensional integrable models that can be solved by the thermodynamic Bethe ansatz\cite{bernard1994note,isakov1994statistical}.  
Several higher-dimensional realizations\cite{SutherlandHES,Hatsugai,Bhaduri1} are also known.

A surprising feature of HES is that, unlike fermionic and bosonic statistics, it appears to require the existence of occupancy constraints involving particles in multiple energy states.  There are two key pieces of evidence supporting this.  First, although the partition function can be well approximated using a distribution function that treats different energy levels as independent\cite{Wu}, this approximation generically assigns negative probabilities to some configurations, indicating that the approximation over-counts the available states\cite{Nayak1,Polychronakos,chaturvedi1997microscopic}.  Second,  HES is realized exactly in the Calogero-Sutherland model\cite{calogero1,calogero2,sutherland1,sutherland2,sutherland3,sutherland4}, as first observed by Ha\cite{Ha-HES}.  Murthy and Shankar\cite{Murthy}  
observed that in this case, HES can be directly attributed to occupancy constraints between particles in different energy levels.  The resulting occupancy constraints dictating which patterns of single-particle orbitals can be occupied represent a qualitatively different class of constraints in many-body quantum systems, whose impact on dynamics has received little attention to date.

In this work, we undertake to study how this new class of constraints affects the dynamics of open quantum systems.  We focus on open quantum systems because there the impact of occupancy constraints in energy space on dynamics is most transparent.  Further, coupling our system to a bath that breaks all conservation laws allows us to work with a specific model (the Calogero-Sutherland model) exhibiting exact HES without having to contend directly with the impact of integrability on the dynamics of our system.  
Even in this simple setting, however, we find that the constraints implied by HES can have a significant impact on relaxation dynamics.    

To elucidate the details of the constraint structure that causes HES, we focus on the Calogero-Sutherland model (CSM)\cite{calogero1,calogero2,sutherland1,sutherland2,sutherland3,sutherland4}, where  HES can be derived explicitly from the exact solution.\cite{Murthy,Isakov96a,Ha,Ha-HES}   
We develop an exact microscopic description of the occupancy constraints that is condusive to developing a second quantized formalism for HES particles.  This second-quantized description allows us to study the Lindblad dynamics of the CSM both numerically and analytically.  
Moreover, by mapping states in the HES Fock space to states in the Fock space of spinless fermions, we use our second-quantized formalism to analytically derive several qualitative features of the relaxation dynamics in these systems, and compare the resulting dynamics to that of fermionic systems. 

Using this approach we are able to prove analytically the following key results.   First, we show that under Lindblad dynamics, the energy relaxation of HES particles coupled to a bosonic bath cannot be faster than that of spinless fermions, and that in generic conditions, it is strictly slower.  Second, we find that the relaxation dynamics is universal for different species of HES particles.  
 Third, though typically our Lindblad dynamics eventually thermalizes the energy of all HES systems, we show that under certain (fine-tuned) conditions it is possible to initialize the system in configurations whose energy will never thermalize, essentially because all relaxation paths are blockaded by the constraint.  

The fact that constraints lead to slower energy relaxation dynamics may not surprise the reader; however, the universality of dynamics for HES systems with very different constraint structures is far from intuitive.  
This result is also different from relaxation rates predicted using Boltzman transport\cite{Bhaduri,Kaniadakis96,Isakov98,Arkeryd}, obtained assuming that the occupations of different energy levels are independent, rather than fully accounting for the constraints.    Interestingly, even in the absence of an exact treatment of the constraints, HES systems have been found to exhibit a universal  1D ballistic thermal conductance\cite{Rego}, though both electrical conductance\cite{Rego} and  shot noise\cite{Gomila} depend on the details of the exclusion statistics.   
 
We emphasize that, unlike many previous investigations of both the thermodynamics\cite{Murthy1,Rajagopal,Sen,Fukui,Joyce,Isakov96b,Potter} and transport\cite{Bhaduri,Kaniadakis96,Isakov98,Arkeryd,Gomila,Rego, Campo2016} of systems of free HES particles, our framework allows us to treat the Fock space constraints exactly.  
Though in the case of thermodynamics this exact treatment leads to small corrections, its impact on dynamics --especially in certain initial states-- is much more significant.

This article is organized as follows. In Sec. \ref{Sec:HESReview}, we review the original definition of Haldane exclusion statistics, as well as Wu's\cite{Wu} approximation to the partition function.  In Sec. \ref{Sec:CSM} we review how Haldane exclusion statistics arises in an exactly solvable model, i.e. the Calogero-Sutherland model, and develop a microscopic description of the associated Fock space occupancy constraints which elaborates on the exact solutions of Refs. \cite{Ha-HES,Murthy}.  We also review the key features of the thermodynamics of free HES particles, and show that treating the constraints exactly has only a minor impact on thermodynamics compared to Wu's results\cite{Wu}.  In Sec. \ref{Sec:Second}, we formulate a second quantized description of HES based on our description of the microscopic occupancy constraints. In Sec. \ref{Sec:Dynamics}, we use this second quantized framework to study the energy relaxation dynamics of an ideal gas of HES particles coupled to a bath, and find a universal relaxation behavior for all species of HES particles.  We summarize our results and discuss its wider implications in Sec. \ref{Sec:Discussion}.

\section{Review of Haldane exclusion statistics} \label{Sec:HESReview}

\subsection{Exclusion statistics: definition and basic considerations}

We will begin with a review of Haldane's original formulation of HES\cite{Haldane}. For a finite many-body system with fixed boundary conditions, the dimension $d_{\alpha}(N)$ of the single particle Hilbert space accessible to the $N^{th}$ particle of species $\alpha$ depends linearly on the total number particles in the system. This dependence is parametrized by a {\it statistical interaction}  $g_{\alpha \beta}$: $\Delta d_{\alpha}=-\Sigma_{\beta} g_{\alpha \beta} \Delta N_{\beta}$, where $\Delta N_{\beta}$ is allowed change of particle number of species $\beta$. In this work, we will primarily be concerned with the case where there is only one species of particle.  In this case we may drop the indices $\alpha$ and $\beta$.  Then, the dimension  $D_{N}(g)$ of the N-particle many-body Hilbert space is\footnote{Here we use the original counting proposed by Haldane; for other generalizations starting from the same statistical interaction see refs. \onlinecite{Polychronakos, Ilinski}.}
\begin{align}\label{DHaldane}
D_{N}(g)=\frac{[d(N)+N-1]!}{N![d(N)-1]!},
\end{align}
 where $d(N)$ is the dimension of single-particle Hilbert space with $N$ particles in the system, which satisfies 
\begin{align}\label{d}
d(N+\Delta N)=d(N)-g\Delta N. 
\end{align}
Here $g$ must be rational (i.e. $g=q/p$, with $q$ and $p$ coprime integers) to ensure that $d(N)$ and $D_{N}(g)$ are integers for suitable $N$ and $\Delta N$.  Taking $g=0$ yields bosonic statistics, since $d(N)=d(N-1) = d(1)$, implying that each added boson can occupy the same number of orbitals as the first one. Fermions are described by taking $g=1$, which gives $d(N)=d(N-1)-1$, indicating Pauli exclusion.  

 In this work, we focus on the case $0 < g < 1$, for which Eq. (\ref{DHaldane}) describes exclusion statistics distinct from those of fermions or bosons.  For example, in the fractional quantum Hall effect with the filling fraction $1/m$, the correct exclusion statistics are obtained by taking $g=1/m$\cite{Haldane}; Eq. (\ref{d}) reflects the fact that the number of single-particle orbitals available to a given quasi-particle is reduced by one when $m$ quasi-particles are added, i.e. $d(N+m)=d(N)-1$. 
In this case,  Eq. (\ref{d}) appears to be valid only when $\Delta N$ is a multiple of $m$, as otherwise the number of available states $d$ is fractional.  In the quantum Hall context this is not unnatural, since this corresponds to requiring that an integer number of electrons is added to the system.  However, this requirement is less natural in other applications, such as in one dimensional systems.  In appendix \ref{Sec:CountingApp}, we show how framing HES in terms of constraints allows us to define an integral $D(N)$ for any $N$.  

\subsection{Approximate treatment of HES systems in equilibrium: Wu's method}

To understand the physical implications of the statistics obtained by taking $g \neq 0,1$, one must first know how to compute the associated partition functions.   The first attempt to solve this problem was made by Wu\cite{Wu}, who considered a system where the dimension of the Fock space at energy $\epsilon_i$ is given by Eq. (\ref{DHaldane}).   This literally describes a system with a discrete spectrum and exact degeneracies; however, it is useful to view it as arising from grouping the single-particle energy levels into``cells", where the $i^{th}$ cell has $d_i$ levels and an average energy of $\epsilon_i$, and the difference in energies within a given cell is small compared to $k_B T$.\cite{Nayak1} In this case, using $d(N_i) = d_i(1) - g(N_i -1)$, Wu found the dimension of the total Fock space
\begin{align}\label{D_N}
D(\{N_i\})=\prod_{\{ N_i \}}\frac{[d_i+(1-g)(N_i-1)]!}{N_i![d_i-1-g(N_i-1)]!} \ .
\end{align}
where $d_i \equiv d_i(1)$ is the number of states at energy $\epsilon_i$ accessible to the first particle in the system.    
Importantly, Eq. (\ref{D_N}) assumes that the occupancy of different energy levels is independent. 
The grand canonical partition function associated with the many-body counting (\ref{D_N}) can be written down exactly; extremizing it with respect to $n_i \equiv N_i / d_i$ gives
 the average number of particles with energy $\epsilon_i$:
\begin{align}\label{Wudisfun}
n_i(\epsilon_i)=\frac{1}{w(e^{(\epsilon_i-\mu)/k_B T})+g}.
\end{align}
Here $w(\zeta)$ satisfies
\begin{align}
w^g(\zeta)[1+w(\zeta)]^{1-g}=\zeta \equiv e^{\beta(\epsilon-\mu)},
\end{align}
where $\beta$ and $\mu$ are the inverse temperature $1/(k_B T)$ and the chemical potential of the system respectively.

Interestingly, Eq.(\ref{Wudisfun}) also follows from a factorized grand canonical partition function:
\begin{align}\label{Zg}
Z_G = \prod_i (1+ w_i^{-1}),
\end{align}
where $w_i \equiv w(e^{\beta(\epsilon_i-\mu)})$.  It is straightforward to check that taking $n(\epsilon_k)= -\partial \ln Z_G/\beta \partial \epsilon_k$ gives exactly Eq.(\ref{Wudisfun}).\cite{Murthy1, Iguchi}  This form is computationally very convenient, since it gives a simple expression from which one can compute quantities independently at each level $i$.  Evidently, a factorization of the form (\ref{Zg}) is possible only when we treat the occupations of different energy levels as independent.

However, Wu's approach is not exact: as first observed by Nayak and Wilczek\cite{Nayak1}, Wu's distribution function cannot be derived from the exact partition function of a physical system, since for $g \neq 0,1$, Eq. (\ref{Wudisfun}) implies the existence of negative Boltzmann weights for certain occupancies $n_i$\cite{Nayak1,Polychronakos,chaturvedi1997microscopic}.
This indicates that for $g \neq 0,1$, HES describes physical systems where the occupation of energy level $i$ is not independent of the occupations of other energy levels, and the grand canonical  partition function cannot be factored.  
We will explore the relevant inter-energy occupancy constraints in detail in the next section.

Though Eq.(\ref{Wudisfun}) is not exact, it nonetheless constitutes a very good approximation to the behavior of generic many-body HES systems in the thermodynamic limit.   Specifically, Wu's description becomes effectively exact for large systems when the number $d_i$ of states in each cell is extensive in the volume.\cite{Nayak1}  
Fortunately in a generic many-body system, where the level spacing at finite energy density is exponentially small in the volume, such a grouping is natural, and we may expect Wu's approximation to provide a good description of HES thermodynamics in these typical cases.   
In Sec. \ref{Sec:thermo}, we will compare Wu's approximation to an exact result in the opposite limit, where the level spacing is not small compared to the temperature.

\section{Haldane exclusion statistics in the Calogero-Sutherland model} \label{Sec:CSM}

\begin{figure*}[tb]
\includegraphics[height=6cm,width=15cm]{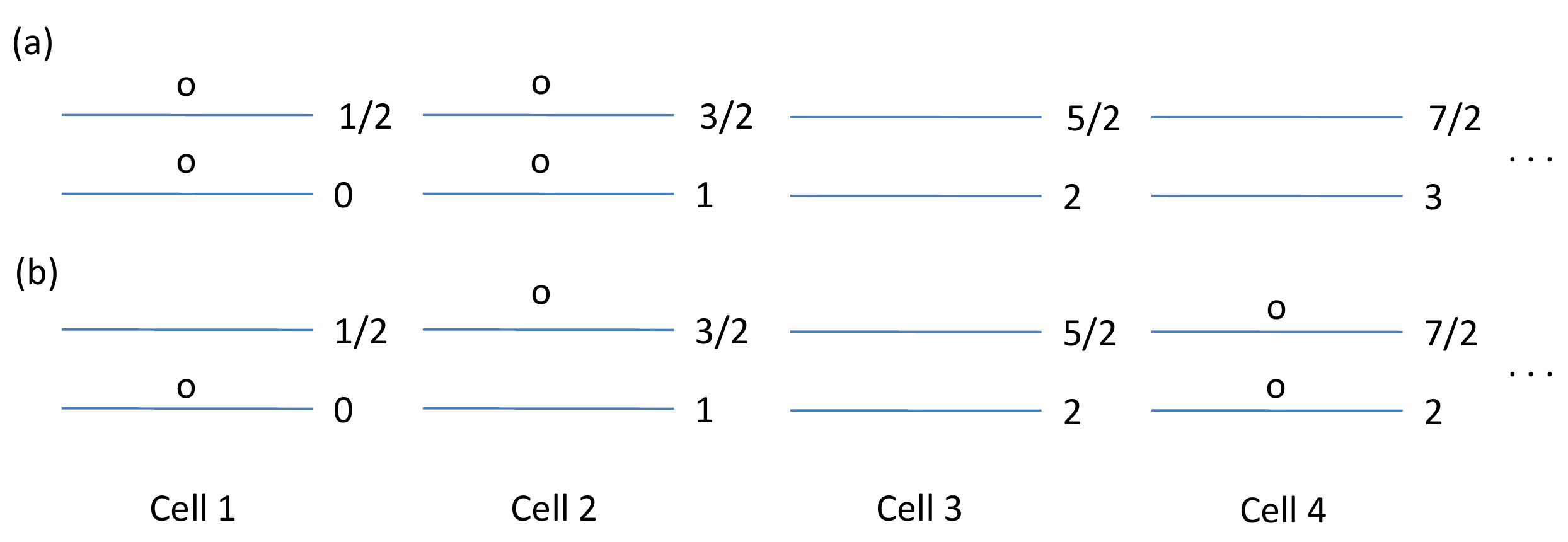}
\caption{  (a) The ground and (b) an excited states with four particles in the spectrum of CSM with $g=1/2$.  Labels to the right of each level indicate the corresponding energies in unit of $\omega$.  All levels are grouped up into cells of size one (i.e. $\omega$).  When particles are enumerated in order of increasing energy, particles with odd ordinal number occupy levels with integer energies, while particles with even ordinal numbers occupy levels with half integer energies.  To map this onto an abstract configuration of HES particles, we ignore the energy differences between states in the same cell, and take particles on the lower chain to have  pseudospins down ($\alpha=1$), while particles on the upper chain  have pseudospins up, labeled by $\alpha=2$.  }
\label{fig1}
\end{figure*} 

In order to explore HES beyond Wu's approximation, it is useful to work with a specific model.  Here we will focus on the Calogero-Sutherland  model (CSM)\cite{calogero1,calogero2,sutherland1,sutherland2,sutherland3,sutherland4}, which exactly realizes the simplest version of HES, with $g_{\alpha \beta} = g \delta_{\alpha,\beta}$\cite{isakov1994statistical,Murthy1,Ha}.  
As discussed above, any physical system that exactly realizes HES must exhibit occupancy constraints between different energy levels.  
Murthy and Shankar\cite{Murthy} showed how constraints of this form arise in the exact solutions of the CSM.  
Here we review in detail the nature of these constraints, and provide a simple explicit description of all states allowed in the resulting Fock space.  

\subsection{The Calogero-Sutherland model}
The CSM describes fermions in a one dimensional harmonic trap, interacting via a $1/r^2$ potential.\cite{calogero1,calogero2,sutherland1,sutherland2,sutherland3,sutherland4} The Hamiltonian for an $N$ fermion system is
\begin{align}\label{openH}
H=\sum_{i=1}^N \left[ -\frac{1}{2} \frac{\partial^2}{\partial x_i^2} + \frac{1}{2}\omega^2 x_i^2   \right] + \frac{1}{2} \sum_{i<j} \frac{g(g-1)}{(x_i - x_j)^2},
\end{align}
where, as we will see, $g$ is the statistical interaction parameter defined by Haldane. The many-body eigenstates of this Hamiltonian are labeled by a set of fermionic occupation numbers ${n_k}$, $k=0,1,...,\infty$ with $n_k=0,1$ and $\sum_{i=0}^{\infty} n_i = N$. \cite{Murthy1} The corresponding energies are 
\begin{align}\label{solOfCS}
E[\{ n_k \}]= \sum_{k=0}^{\infty} \epsilon_s(k,g) n_k,
\end{align}
where the shifted single particle energy is
\begin{equation} \label{Eq:energies}
\epsilon_s(k,g)=\epsilon_k-(1-g)N(\epsilon_k)\omega \ .
\end{equation}
Here $\epsilon_k= k\omega $ is the energy of the $k^{th}$ eigenstate of the single-particle 1D harmonic oscillator, and $N(\epsilon)$ is the number of particles with energy less than $\epsilon$, i.e. $N(\epsilon)=\sum_{j=1}^{\infty}\theta(\epsilon-\epsilon_j)n_j$.  (Here $\theta(x)$ is step function: $\theta(x)=1$ for $x>0$ and $0$ for $x\leq0$.)\footnote{We note that strictly speaking the $g=0$ result of the above formula should be interpreted as obtained from the limit $g \rightarrow 0$.}

There are several properties of the shifted energies (\ref{Eq:energies}) that are worth emphasizing.  First, inspecting Eq. (\ref{Eq:energies}), we see that the spacing between energy levels is $g \omega$.  Correspondingly there are no more than $\text{ceil}(1/g)$ states in an energy cell of size $\omega$, and the occupation number $n_j$ of the $j^{th}$ cell must satisfy $n_j \leq \text{ceil}(1/g)$.  Second, we may describe the many-body eigenstates by a set of integers $\{k_i\}, i=1,2,...,N$ indicating which of the shifted single-particle energy levels are occupied.  If we choose the labels $i$ in order of increasing energy (i.e. $k_1 < k_2 < ... < k_N$), we have $N(\epsilon_{k_i})=i-1$ for $i$th particle.   Note that if we add particles into the system, we must re-define $N(\epsilon)$ so as to include all of the particles ultimately added with energy less than $\epsilon$, thereby shifting the corresponding energies.  In constructing many-body states, we will therefore always think of adding particles to the vacuum in order of increasing $k$ value.

\subsection{Hilbert space dimension and HES in the Calogero-Sutherland model}

In order to see how HES emerges in the CSM, we must divide the shifted energy levels into cells.  Here we will show that by choosing cells of size $\omega$, we can exactly reproduce the counting in Eqs. (\ref{DHaldane}) and (\ref{d}).

Consider first the simplest case, $g=1/2$.  Figure (\ref{fig1}) shows how to group the spectrum into cells of size $\omega$: the $j^{th}$ cell contains $1/g=2$ distinct fermionic states, with energies $j \omega$ and $(j+ 1/2) \omega$.  Each of these states can either be occupied or vacant, such that the maximum number of particles in each cell is 2.  
To make the connection to HES, observe that if we identify the single-particle Hilbert space dimension $d(n)$ with the number of {\it cells} (rather than distinct energy levels) that any one of the $n$ particles can occupy,  we see that adding two particles reduces $d(n)$ by one -- i.e. this choice of $d(n)$ satisfies Eq. (\ref{d}).  


More generally, following  Murthy and Shankar\cite{Murthy}, we can argue that Eq. (\ref{D_N}) correctly describes the dimension of the many-body Hilbert space of the CSM for $N=1+np$, if we take $d(N)$ to be the number of cells of size $\omega$  available to a particle added to a system with $N$ particles.  To match Haldane's assumption of a system with a finite Hilbert space, we constrain the maximum single-particle energy  to be $\epsilon_{0 \text{max}} = k_{\text{max}} \omega$. 
 This gives a maximum shifted single particle energy of $\epsilon_{\text{max}}=\omega\left(  k_{\text{max}}-(1-g)(N-1) \right)$.  
If we divide these shifted energies into cells of size $\omega$, we find a total of $d= k_{\text{max}}-(1-g)(N-1) $ such cells.  (Here, we choose $N$ such that  $(1-g)(N-1) \in \mathbb{Z}$, i.e. $N=1+np$; the cases that $k_{\text{max}}$ or $(1-g)(N-1)$ is not an integer are discussed in Appendix \ref{Sec:CountingApp}). It follows that 
\begin{equation}  \label{Eq:kmax}
k_{\text{max}}= d + (1-g)(N-1)  \ .
\end{equation}
The total number of many-body states must therefore be given by the number of ways to arrange $N$ fermions into the first $k_{\text{max}}$ eigenstates of the 1D simple harmonic oscillator, which is given by Eq. (\ref{D_N}), with $d(N) = d - g(N-1)$, where $d$ describes the number of energy cells available for the first particle to occupy.

Thus quite generally, we recover the counting appropriate to HES by treating each energy cell of size $\omega$ as comprising a single quantum state.     As noted above, this state can contain a maximum of $\text{ceil}(1/g)$ particles; hence if $g=q/p$, adding $p$ particles decreases the number of available states by $q$, in agreement with Haldane's ansatz.  

One might wonder whether choosing a cell size of $\omega$ is a fundamental requirement for obtaining a description of the CSM in terms of particles that obey HES. 
Suppose we divide the spectrum into cells with $n$ levels, each of size  $n g \omega$, with $g= q/p$.  
There are a total of $n_{\text{cell}} = \text{floor}[\epsilon_{\text{max}}/ (n g \omega) ]+1 = \text{floor}[\left(  k_{\text{max}}-(1-g)(N-1) \right)/ng]+1$ cells available to the first particle in the system.   Thus with this cellulation, we obtain:
\begin{equation}
k_{\text{max}}= n g (n_{\text{cell}}-1 + R) + (1-g)(N-1) \ \ .
\end{equation}
where R is the remainder of the quantity $(k_{max} - (1-g)(N-1))/ ng $. 
In this case, we see that the total Hilbert space size does {\it not} satisfy Eq. (\ref{DHaldane}) with $d(N) = n_{\text{cell}} - g(N-1)$; instead, we must take $d(N) = n g (n_{\text{cell}} - 1 + R) - g(N-1) $.    If we choose a cell size that is not an integer multiple of $\omega$, such that $ng$ is non-integral, then $n g ( n_{\text{cell}} -1 +R)$ is not an integer in general, and cannot be identified  with $d(1)-1$; hence we cannot  obtain a description of our counting in terms of free HES particles.  

Moreover, suppose we choose cells of size  $m \omega, m \in \mathbb{Z}$, and, as above, consider particle numbers $N$ where $(N-1)g $ is an integer.  (In this case  $m R = \text{mod}((k_{max} - (1-g)(N-1)), m)$ is also an integer).   Then, formally, we recover Eq. (\ref{d}) with $d(1) -1= m( n_{\text{cell}} -1)+ m R$, and $d(N) = d(1) - g(N-1)$.  In particular, the statistical parameter is still $g$, even though the maximum number of particles in a cell is now $m$ ceil$(1/g)$, in contradiction with our basic expectation of exclusion for HES particles.\footnote{One can easily check that  choosing $m$ species of particle in each cell does not remedy this: taking $g_{\alpha \beta}\equiv g$ for all $\alpha$ and $\beta$, we would have $\sum_\alpha \Delta d_\alpha = - \sum_{\alpha \beta} g_{\alpha \beta} \Delta N_\beta = - m g\sum_{ \beta}  \Delta N_\beta$, and adding one particle would remove too many states.  Choosing $g_{\alpha \beta } = g \delta_{\alpha \beta}$ we recover the correct change in the total number of available states; however, when calculating the total Hilbert space dimension for a given choice of $\sum_{\alpha} N_{\alpha}$, we necessarily over-count, since for example we separately count configurations with $N_1 = N-k, N_2 =k$ and configurations where  $N_2= N-k, N_1 =k$.  Taking $g=1/2$ and $m=2$ these choices exhaust our options, and we conclude that in general cells of size $m \omega$ do not lead to a simple correspondence with HES. }  
Thus we see that a cell size of $\omega$ is the unique correct choice to interpret excitations in the CSM as a gas of free HES particles.

\subsection{Occupancy constraints of HES particles in the CSM} \label{Sec:Constraints}

As observed above, choosing HES quantum states to correspond to energy cells of size $\omega$ automatically produces one constraint on the occupancy of our quantum states: namely, we can have no more than $\text{ceil}(1/g)$ particles in each state.  However, as emphasized by Murthy and Shankar\cite{Murthy}, this is not the only constraint governing how these states can be occupied.  
To see why this is so, consider the case of semions ($g=1/2$), in which case the shifted single-particle energies are $\epsilon(k)=k-\frac{1}{2}N(k)$, where $N(k)$ is the total number of particles with energy less than $\epsilon(k)$.  
If we enumerate the particles in terms of increasing energy, then particles $1,3, 5,...$ have integer energies (in units of $\omega$), while particles $2,4,6,... $ have half-integral energies. 
Thus if there is only one particle in the first energy cell, then the lowest energy particle in the second cell must have half-integral energy -- and thus the second cell can only contain a single particle.  

For the case $g=1/m$, these constraints can be solved exactly in the following way.\cite{Ha-HES,Murthy}  If $n_i$ represents the number of particles in cell $i$, the allowed many-body states of the CSM correspond to sequences of partitions of $m$, with an arbitrary number of zeros inserted between each non-zero element.  For  example, with $g=1/2$ in a system of
$d=4$ cells and $N=5$ particles, we may have $(n_1,n_2,n_3,n_4)= (2210), (2201), (2021), (0221), (2111),$ or  $(1121)$, since the partitions of $2$ are $[2], [11]$. One can check the Haldane formula (\ref{D_N}) gives exactly $D_5(1/2,4)=6.$

\begin{figure*}[tb]
\includegraphics[height=8cm,width=17cm]{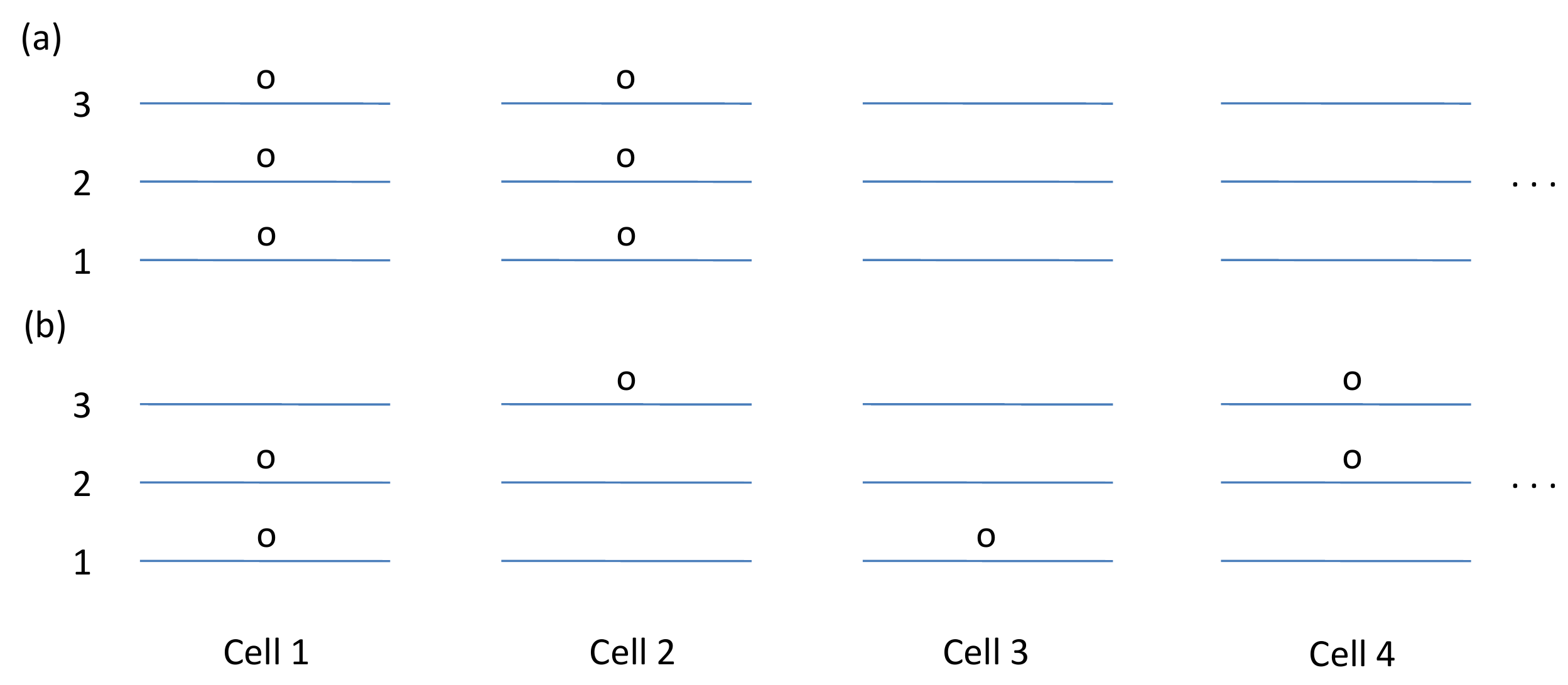}
\caption{ The ground state $(3300...)$ (a) and an excited state $(211200...)$ (b) of the ideal gas of HES particles with $g=1/3$.  Other excited states can be generated by moving one or more particles up in energy, while respecting the rule that the single-particle energies must be ordered by increasing pseudospin, as described in the main text. }
\label{fig3}
\end{figure*}

For our purposes, it will be useful to introduce a different representation of the allowed occupancy configurations, which is convenient for the second-quantized formalism which we will use to study the dynamics of HES particles.   Figure (\ref{fig1}) summarizes this picture for the case $g=1/2$.  
We replace the two energy levels in each cell with a pseudospin index $\alpha=0(1)$, represented in Figure \ref{fig1} by the upper and lower rows of states. Thus pseudospin-1 (0) particles have half-integer (integer) energies.     If we enumerate the particles $i=1,2, ...$ in order of increasing energy, then the constraints can be summarized as follows. (i) If the $i^{th}$ particle is in the $j^{th}$ cell and has pseudospin 0, the $(i+1)^{st}$ particle, which by definition is in the $k^{th}$ cell with $k\geq j$, necessarily has pseudospin 1; (ii) If the $i^{th}$ particle occupies the $j^{th}$ cell and has pseudospin 1, the $(i+1)^{st}$ particle, which occupies the $k^{th}$ cell with $k > j$, necessarily has pseudospin 0.

\begin{figure*}[tb]
\includegraphics[height=9cm,width=17cm]{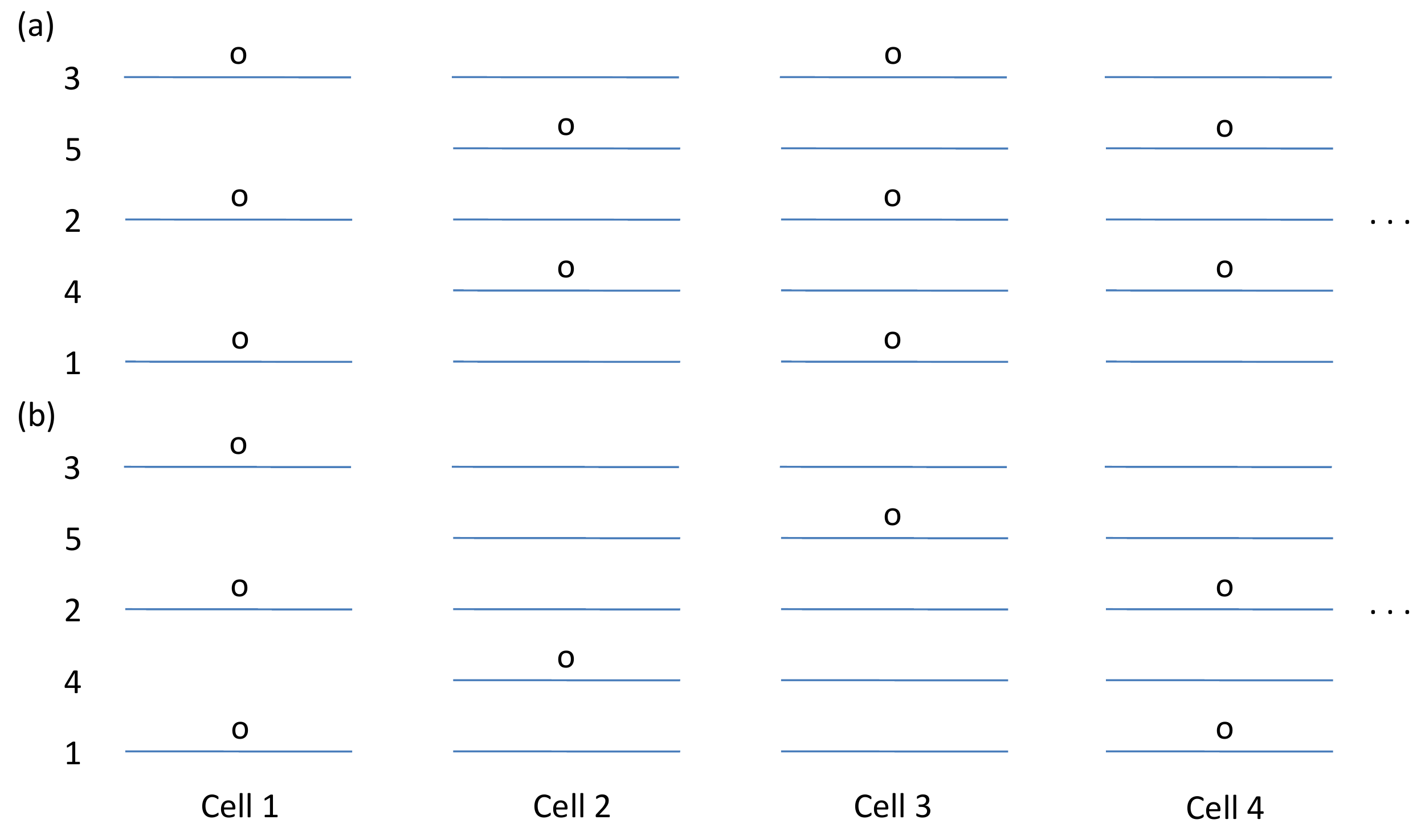}
\caption{ The ground state $(3232...)$ (a) and an excited state $(3112...)$ (b) of the ideal gas of HES particles for $g=2/5$. In this case there are five pseudospin chains , or $\alpha=1,2,...,5$, which are labeled in the order of first five particles' energies in the ground state. The first cell consists of only three levels, compared to five levels in the case of $g=1/5$. }
\label{fig4}
\end{figure*}

For $g=1/m$, with $m$ an integer, we can generalize this picture as follows.  We replace the $m$ distinct energies within each cell of size $\omega$ with a pseudospin index $\alpha = 0, 1, ... m-1$, where particles of the same pseudospin have energies that differ by an integer multiple of $\omega$.  Diagramatically, we represent this with $m$ chains, instead of the two shown in Figure (\ref{fig1}).   Enumerating the particles in order of increasing energy, we find that if the $i^{th}$ particle has pseudospin $\alpha$, the $i+1^{st}$ particle must have pseudospin $\alpha+1 \ \text{mod } m$.   This implies that if the $i^{th}$ particle has pseudospin $\alpha =1$, then the $i-1^{st}$ particle must be in a cell of lower energy.  (By definition, it cannot be in a cell of higher energy).   
Similarly, if the $i^{th}$ particle has pseudospin $\alpha =m$, then the $i+1^{st}$ particle must be in a cell of strictly higher energy.     For $g=1$ ($g= 1/m, m \rightarrow \infty$), this reproduces fermionic (bosonic) exclusion statistics. 

Evidently, these constraints ensure that the ground state has $n_1 =  n_2 = n_{\text{floor}(N/m)}= m$, with the remainder of the particles in the $\text{floor}(N/m)+1^{st}$ cell.  As for the excited states, they are constructed by moving particles to higher-energy cells, while respecting the energetic ordering of the pseudospin indices.  In terms of the occupancies $n_i$, this gives exactly the configurations identified by Murthy and Shankar. Figure (\ref{fig3}) shows the example of $g=1/3$, $d=3$ and $N=4$, for which the allowed states are  $(310), (301), (031), (211), (121)$, since the partitions of $3$ are $[21], [12]$ and $[111]$.  One can check that this agrees with the formula in Eq.(\ref{D_N}), which gives $D_4(1/3, 3)=5$.

Finally, we extend our description to $g=q/p$, for which Ref. \onlinecite{Murthy} did not explicitly solve the constraints.\footnote{See, however, Ref. \cite{Ha-HES} for a general solution to the periodic Calogero-Sutherland model, in which eigenstates are labeled by pseudo-momenta rather than oscillator states.}  Here $q,p$ are coprime integers with $q<p$.  
 In the ground state, the occupancies of the cells first $\text{floor}(g N)$ cells are  $(x_1, x_2,..., x_q, x_1, x_2...,x_q, x_1, x_2 ......)$.
 Here the repeated block $[x_1, x_2,..., x_q]$ is the solution to the $q$ linear equations:
\begin{align}
\sum_{i=1}^{n} x_i = \text{ceil}(n\frac{p}{q}), \;  n=1,2,...,q.
\end{align}
For example, if $g=2/5$, we have $x_1=\text{ceil}(5/2) =3$, and $x_1 + x_2=\text{ceil}(5)$, implying that $x_2=2$. Thus the ground state is $(323232...)$.

Excited states are obtained by increasing $\epsilon_s(k,g)$ for one or more particles.   Since the particles are identical, we can generate all such configurations by increasing $k$ for some number of particles in a way that preserves the particle ordering in energy space.  As a consequence, we need think only of processes in which a particle changes its energy by an integer amount --i.e. where both the pseudospin of each particle, and the ordering of different pseudospins in energy, are preserved.

The resulting allowed occupancy patterns of a 1D ideal gas of HES particles with $g=q/p$  can be constructed as follows.  
Our elementary building blocks consist of all partitions of the basic occupancy pattern $(x_1, x_2,..., x_q)$, with zeros inserted at arbitrary positions in the sequence.  Excited states are obtained by concatenating these elementary building blocks.  For example, if $g=2/5$ above, the elementary building blocks are formed by first choosing one of the four partitions $\{ [3],[12],[21],[111] \}$ of 3, followed by one of the two  partitions of 2: $\{ [2],[11] \}$.  Explicitly, this gives $[32], [311], [122],[1211],[212],[2111],[1112],[11111]$.  Excited states are described by concatenating these building blocks, with zeros inserted at arbitrary locations in the sequence.  
For $g=2/5$ and $N=6$, the possible concatenations are $[321], [2 1 2 1],[1221],[11121],[3111], [2 1 11 1],[12111]$, and $[111111]$.   If we take the number of cells $d$  to be $5$ the last partition is not allowed, and there are $28$ ways of inserting zeros in the remaining partitions to fully describe the occupancies of all 5 cells, matching the prediction of Eq.(\ref{D_N})that $D_6(2/5, 5)=28$.

As for the $g=1/m$ case described above, it is convenient to introduce a diagrammatic visualization of the allowed occupancy patterns.  For $g=q/p$, for all cells except the first there are $p$ distinct energies (modulo $\omega$), which we represent with $p$ possible pseudospin values.   The first cell has only $\text{ceil}(1/g)$ possible energies, due to the fact that there are no cells below it that can be occupied; hence we assign it only $\text{ceil}(1/g)$ pseudospin values.
In this representation, the constraints are as follows.  First, when we enumerate the particles in order of increasing energy, if the $i^{th}$ particle has pseudospin $\alpha$, then the $i+1^{st}$ particle has pseudospi $\alpha +1\ \text{mod } p$.   In general, the $i+1^{st}$ particle cannot be in a lower-energy cell than the $i^{th}$ particle, i.e. if the $i^{th}$ particle is in the $j^{th}$ cell, then the $i+1^{st}$ particle is in the $k^{th}$ cell with $k \geq j$.  Further, we may have $k=j$ only if
in the ground state the $i^{th}$ and  $i+1^{st}$ particles are in the same cell. 
The ground state and one excited state are shown for the case $g=2/5$ in Figure (\ref{fig4}).

\begin{figure*}[htb]
\includegraphics[height=11cm,width=17cm]{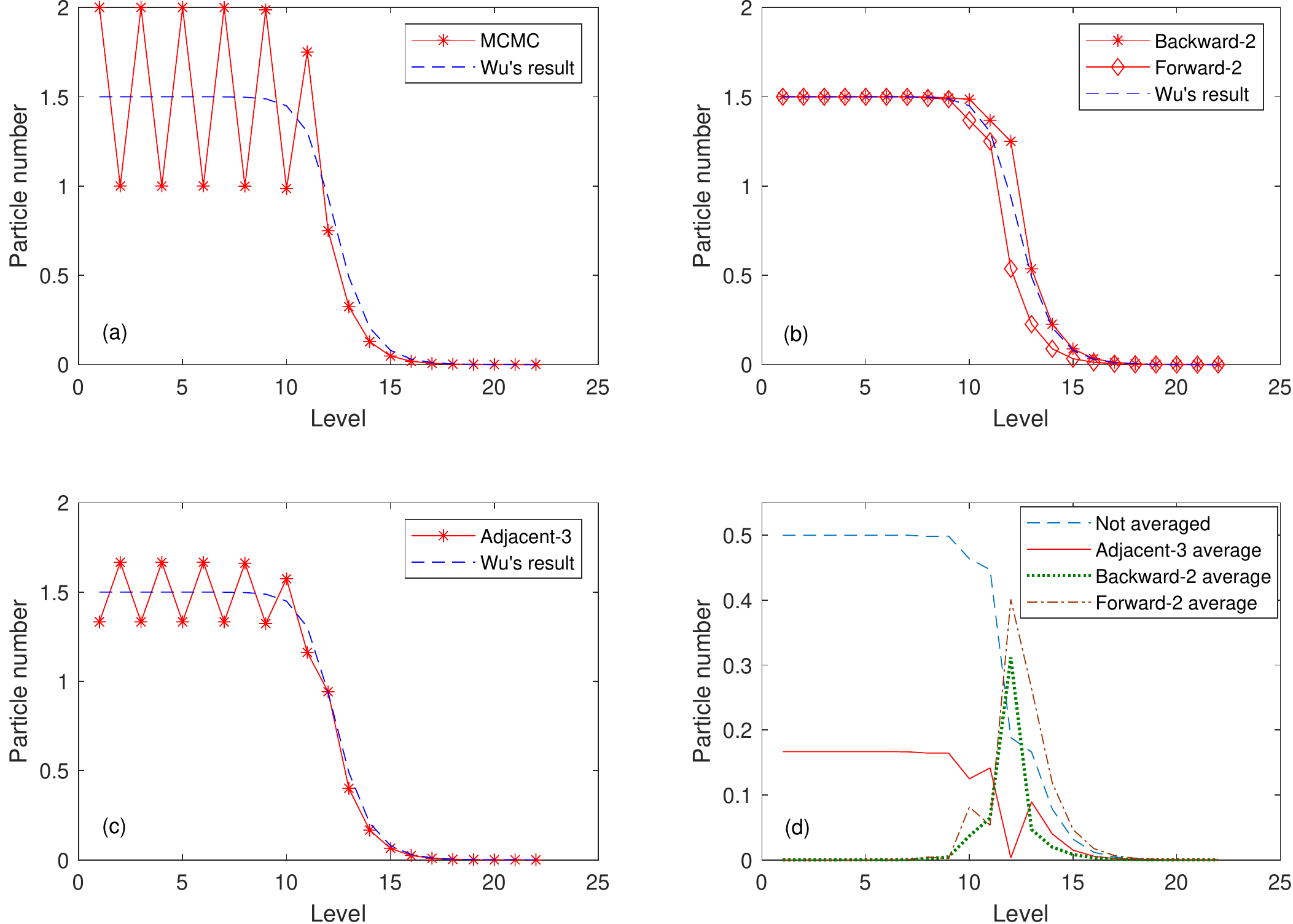}
\caption{(a) Exact distribution function of an ideal gas of $g=2/3$ HES particles at temperature $T= \omega$ compared to Wu's approximation.  The two-cell structure of the ground state (212121......) is not captured by Wu's approximation.  (b) Averaging the exact distribution function over two cells with both a forward ($\bar{n}_i = 1/2(n_i + n_{i+1}$) and backward ($\bar{n}_i = 1/2(n_i + n_{i-1}$) two-cell average gives a distribution that matches well to Wu's result away from the Fermi level.  (c) Averaging over three adjacent cells reduces the magnitude of the oscillations below the Fermi surface, but does not eliminate them.  (d) Difference between the various MCMC averages and Wu's approximation. The error of each MCMC data point is $~0.003$ with $90\%$ confidence.}
\label{fig7}
\end{figure*}

\begin{figure*}[htb]
\begin{tabular}{cc}
\includegraphics[height=5cm]{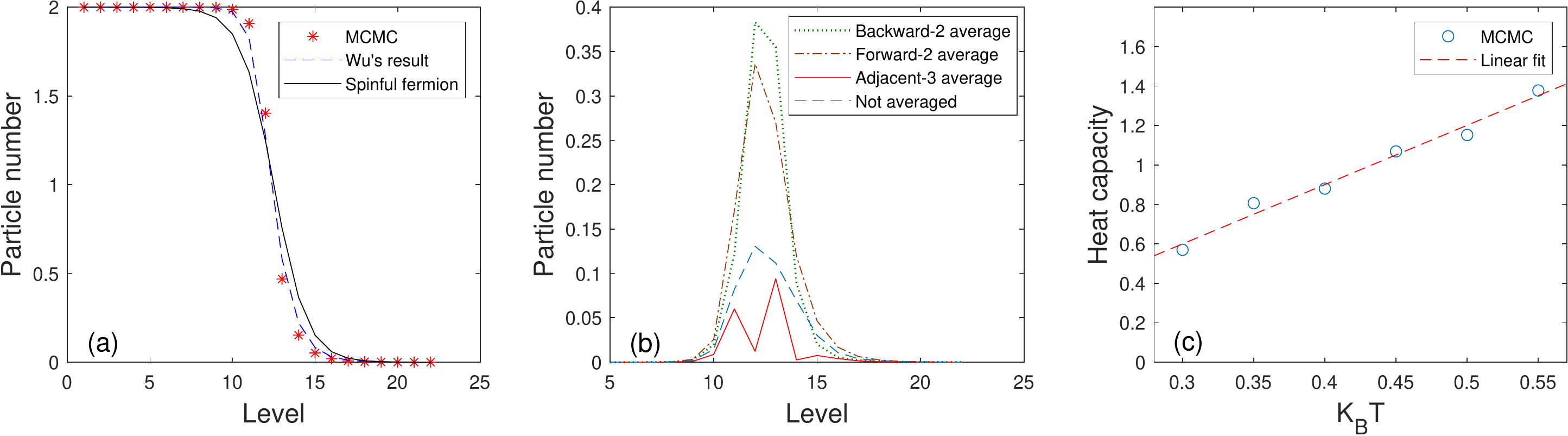} 
\end{tabular}
\caption{(a) Exact distribution functions of $g=1/2$ HES particles with $T = \omega$ calculated using MCMC, compared to Wu's distribution function, and the distribution function of spinful fermions. (b) The absolute values of the differences between the MCMC distribution function of HES particles and Wu's approximation, shown for the exact MCMC distribution, as well as forward ($\bar{n}_i = (n_{i-1}+n_{i})/2$) and backward ($\bar{n}_i = (n_{i}+n_{i+1})/2$) two-cell averages and a three-cell  average ($\bar{n}_i = (n_{i-1}+n_{i}+n_{i+1})/3$). 
(c) Heat capacity vs. temperature (both in units of $\omega$) with linear fit ($R^2=0.985$) for $g=1/2$.   The error of MCMC in this case is $~0.003$ [(a) and (b)] and $~0.06$ (c) with $90\%$ confidence.  
 }
\label{fig6}
\end{figure*}

 In summary, our diagrammatic representation of the many-body eigenstates of the CSM captures the physics underpinning HES as follows.  First, although the underlying particles in the CSM are fermions, by grouping the energy levels into cells of size $\omega$, and identifying $n_i$ as the occupancy of the $i^{th}$ cell, we obtain many-body occupancy patterns that exhibit exact HES.  In this description, HES arises due to inter-cell constraints in the allowed occupancy patterns.  Second, we may conveniently represent the allowed occupancy configurations using a $p$-leg fermionic ladder, with a pseudospin index $\alpha$ to keep track of which leg each particle occupies.  In this representation the contstraints can be described straightforwardly: for example, an  $\alpha =1$ particle must be followed by an $\alpha=2$ particle occupying either the same cell or a cell of higher energy, while an $\alpha = p$ particle must be followed by an $\alpha=1$ particle in a cell of strictly larger energy.

\subsection{Thermodynamics in the limit $K_B T \sim \omega$}\label{Sec:thermo}

\begin{figure}[tb]
\includegraphics[height=7cm,width=8cm]{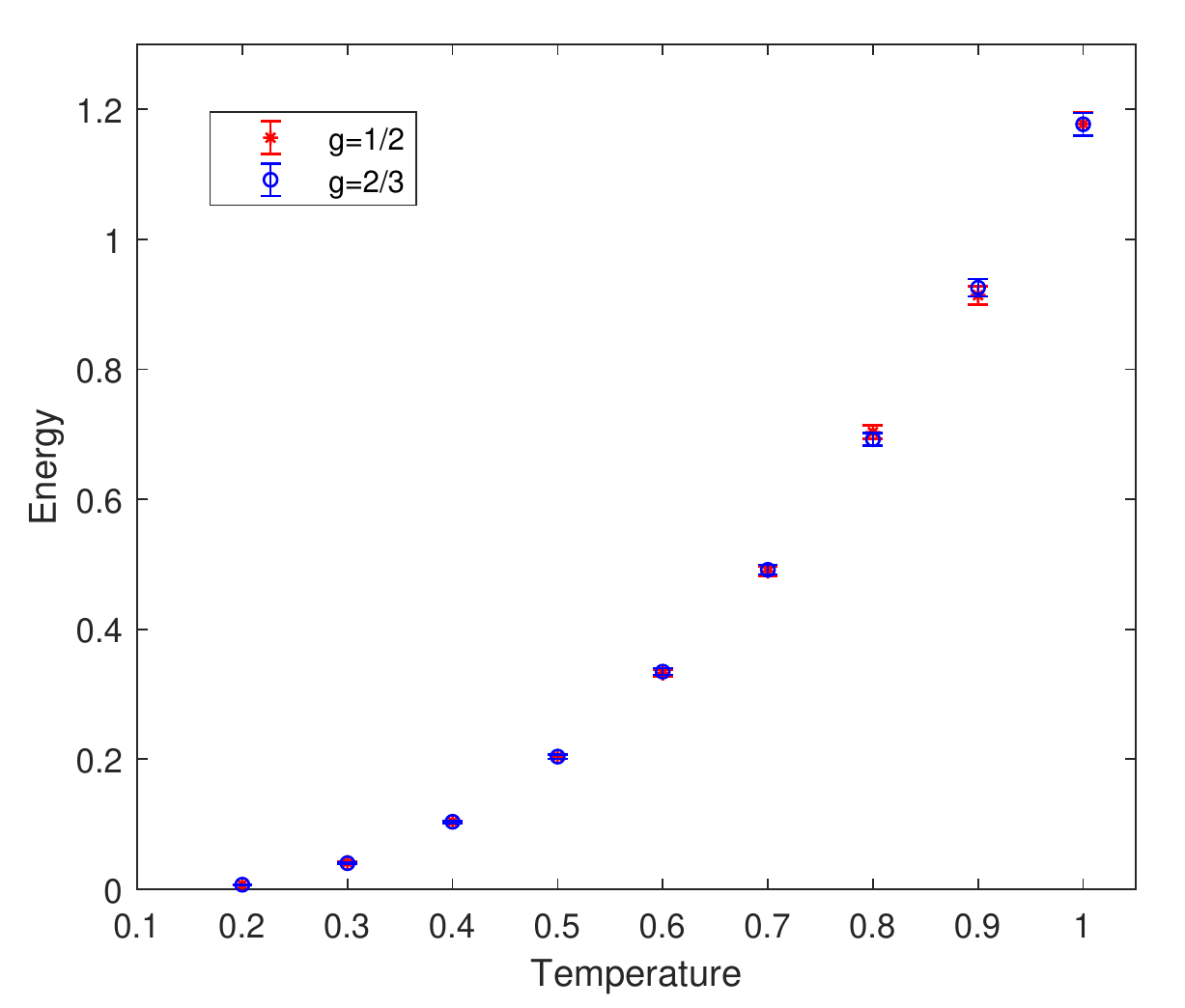}
\caption{Energy expectation values of $g=1/2$ and $g=2/3$ vs temperature, both measured in units of $\omega$.  The results for both values of $g$ agree to within the statistical error.  Error bars indicate a $90\%$ confidence interval. }
\label{fig21}
\end{figure}

The description of the HES Fock space outlined above allows us to evaluate the partition function for HES particles exactly at low temperatures $K_B T \approx \omega$ using a Markov Chain Monte Carlo (MCMC) method. Note that our algorithm assumes that all pseudospin levels in the same cell have the same energy, and is thus a slight simplification of the exact Calogero-Sutherland model. 
We now use this to compare the exact thermodynamics of HES particles with that described by Wu's approximation, in the regime $K_B T \sim \omega$ where the latter is not well-controlled.

Before discussing our exact results, it is worth reviewing what can be deduced about thermodynamics of HES particles on general grounds, and from Wu's approximation.\cite{Nayak1,Sen,Joyce,isakov1994statistical,Iguchi,Potter}   First, 
 the distribution function of HES particles (with $0<g\leq 1$) is necessarily similar to the Fermi-Dirac distribution function, in the sense that both have a Fermi level, with states far below Fermi level fully occupied, and states far above it empty. This simply follows from the existence of a maximum number of particles per cell.    Second, the existence of this Fermi level ensures that the low temperature specific heat scales linearly with  $T$, since only states within $K_BT$ of the Fermi surface can be thermally occupied, with an average energy increase proportional to $K_B T$.\cite{Sen, Iguchi, Iguchi1, Iguchi2}  Both of these features are apparent in the numerical results for $g=1/2$ and $g=2/3$, shown in Figs. (\ref{fig7})-(\ref{fig21}).
 
 However, our numerical results also highlight some features not captured by Wu's approximation, as it assumes that the number of particles in each cell is large.  Notably, for $1/g$ non-integral,  the occupancy of energy cells below Fermi energy is not uniform, because the ground state does not consist of a uniform occupation within each cell.   Figure (\ref{fig7}a) shows this difference for $g=2/3$,  where the ground state is given by $(2,1,2,1,2,1,...)$.  Wu's approximation cannot resolve this fine structure in the occupancy of adjacent energy levels, and instead predicts a uniform plateau at the height of the average occupation number.  We can resolve the discrepancy for states below the Fermi level by averaging the MCMC results over an even number of cells [Figs. \ref{fig7}(b) and \ref{fig7}(c)]; this leads to a distribution that resembles Wu's both well below and well above the Fermi surface, and shows quantitative differences near the Fermi energy.  (Averaging over an odd number of adjacent cells [Fig. (\ref{fig7}c)], in contrast, does not remove the discrepancy.)  
 
Moreover,  for every $g \neq 0,1$, in the regime $K_BT \sim \omega$ we expect quantitative differences with Wu's prediction, as this is precisely the regime where the assumption that the number of particles in each cell (whose energetic extent must be small relative to $K_BT$) is invalid.   Figure \ref{fig6}(b) shows the differences between the two distributions at energies near the Fermi level for $g=1/2$, which is not greatly reduced by averaging over multiple cells.   However, even in this regime such differences are small for all values of $g$ that we have examined.   
 
A final interesting feature not captured by Wu's approximation is that for fixed temperature $T$, the energy expectation values are the same for different species of HES particles (shown for $g=1/2$ and $g=2/3$ in  Figure (\ref{fig21})), where we fix the ground state energy to be zero.   We will elaborate on the origin of this phenomenon in Sec. \ref{Sec:Dynamics}, where we show that this holds for all different values of $g$.   This leads to a surprising result: the thermal energy of the ideal gas of HES particles, relative to the ground state, doesn't depend on $g$.

\section{Second quantization of HES particles}\label{Sec:Second}

In order to study the dynamics of (open) systems with HES, it is convenient to use a second-quantized formalism appropriate to the constrained Hilbert space.  Here we develop such a formalism, based on the exact description of the constrained Hilbert space described in Sec. \ref{Sec:Constraints}.   Several protocols for second quantization of HES particles \cite{karabali,Ilinski,Speliotopoulos97,Meljanac_1999} have been proposed; however many of these do not fully capture the constraints on Fock space.   Specifically, Ref. \cite{Ilinski} introduces a probabilistic treatment of the exclusion constraints or $g=1/m$, treating the occupancy probabilities of different energy levels as independent, and hence does not exclude configuration such as $|... m, m-1, m ... \rangle$, which we have shown are not in the true many-particle Fock space.  The construction of reference \cite{karabali} and \cite{Speliotopoulos97} similarly respects the generalized Pauli exclusion, but does not discuss the structure of the inter-level occupancy constraints.  Ref. \cite{Meljanac_1999} discusses a second quantization specific to the Calogero-Sutherland model, which does fully capture the constraints in Fock space.   
The framework we present here is equivalent to theirs, but has the advantage that it can be interpreted straightforwardly in terms of free fermion operators and constraints, which we will use to understand the impact of constraints on relaxation dynamics in the next section.

\subsection{Basic formulation: definitions of states and  operators}

As described above, the HES particles of the Calogero-Sutherland model with $g=q/p$ can be described using a $p$-leg (or $p$ pseudospin) fermionic ladder.  
Our second quantization procedure thus begins with the set of fermion creation and annihilation operators $f_i^{\alpha}, (f_i^{\alpha})^\dag$, where $i$ indexes the energy of the cell ($\epsilon_i = i \omega$), and $\alpha$ is the pseudospin.   To obtain a second-quantized representation of HES particles, we must account for two differences relative to ordinary fermionic ladders.  First, the number operator $N_i$ associated with HES particles in cell $i$ is described by ignoring the fermion pseudospin $\alpha$, and simply counting the total number of fermions in a given cell $i$:
\begin{equation} \label{Eq:NHES}
N_i=\sum_{\alpha=1}^{p} n^{(\alpha)}_i  \ ,
\end{equation}
where $n^{(\alpha)}_i =  f_i^{\alpha \dag} f_i^{\alpha}$ is the fermion number operator for the state in cell $i$ with pseudospin $\alpha$.

Second, we must impose the occupancy constraints on our Fock space, which we do by projecting states in the fermionic Fock space onto states in the constrained Fock space. 
Thus a state $\ket{\Psi}_h$ in the Fock space of HES particles can be written as, for $g=q/p$, 
\begin{align}\label{Psih}
\ket{\Psi}_h=\prod_{\otimes i}\ket{N_i}=P \prod_{\otimes i}\ket{n_i^{(1)},n_i^{(2)},...,n_i^{(p)}},
\end{align} 
where $P$ is a projector onto states obeying the HES occupancy constraints.

To describe the possible operators acting on this Fock space, it is useful to introduce the creation and annihilation operators  $\hat{h}_i^{\dagger}$ ,$\hat{h}_i$: 
\begin{align}\label{h}
h_i= \sum_{\alpha=1}^{p}  f_i^{(\alpha)} , \;\;\; h_i^{\dagger}= \sum_{\alpha=1}^{p}  f_i^{(\alpha) \dagger}.
\end{align}

The operator $h_i$ creates a superposition of all possible ways that one fermion can be added to cell $i$, {\it before} imposing the constraints.  Physical operators acting within the constrained Fock space of HES particles can be expressed in terms of linear combinations of operators $O_{n,m}$, 
where $O_{n,m}$ is a product of $n$ fermion creation operators and $m$ fermion annihilation operators, projected to the constrained Hilbert space:
\begin{align}\label{hhhh}
O_{n,m} &= P h_{i_1}^{\dagger} h_{i_2}^{\dagger} ... h_{i_n}^{\dagger} h_{j_1} h_{j_2} ... h_{j_m} P \nonumber \\
& =  \sum_{\{\alpha_k\}} P f_{i_1}^{(\alpha_{i_1}) \dagger} f_{i_2}^{(\alpha_{i_2}) \dagger} ... f_{i_n}^{(\alpha_{i_n}) \dagger} f_{j_1}^{(\alpha_{j_1})} ... f_{j_m}^{(\alpha_{j_m})} P.
\end{align} 
Here the sum in the second line is over all possible pseudospins, and as above $P$ projects onto states satisfying the constraint.
 For example, the HES number operator $\hat{N}_i$ can be expressed:
\begin{align}\label{Nocc}
N_i \equiv P h_i^{\dagger} h_i P&=&\sum_{\alpha=1}^{p} P f_i^{(\alpha)\dagger} f_i^{(\alpha)} P + \sum_{\alpha \neq \beta} P f_i^{(\alpha)\dagger} f_i^{(\beta)} P \ .
\end{align}
The second term is zero because the constraints do not allow pseudospin- changing processes.  The first term does not change the occupancy of any pseudospin level, and thus is always non-zero provided it acts on a state within the constrained Fock space.  Thus we recover the expression above,  $N_i = \sum_{\alpha=1}^{p} n^{(\alpha)}_i P$.  [In Eq. (\ref{Eq:NHES}), we have dropped the projector, as we are implicitly assuming that $N_i$ acts only on states within the constrained Fock space.]

\subsection{Tools for calculating matrix elements of HES operators}

For practical purpose, we would like to derive a general expression describing how an arbitrary second-quantized operator of HES particle acts on states in the constrained Fock space. This must be done with care, since 
\begin{align}
& \sum_{\{\alpha_k\}} P f_{i_1}^{(\alpha_1) \dagger} f_{i_2}^{(\alpha_2) \dagger} ... f_{i_{n-1}}^{(\alpha_{n-1})} f_{i_n}^{(\alpha_n)} P \nonumber \\
  \neq & \sum_{\{\alpha_k\}} P f_{i_1}^{(\alpha_1)  \dagger} P f_{i_2}^{(\alpha_2) \dagger} P ... P f_{i_{n-1}}^{(\alpha_{n-1})} P f_{i_n}^{(\alpha_n)} P.
\end{align} 
For example, the $O_{1,1}$ operator $P h^{\dagger}_i h_i P$  cannot be expressed as a product of an $O_{1,0}$ operator and an $O_{0,1}$ operator.  This is because such products have the form  $P h^{\dagger}_i P h_i P$, but  $P h_i P=\sum_{\alpha}P f_i^{\alpha} P$ is zero unless  $h_i$ annihilates a particle in the highest energy occupied cell, since in all other cases it fails to preserve the pseudospin-energy ordering.  On the other hand $P h_i^{\dagger} h_i P$ is non-zero acting any state in which there is at least one particle in the $i^{th}$ cell, irrespective of the occupancy of cells with higher energy.  

Here we describe some tools to quickly  determine whether an arbitrary operator $P f_{i_1}^{(\alpha_1) \dagger} f_{i_2}^{(\alpha_2) \dagger} ... f_{i_{n-1}}^{(\alpha_{n-1})} f_{i_n}^{(\alpha_n)} P$ acting on any state $\ket{\Psi}_h$  is zero or not.  If it's non-zero, the associated matrix elements are the same as those of the relevant fermionic operator.  
We will first consider single-particle processes, described by operators of the form $P f_{i}^{(\alpha) \dagger} f_{j}^{(\beta)} P$.  These are the processes relevant to relaxation dynamics, which we will study in Sec. \ref{Sec:Dynamics}.  We then briefly comment on the more general case.

\begin{figure*}[tb]
\includegraphics[height=3.5cm,width=16cm]{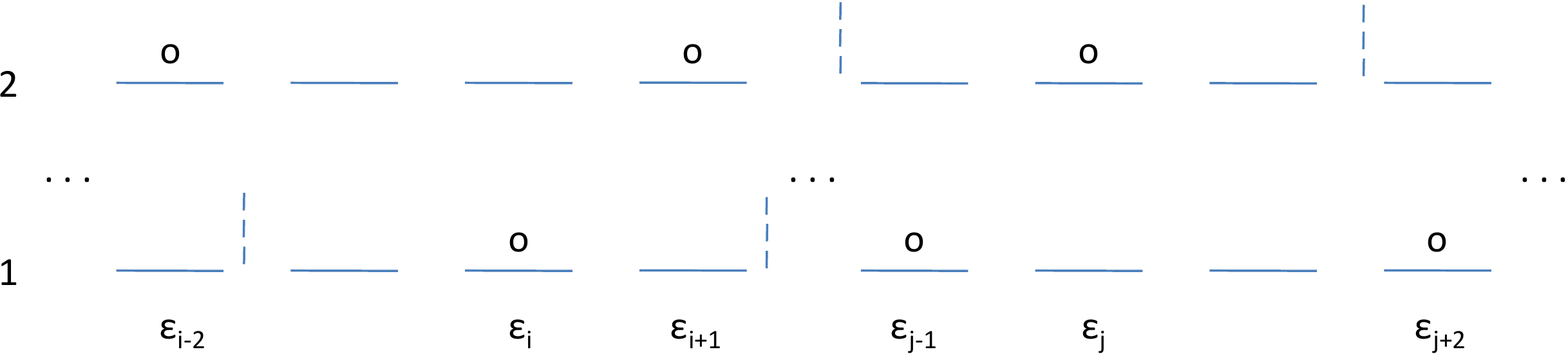}
\caption{ In single-particle processes with $g=1/2$, the $\alpha=1$ particle with energy $\epsilon_i$ can only move inside a box whose boundary is fixed by the nearest $\alpha=2$ particles.  In the example shown these have energies $\epsilon_{i+1}$ and $\epsilon_{i-2}$, and the box is given by $\epsilon_{i-2}< \epsilon \leq \epsilon_{i+1}$. Similarly, an $\alpha=2$  particle of energy $\epsilon_j$ is constrained to move inside a box whose boundaries are fixed by the positions of the nearest $\alpha=1$ particles.  In a configuration where these have energies $\epsilon_{j-1}$ and $\epsilon_{j+2}$, this box includes the levels $\epsilon_{j-1} \leq \epsilon < \epsilon_{j+2}$. }
\label{fig9}
\end{figure*} 

\subsubsection{Single-particle processes}

 We begin by considering operators of the form $P f_{i\pm 1}^{(\alpha) \dagger} f_{i}^{(\beta)} P$, with $(g=1/2, \alpha, \beta =1,2)$, which describe processes that ``hop" fermions between cells $i$ and $i+1$, taking a many-body state of energy $E$ to one with energy $E\pm \omega$.  
 Within the constrained Fock space, such processes must satisfy two conditions.  First,  we require that $\alpha= \beta$.  
 This is because for $\alpha \neq \beta$ the resulting final states do not obey the correct ordering of pseudospin relative to energy.   
 Second,  in order to maintain the correct pseudospin order, an $\alpha=1$ fermion cannot hop past an $\alpha=2$ fermion, and vice versa.  More specifically, an $\alpha=1$ fermion 
 in the $i^{th}$ cell can move to the $i+1^{st}$ cell only if  $n_i^{(2)}=0$; it can move to the $i-1^{st}$ cell only if  $n_{i-1}^{(2)} = 0$.   Similarly an $\alpha=2$ fermion in the $i^{th}$ cell can move to the $i+1^{st}$ cell only if  $n_{i+1}^{(1)}=0$, and to the $i-1^{st}$ cell only if  $n_{i}^{(2)} = 0$.  For example, in Fig.\ref{fig1}(a) the $\alpha=1$ particle in  cell $2$ cannot hop to cell $3$, while in Fig.\ref{fig1}b), the $\alpha=2$ particle in cell  $4$ cannot hop to cell $3$.  (In contrast, the $\alpha = 2$ particle in Fig.\ref{fig1}(a) [$\alpha = 1$ particle in Fig.\ref{fig1}(b)] can hop up (down) in energy.)  

The remaining matrix elements of the operator $P f_{i\pm 1}^{(\alpha) \dagger} f_{i}^{(\alpha)} P$, when acting on states 
 $\ket{\Psi}_h \equiv P \ket{..., N_i, N_{i+1}, ... } =P \ket{..., n_i^{(1)}, n_i^{(2)}, n_{i+1}^{(1)}, n_{i+1}^{(2)}, ... } $, can be obtained by dropping the projectors $P$ and using the usual fermionic operator relations.  For example,
\begin{widetext}
\begin{align}\label{matrix1_2}
P f_{i+1}^{(1) \dagger} f_{i}^{(1)} P \ket{..., n_i^{(1)}, n_i^{(2)}, n_{i+1}^{(1)}, n_{i+1}^{(2)}, ... } &=  (-1)^{n_i^{(2)}} n_i^{(1)} (1-n_{i+1}^{(1)})  P \ket{..., n_i^{(1)}-1, n_i^{(2)}, n_{i+1}^{(1)}+1, n_{i+1}^{(2)}, ... }\\
     &= n_i^{(1)}  (1-n_{i+1}^{(1)})(1-n_i^{(2)})  P \ket{..., n_i^{(1)}-1, n_i^{(2)}, n_{i+1}^{(1)}+1, n_{i+1}^{(2)}, ... }.
\end{align}
\end{widetext}
In the last line, we have used the fact that, if the initial configuration is allowed by the constraints, the hopping process creates an allowed configuration only if $n_i^{(2)} =0$; this ensures that the matrix element is always positive.
Similar expressions hold for the $\alpha=2$ process, and for hopping processes to cells of lower energy (see Appendix \ref{Sec:Appe}).

We can now straightforwardly generalize this analysis to processes of the form $P f_{j}^{(\alpha) \dagger} f_{i}^{(\beta)} P$, and general $g$.   First, as above, the operator is non-zero only if $\alpha=\beta$.  Second, processes in which a particle of pseudospin $\alpha$  hops past a particle of  pseudospin $\alpha\pm 1$ fail to preserve  the order of pseudospins relative to energy.   Here our definition of ``past" includes a particle of lower (higher) pseudospin in the same cell for processes that decrease (increase) the energy.   In other words, we find that single-particle processes can never hop a particle past any other particle.  As a consequence, the matrix elements of non-vanishing single-particle operators are always 1, as they never exchange fermions.   
In Appendix \ref{Sec:Appe}, we give an example of how the above rules can be used to compute $P h_{p+1}^{\dagger} h_pP \ket{\Psi}_h$ for $g=1/3$.

\begin{figure*}[tb]
\includegraphics[height=11cm,width=15cm]{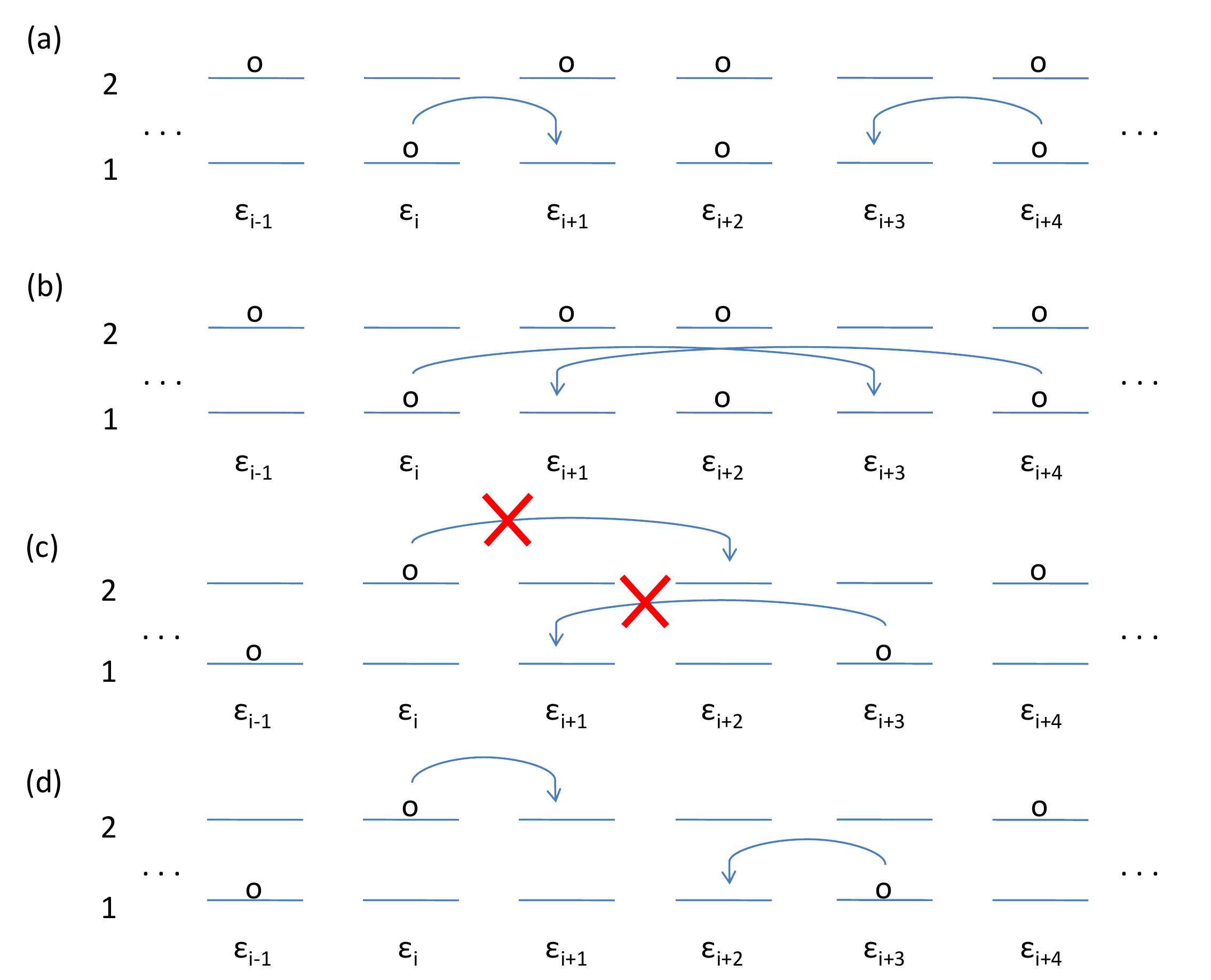}
\caption{  
Allowed (a,b, and d) and forbidden (c) two-particle hopping process. (a) $ P f_{i+1}^{(1)\dagger} f_{i+3}^{(1)\dagger} f_{i+4}^{(1)}  f_{i}^{(1)} P$ and (b) $P f_{i+3}^{(1)\dagger} f_{i+1}^{(1)\dagger} f_{i+4}^{(1)}  f_{i}^{(1)} P$ are allowed two-particle processes connecting the same initial and final states, which differ by a relative $-$ sign.   
(c) represents an illegal two-particle hopping, because the final state is not in the Fock space of HES particles; in contrast, process (d) $P f_{i+1}^{(2)\dagger} f_{i+2}^{(1)\dagger} f_{i+3}^{(1)}  f_{i}^{(2)} P$, which preserves the pseudospin-energy ordering, is allowed.  
}
\label{fig10}
\end{figure*}

A convenient way to describe the allowed single-particle hopping processes is to define a "box" for each particle.  The left (right) boundary of the box is fixed at the position of  the nearest particle on the left (right), which necessarily has pseudospin $\alpha-1$ ($\alpha+1$) modulo $p$ for $g = q/p$.  The hopping constraints for single particle processes can then be viewed as a constraint that the particle cannot hop past the boundaries of this box.  
  (An example is shown in Fig.(\ref{fig9}).) 
We define the {\it box number} $\ell$ of a particle of pseudospin $\alpha$ by counting the number of particles with the same pseudospin, but lower energy.  Enumerating the particles in order of increasing energy,  the 1st $p$ particles  have box number $\ell =1$, the 2nd $p$ particles have box number $\ell =2$, and so on.   Because each particle can hop only within its box, these box numbers -- as well as the pseudospin ordering within each box -- are necessarily conserved under any dynamics that respects the constraints.   

It is straightforward to see that this description captures the structure of excited states described in Sec. \ref{Sec:Constraints}.  Recall that the ground state of  ideal gas of $g=q/p$ HES particles is described by the occupancy pattern $(x_1, x_2, ... , x_q )$ ($x_1 + x_2 +...+ x_q=p$), repeated for all cells below the Fermi level (except possibly the last, which may be partially occupied).   The excited states are obtained by moving particles above the Fermi level while conserving both the pseudospin ordering in each box, and the box number of each particle -- in other words, they are obtained by moving the boundaries of each box while preserving the particle ordering.  This leads to occupancy patterns obtained by concatenating the integer partitions of $(x_1, x_2, ... , x_q )$, and inserting zeros at arbitrary points in the sequence.

\subsubsection{Multi-particle processes}

We now turn to more generic particle-number-conserving operators, of the form $O_{n,n} =
 P \prod f_{i_1}^{(\alpha_1) \dagger} ... f_{i_n}^{(\alpha_n) \dagger} f_{j_{1}}^{(\beta_{1}) } ...  f_{j_n}^{(\beta_n) } P$. 
For general $g=q/p$, the number of particles of each pseudospin must be conserved; otherwise, the pseudospin-energy ordering cannot be preserved.  
Further, suppose that our operator destroys a set of particles with pseudospin $\{ \alpha_k \}$, and box numbers $\{\ell_k\}$.    There are two possibilities: first, our operator could delete an integer number of boxes.  The resulting occupancy pattern will also satisfy the constraints, and creation operators can then re-insert complete boxes at arbitrary energies, as long as they do not ``chop up" any of the existing boxes.  Second, our operator could annihilate only some of the particles in box $\ell_i$.  In this case the resulting occupancy pattern is not in the constrained Fock space.    To ensure the final state satisfies the occupancy constraints, acting with the creation operators must fill the holes in box $\ell_i$.   In other words, a general particle-number-conserving operator can be viewed as describing multi-particle hopping processes in which each particle's pseudospin and box number are conserved. However, unlike the case of single-particle-hopping operators $P f_{j}^{(\alpha) \dagger} f_{k}^{(\alpha)} P$, the matrix elements of multiple-particle-hopping operators can involve an odd number of fermion exchanges, and thus can be negative.  Figure (\ref{fig10}) shows some examples of two-particle-hopping processes for $g=1/2$.

Operators that do not conserve particle number can be expressed as a product of a particle-number-conserving operator, times some numbers of either only creation or only annihilation operators, such as  $f_{i}^{(\alpha) \dagger} f_{j}^{(\beta) \dagger}$ and $f_{i}^{(\alpha)} f_{j}^{(\beta)}$. The whole operator acting on a state is non-zero only  if both components of the operator acting individually on the state give non-vanishing results.   For the part consisting of only creation (only annihilation) operators to be non-zero acting the state, it must either  create (annihilate) 
a set of particles with energy strictly greater than the existing (remaining) particles in the system, or insert (delete) a complete box into the system.

The above rules are sufficient to determine whether a general operator of the form $P f_{i_1}^{(\alpha_1) \dagger} f_{i_2}^{(\alpha_2) \dagger} ... f_{i_{n-1}}^{(\alpha_{n-1})} f_{i_n}^{(\alpha_n)} P$ annihilates a state $\ket{\Psi}_h$.  If $\ket{\Psi}_h$ is not annihilated, the relevant matrix element can be calculated from the usual calculus of the fermionic operators $f_{i}^{(\alpha)}, f_j^{(\beta) \dag}$.

\section{Dynamics}\label{Sec:Dynamics}

\begin{figure*}[tb]
\includegraphics[height=10.5cm,width=16cm]{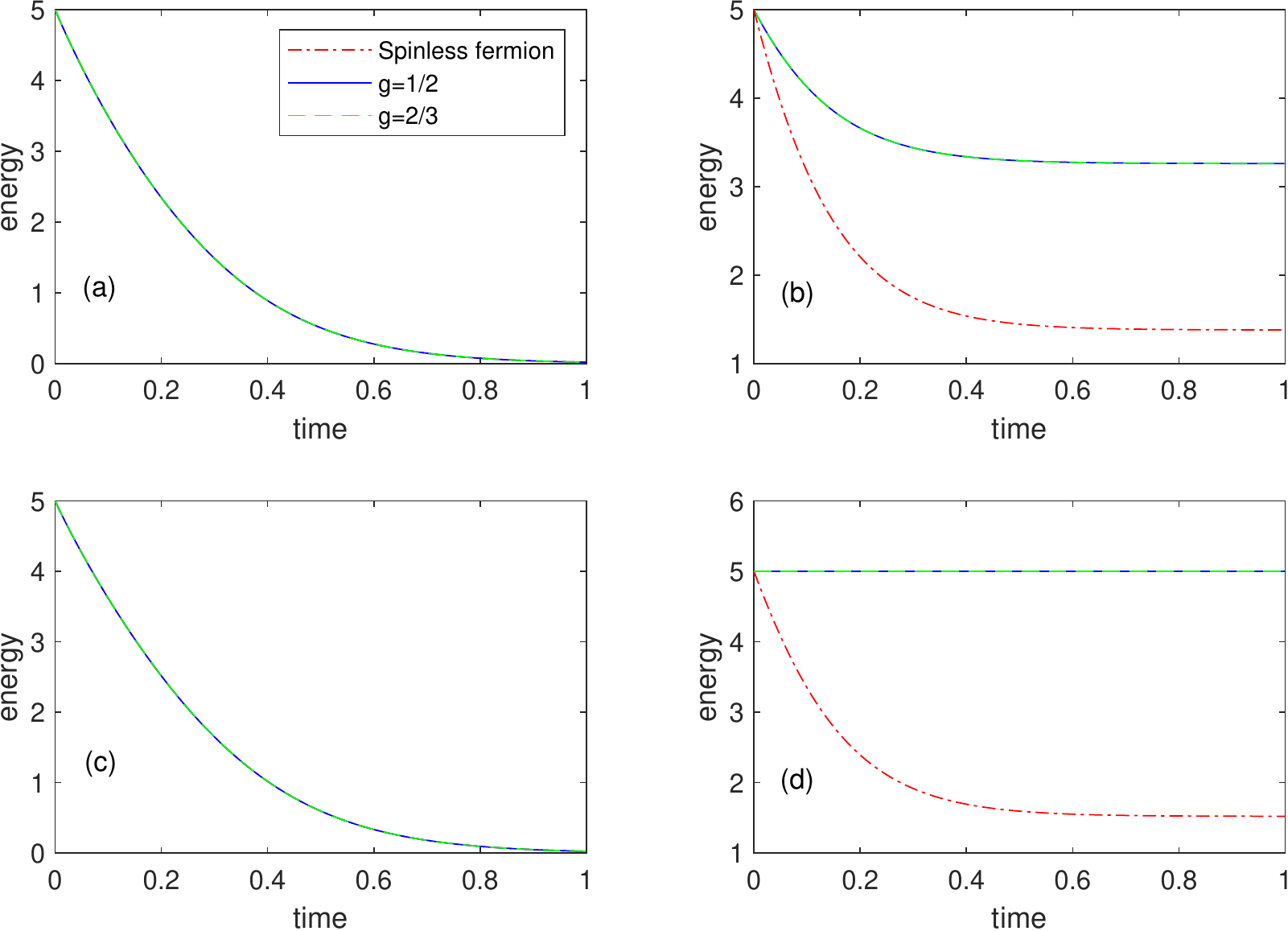}
\caption{
Energy expectation value versus time at temperature $T=0.01\omega$.  In the top two panels, the initial density matrix is a random one in the subspace of $E=5$. In the bottom two panels, we start with a random density matrix in the subspace spanned by the three states $[221]_e$, $[2111]_e$ and $[11111]_e$.(This notation of states are introduced in Sec.(\ref{Sec:UniLin})).  In (a) and (c), the only interaction channel turned on is $k=1$. In (b) and (d), the only interaction channel turned on is $k=2$. In all four cases, the curve for $g=1/2$  exactly falls on the top of the curve for $g=2/3$. Time and energy  are in units of $1/\omega$ and $\omega$ respectively.   In all four cases, the curve for $g=1/2$  exactly falls on the top of the curve for $g=2/3$. Time and the energy  are in units of $1/\omega$ and $\omega$ respectively,  and the rate is taken to be $\Gamma_0=10\omega$. }
\label{fig28}
\end{figure*}

\begin{figure*}[tb]
\includegraphics[height=10.5cm,width=16cm]{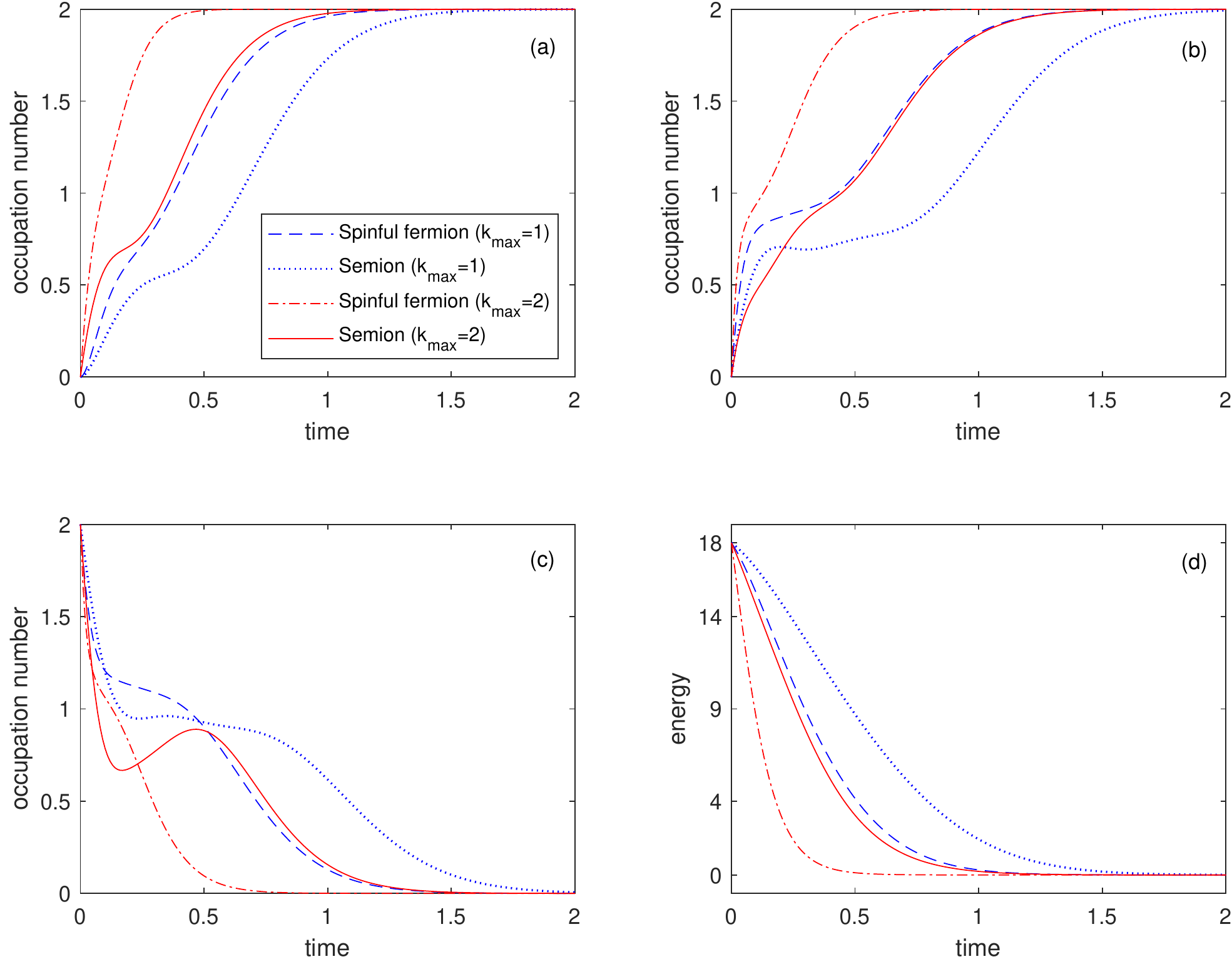}
\caption{Time evolution of energy and occupation numbers near the Fermi level in the low- temperature limit ($T=0.01\omega$), for both $k_{\text{max}}=1$ (a single relaxation channel), and $k_{\text{max}}=2$ (energy changes of $\omega$ and $2 \omega$ are allowed). Time and energy  are shown in units of $1/\omega$ and $\omega$ respectively.   The initial state has six particles fully occupying the cells of energy $3\omega$, $4\omega$ and $5\omega$, with the lowest energy level being $0$.   The rate $\Gamma_0=10\omega$. (a) Occupation number of the level just below the Fermi level vs. time.  (b) Occupation number of the Fermi level vs. time. (c) Occupation number of the level just above the Fermi level vs. time.  (d) Total energy vs. time. }
\label{fig22}
\end{figure*}

\begin{figure*}[tb]
\includegraphics[height=10.5cm,width=16cm]{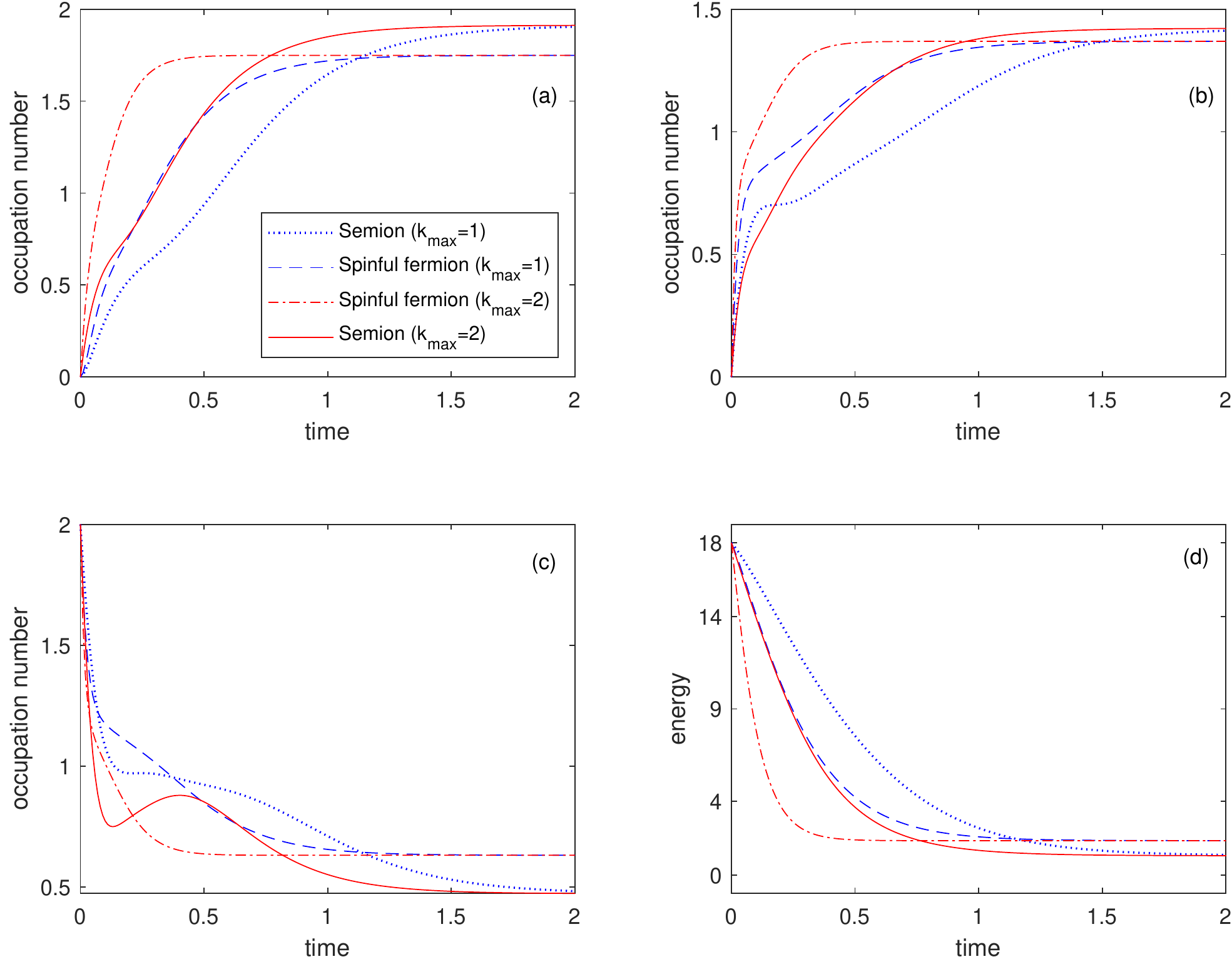}
\caption{Time evolution of energy and occupation numbers near the Fermi level in the intermediate- temperature limit ($T=\omega$), for both $k_{\text{max}}=1$ and $k_{\text{max}}=2$. (a) Occupation number of the level just below the Fermi level vs. time.  (b) Occupation number of the Fermi level vs. time. (c) Occupation number of the level just above the Fermi level vs. time.  (d) Total energy vs. time. The initial state and all other parameters are the same as in Fig.(\ref{fig22}).}
\label{fig24}
\end{figure*}

In the previous section, we have shown how HES in the Calogero-Sutherland model is intimately linked to the emergence of constraints in the allowed occupancy configurations of the many-body Hilbert space.  We have also shown that these constraints have a minor impact on properties in thermodynamic equilibrium.

We now turn to an area where the constraints can be expected to have a more noticeable physical impact -- namely, dynamics.  Because the Calogero-Sutherland model is integrable, here we will study open-system dynamics, in which an ideal gas of HES particles is coupled to an external bath.  Our main focus will be on understanding how the constraints affect relaxation times in our HES system, relative to systems of free fermions.  As one might expect, we find that the constraints typically slow relaxation times relative to fermionic systems-- though, surprisingly, the magnitude of this effect is independent of $g$.  We also find an interesting exception: when the bath can only increase or decrease the energy of the system by $\omega$, we show that for any rational $g \in (0,1]$, the relaxation rate is equivalent to that of an ideal gas of spinless fermions.

Here, we analyze energy relaxation in HES systems using the Lindblad formalism, which  successfully models relaxation of energy (among other observables) provided that (1) the system-bath coupling is weak compared to the scales of both system and bath Hamiltonains, and (2) the bath is well-approximated as a thermal resevoir, meaning that the relaxation time-scale of the system is long compared to the correlation time of the bath.  In practice, these conditions are experimentally realized in a number of cold-atom systems (see, e.g., Refs. [\cite{Raitzsch2009,Barreiro2011}]); here we assume a bath and system-bath coupling with these properties, and investigate the resulting energy relaxation dynamics of HES particles.  


\subsection{Review of the Lindblad formalism} 

We first briefly review the Lindblad formalism (see Ref. \cite{Manzano2020} for a pedagogical overview).  Suppose that  our ideal gas of HES particles is coupled to an external (bosonic) bath, via the Hamiltonian:
\begin{align} \label{Hdynamics}
H=& H_{\text{S}} + H_{\text{B}} + H_{\text{BS}} \nonumber \\ 
= &\sum_{i} \epsilon_i P h_i^{\dagger} h_i P  + \sum_{j} \epsilon_j b_j^{\dagger} b_j + \lambda \sum_{i,k} ( b_k^{\dagger} P h_i^{\dagger}  h_{i+k} P + H.c. ),
\end{align}
where $\epsilon_n = n \omega$ ($n=0,1,2...$), $P h_i^{\dagger} h_i P$ and $P h_i^{\dagger}  h_{i+k} P$  are given by Eq.(\ref{hhhh}), and $b_j^{\dagger}$ ($b_j$) creates (annihilates) a boson with energy $\epsilon_j$. For simplicity, we assume that the coupling constant $\lambda$ is independent of the energy level.  It is convenient to split the interaction between the system and the bath into different channels according to the value of $k$. For instance, for the channel of $k=1$, particles in the ideal gas system can only move up or down by one energy unit $\omega$.

 In general, when  the coupling is weak and the bath can be treated as a thermal resevoir (which we will assume here), the time evolution of the density matrix of our system can be described using the Lindblad formalism.\cite{Breuer}   For a Hamiltonian of the form in Eq. (\ref{Hdynamics}), the Lindblad equation has the form 
\begin{align} \label{LindbladEq}
\frac{d}{d t} \rho =& -i [H_0, \rho] \nonumber \\
&+ \sum_{k,j,a = \pm} \Gamma_{ka}  \left[  L_{kja} \rho L_{kja}^{\dagger}  - \frac{1}{2} \{ L_{kja}^{\dagger} L_{kja},  \rho \}  \right],
\end{align}
where $H_0$ is the Hamiltonian of the ideal gas system, and the sum runs over all energy levels $j$ and interaction channels $k$.  Here $a=+$ for a process that creates excitations in the system, while $a=-$ for a relaxation process.  Explicitly, the associated Lindblad operators are:
\begin{align}
 L_{kj+}= P h_{j+k}^{\dagger} h_j P  \ , \ \ L_{kj-}= P h_{j}^{\dagger} h_{j+k} P \ .
 \end{align}     
 The rates $\Gamma_{k+}$ and $\Gamma_{k-}$ are given by
\begin{align}
 \Gamma_{k+}= \Gamma_0  n_B(k\omega) \ , \ \ \Gamma_{k-}= \Gamma_0 ( 1+ n_B(k\omega))  \ ,
 \end{align} 
 where $\Gamma_0 = 2\pi |\lambda|^2 D$ ($D$ is the density of states in the bath), and $n_B(\epsilon)$ is the Bose-Einstein distribution function.  

In practice, we will solve the Lindblad equation (\ref{LindbladEq}) by considering a finite set  $\{ k\leq k_{\text{max}} \}$ of interaction channels between the  system and  bath, such that the allowed transitions have energy $k \omega$, with $k \leq k_{\text{max}}$.  For systems where only energy cells within a finite distance of the Fermi level can deviate from their ground-state occupation numbers, we can solve Eq. (\ref{LindbladEq}) numerically by simulating a system with a finite number  of energy cells. 
We expect that the dynamics will be well-approximated by such a finite-level system when $\beta \omega$ is larger than unity, such that excitations in the bath are exponentially suppressed with $k$, and the equilibrium distribution of our HES particles deviates from the ground state only on a finite number of levels near the Fermi energy.

\begin{figure*}[tb]
\includegraphics[height=10.5cm,width=16cm]{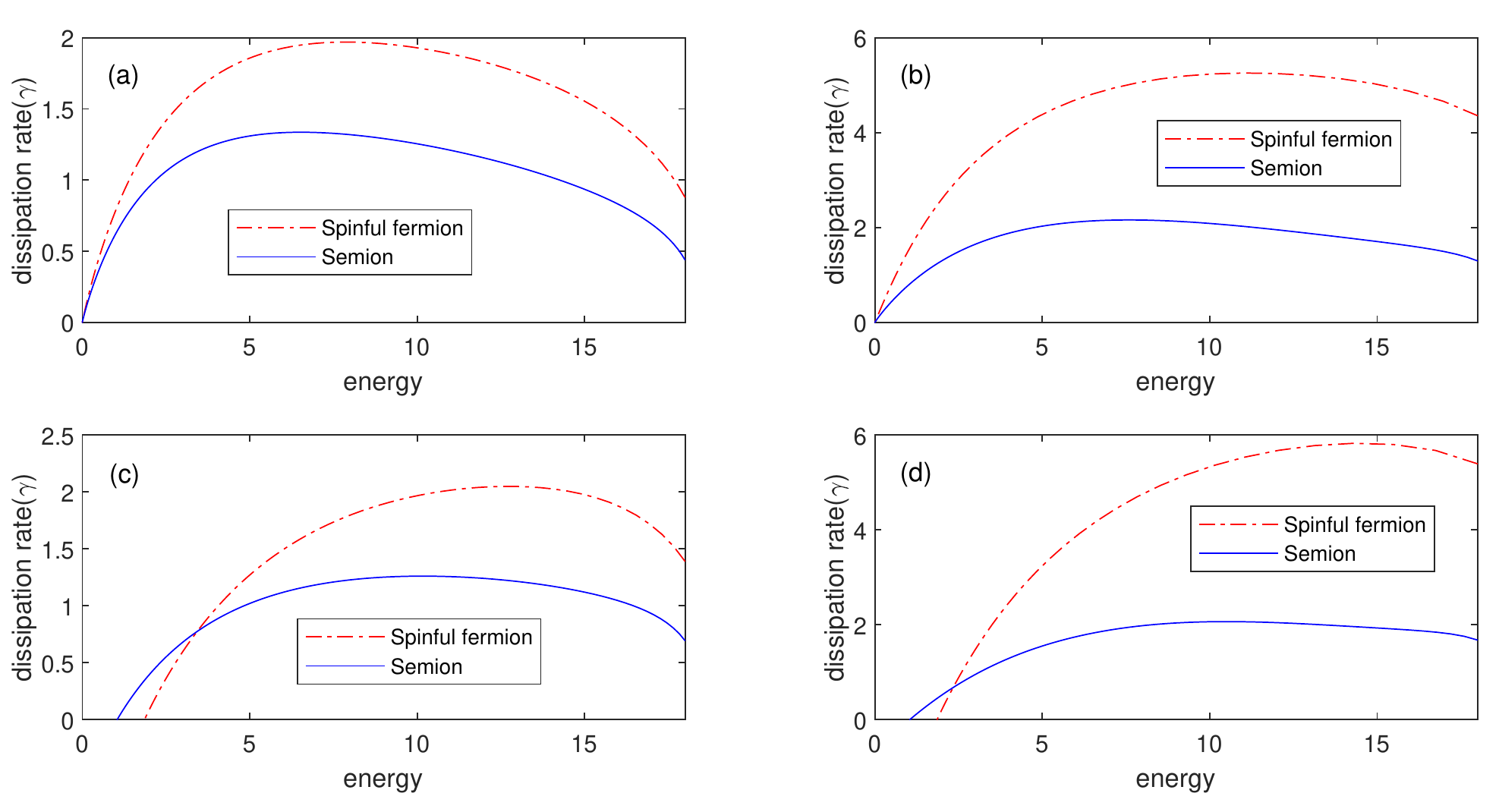}
\caption{  Energy dissipation rate $\gamma \equiv - d \log (E(t)) /dt$ vs. energy $E(t)$, as extracted from the time-dependent relaxation dynamics of both spinful fermions and semions. (a) $T=0.01\omega$, $k_{\text{max}}=1$. (b) $T=0.01\omega$, $k_{\text{max}}=2$. (c) $T=\omega$, $k_{\text{max}}=1$. (d) $T=\omega$, $k_{\text{max}}=2$. The energy of the many-body ground state is assumed to be zero. Both quantities are plotted in units of $\omega$. The initial state and all other parameters are the same as Fig.(\ref{fig22}).}
\label{fig26}
\end{figure*} 

\begin{figure*}[htb]
\includegraphics[height=10.5cm,width=16cm]{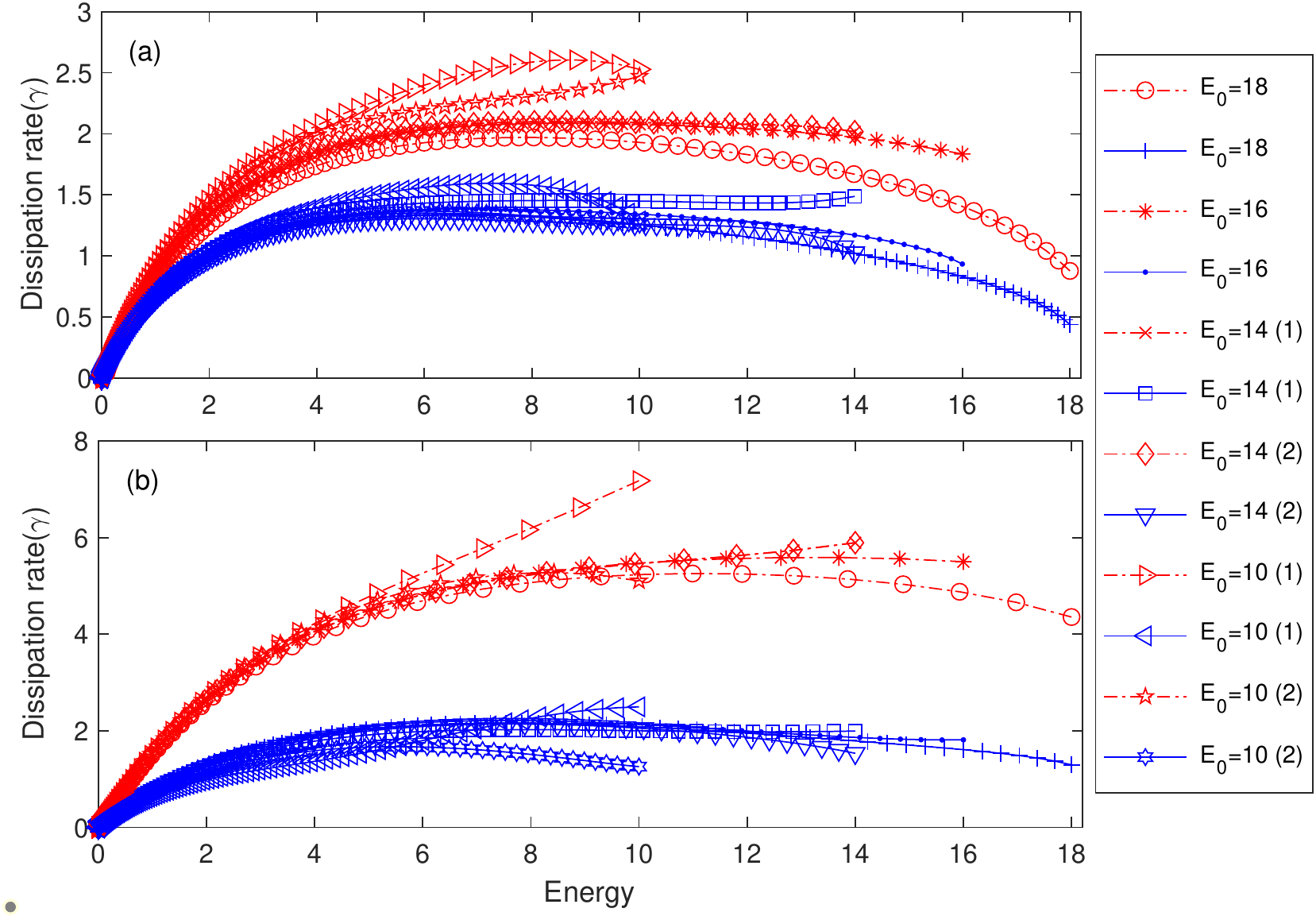}
\caption{Energy dissipation rate $\gamma \equiv - d \log (E(t)) /dt$ vs. energy $E(t)$, for  $g=1/2$ (blue) and spinful fermions (red), with (a) $k_{\text{max}}=1$  and (b) $k_{\text{max}}=2$ at $T=0.01\omega$.   Here we compare curves resulting from the time evolution of several different initial states, including states with different initial energies (corresponding to the largest projection of data points onto the $x$-axis), and different initial states with the same total energy.  Note that the $y$-axis scale in (a) goes from 0 to 3, while in (b) it goes from 0 to 8. The rate $\Gamma_0=10 \omega$.}
\label{fig30}
\end{figure*}

\subsection{Lindblad relaxation dynamics of HES systems}

To illustrate how constraints can have a dramatic qualitative impact on relaxation processes, we begin with a somewhat contrived example.   Suppose that, rather than coupling our system to a many-body thermal bath, we couple it to a system of oscillators with a discrete spectrum, such that the allowed transitions in the bath can change the energy only by multiples of $2 \omega$.  
We initialize our HES system in a high-energy many-body eigenstate for which the difference  in energies between any two successive occupied levels is at most $2 \omega$.    In terms of the underlying fermions, such states correspond to clusters of some number $l$ of consecutive occupied orbitals, separated by a single unoccupied orbital.  Since any given fermion can change its energy only by a multiple of $2 \omega$, relaxation processes require ``hopping" a fermion past another fermion in energy space -- i.e. 
 by having the second-lowest energy particle in a given cluster decrease its energy by $2\omega$.   However, this transition is not allowed for HES particles, since it fails to respect the pseudospin-energy ordering; hence this particular initial configuration is stable.  Thus, if we allow for an appropriately fine-tuned bath (or fine-tuned couplings to the bath), the constraints associated with HES can lead to high-energy excited states that cannot relax at all.  

This phenomenon is illustrated for $g=1/2$ and $g=2/3$ in Figs.  \ref{fig28}(b) and \ref{fig28}(d), which compare the time evolution of these two HES systems to that of spinless fermions $(g=1)$, with only the $k=2$ channel (i.e. only transitions that change the energy by $2 \omega$).  In Fig.  \ref{fig28}(b), the system is initiated in a random initial state with energy $5\omega$.  In Fig.  \ref{fig28}(d),  we show the time evolution of three of these states, which for $k=2$ are completely unable to relax for HES particles due to the constraint. For $g=1/2$, in our particle occupancy representation, these excited states are $[...2221102000...]$, $[...2221111000...]$ and $[...222122000...]$, with the ground state being $[...22222000...]$.  For $g=2/3$, with ground state $[...212121000..]$, they are $[...21210111000...]$, $[...21201201000...]$ and $[...2111121000...]$.  Though the similarities between these states are not apparent in the occupancy representation that we use here, in Sec. (\ref{Sec:UniLin}) we will introduce an energy-based representation of the many-body states, in which these correspond to the same states, $[221]_e$, $[2111]_e$ and $[11111]_e$.  We also show the time evolution of the corresponding states for spinless fermions, which in occupancy representation are $[...11101011000...]$, $[...111011101000...]$ and $[...111011111000...]$.   In the channel $k=2$, the fermionic states can respectively relax to $[...11101000...]$, $[...1110101000...]$ and $[...11101000...]$ by hopping a fermion past another fermion of lower energy. (Since the initial energy is an odd multiple of $\omega$, clearly for $k=2$ the system cannot relax to its true ground state regardless of the dynamics).  For HES particles, in contrast, there are no allowed relaxation transitions of energy $2 \omega$, and the configurations are stable.  Evidently, there are allowed transitions of energy $\omega$, and coupling to such transitions allows our HES systems to relax, as seen in Figs.  \ref{fig28}(a) and \ref{fig28}(c).  In this sense, the stability of these configurations results from a fine-tuning of the system-bath coupling.

 Figure (\ref{fig28}) also illustrate a second striking feature of relaxation in HES systems.  Panels (a) and (c) show relaxation for two different initial configurations, for $g=1, 1/2$, and $2/3$, in the case that the only allowed transitions have energy $\omega$ (i.e. $k=1$).  Strikingly, the relaxation dynamics is identical for all three systems, in spite of their very different constraints.  Panels (b) and (d) show relaxation for the same values of $g$ and choices of initial states, but now with the only allowed transitions having energy $2 \omega$ (i.e. $k=2$).  In this case, we see that spinless fermions $(g=1)$ relax more quickly than HES particles, but that the relaxation dynamics is identical for $g=1/2$ and $g=2/3$.  In the remainder of this section, we will show that this universal relaxation dynamics is a ubiquitous feature of HES particles, and explain why, for $k=1$, it corresponds to that of spinless fermions.

Before doing this, however, it is worth emphasizing in what sense the relaxation dynamics of HES particles is ``slow".  
A useful benchmark for the effect of particle statistics on  dynamics is to compare the relaxation time of a spinful fermion system  and a HES system of semions (i.e. $g=1/2$), since both allow at most two particles in a given energy cell.  This is shown for $T \ll \omega$ in Fig. (\ref{fig22}), for $k_{\text{max}} =1$ and $2$ (note that the latter includes both $k=1$ and $k=2$ channels), which also show the time evolution of the occupancy of a few levels near the Fermi level.   Figure (\ref{fig24}) shows the intermediate temperature regime, $T = \omega$, again for $k_{\text{max}} =1$ and $2$.  
 In all cases, energy relaxes more slowly for semions than spinful fermions.  
 In addition, these figures again exhibit the qualitative behavior described above.
For $k_{\text{max}} =1$, the relaxation rates of semions are identical to spinless fermions (which in turn necessarily relax more slowly than spinful fermions).  
For $k_{\text{max}} =2$, semions relax more slowly than spinless fermions, which in turn relax more slowly than spinful fermions.

\subsection{ Relaxation rates from relaxation paths}

To explain the qualitative features of Figs. (\ref{fig28})-(\ref{fig24}), we now review how the number of relaxation paths available to a system at a given time $t$ determines its instantaneous relaxation rate $\gamma(t)$.  Our analysis focuses on the low-temperature regime $T \ll \omega$, where excitation processes can be neglected. In the next section, we will compare how HES occupancy constraints affect the number of such relaxation paths, relative to those available for spinless fermions, thereby quantifying the impact of constraints on relaxation dynamics.

We can see that
the energy relaxation shown in Figs.(\ref{fig28})-(\ref{fig24}) depends on the number of available relaxation paths in two ways.  First, the dynamics is not described by a single relaxation rate.  Rather, it has the form $E(t)=E_0 \exp[-\beta(t)]$, where the instantaneous dissipation rate $\gamma \equiv d\beta/dt$ varies with time.  It is convenient to convert this time-dependence to a dependence on the energy $E(t)$, and plot $\gamma = -d \log(E(t))/ dt$ vs $E(t)$ for the various conditions, as shown in Fig.(\ref{fig26}).     Moreover, the effective dissipation rate at a given energy depends not only on the energy, but also on the initial state of the system.  This  is shown in Fig. \ref{fig30}, which plots $\gamma = -d \log(E(t))/ dt$ vs $E(t)$ for $k_{\text{max}}=1$  and  $k_{\text{max}}=2$ at $T=0.01\omega$ with different initial states.  Each line represents the dynamics of a single initial many-body state.  We see that the curves tend to agglomerate after some dependence on the initial state, but do initially show a significant dependence on the specific choice of initial state.
Together, Figs. (\ref{fig26}) and (\ref{fig30}) clearly show that how quickly the system relaxes depends on its current configuration.  

\begin{figure}[tb]
\includegraphics[height=5cm,width=8cm]{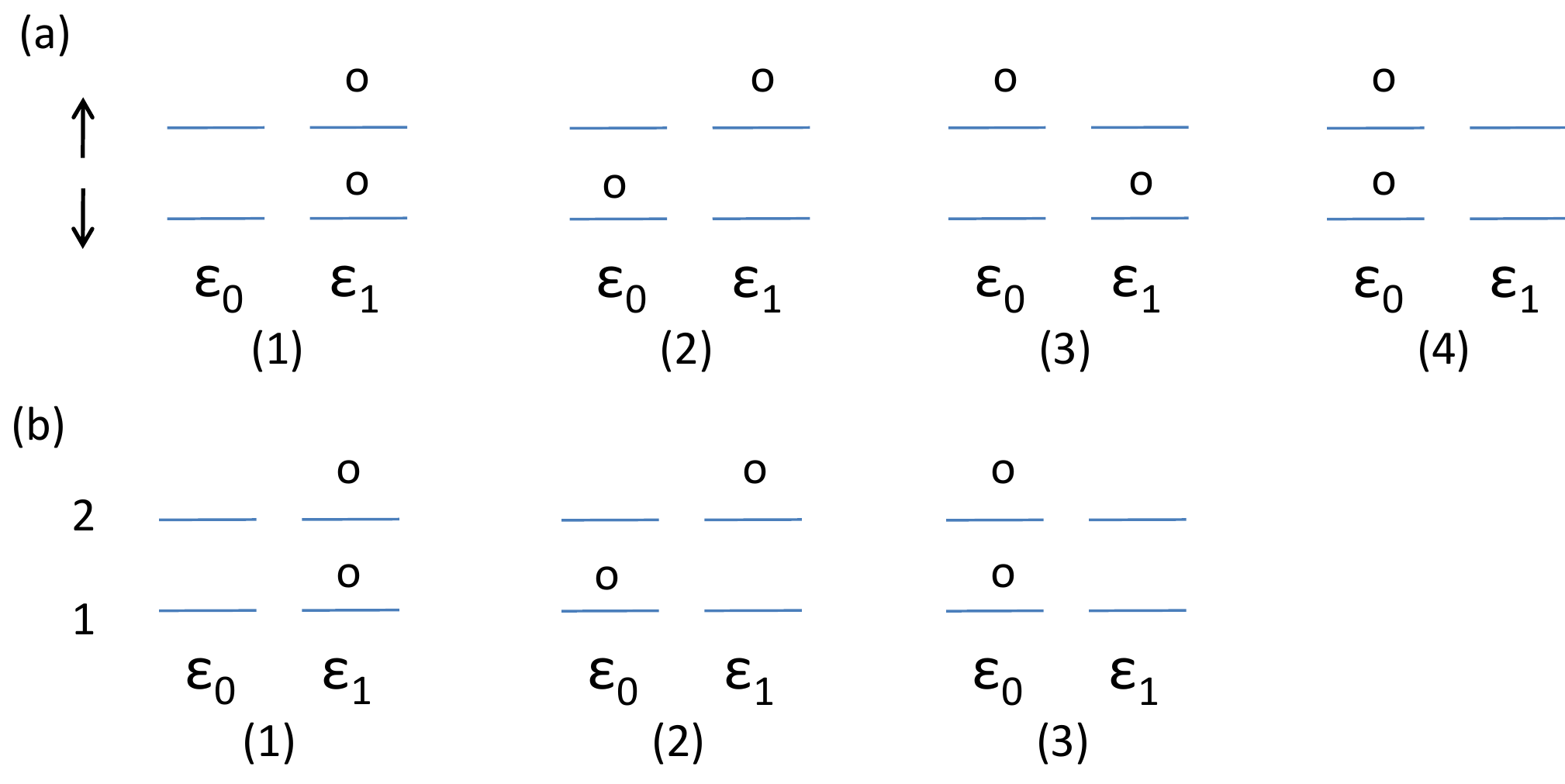}
\caption{ (a) Four possible states of two spinful fermions in a two-level system. (b) Three possible states of two semions in a two-level system. Starting with the excited state (1), spinful fermions can go through either state (2) or state (3) to relax to ground state (4), while semions can only reach ground state (3) through  state (2).}
\label{fig13}
\end{figure}

This dependence arises because relaxation processes must satisfy the occupancy constraints, which restrict the number of paths available between a given state of total energy $E$, and states of total energy $E- \omega$.  
We illustrate this with a simple example.  Consider a two-level system (with single-particle energy levels $0$ and $\epsilon$), containing either two spinful fermions, or two semions with $g=1/2$ (see Fig. \ref{fig13}).  Suppose both systems are initially in the state of energy $2 \epsilon$, with both particles in the excited state.    For spinful fermions there are two available relaxation paths between this initial state and the ground state, as either the spin up or the spin down fermion may relax first.  For semions, on the other hand, the semion with pseudospin $\alpha =1$ must relax first, and there is only one possible relaxation path.  If the probability of interacting with the bath per unit time is the same for all particles, then the energy of the fermionic  system will initially approach the ground state at twice the rate as for the semionic system.

To quantify the relationship between relaxation paths and the instantaneous relaxation rate $\gamma(t)$ more precisely, 
we consider initializing our system in an eigenstate.  In this case the density matrix is diagonal in the energy eigenbasis, and the time evolution of the vector $\vec{\rho}$ of diagonal elements of density matrix (defined by $\vec{\rho} \cdot \hat{i} \equiv \rho_{ii}$) is described by
\begin{align}\label{Pequation}
\frac{d \vec{\rho}}{d t} =  
 (\sum_{k, a=\pm} \Gamma_{k,a} \mathbf{A}_{k,a} )\vec{\rho}.
\end{align}
Here, $\mathbf{A}_{k,\pm}$ is a lower (upper) triangular matrix encoding the relaxation (excitation) paths for channel $k$. The off-diagonal entry $(\mathbf{A}_{k,\pm})_{ \beta \alpha} = 1$ if a process in which a single particle moves $k$ units down (up) in energy leads to a  transition between eigenstate $\alpha$ and eigenstate $\beta$, and is $0$ otherwise; these correspond to processes described by $L_{kja} \rho L^\dag_{kja}$ in Eq. (\ref{LindbladEq}).   The diagonal entry $- (\mathbf{A}_{k,\pm})_{\alpha \alpha}$ describes the total number of  transitions out of eigenstate $\alpha$ by moving a single particle down (up) in energy by $k$ energy levels; such processes are described by the anti-commutator in Eq. (\ref{LindbladEq}).
This ensures that all columns in $\mathbf{A}_{k,\pm}$ sum to zero, and Eq. (\ref{Pequation}) conserves $|\vec{\rho}|$ (i.e. conserves probability).

If the matrix $\mathbf{B}\equiv 
 \sum_{k, a=\pm} \Gamma_{k,a}  \mathbf{A}_{k,a} $, and  $\{ \vec{x}_a \} $ is non-degenerate, such that its eigenvectors span the whole Hilbert space, Eq. (\ref{Pequation}) is solved by 
\begin{equation} \label{Psolution}
\vec{\rho}(t)  = \sum_a c_a e^{-  \lambda_a t} \vec{x}_a \ ,
\end{equation} 
where $\{ c_a \}$ specifies a choice of initial eigenstates, $\{-\lambda_a \}$ are the eigenvalues of $\mathbf{B}$ are the corresponding eigenvectors.   
Since $\mathbf{B}$ conserves probability, one of the eigenvalues is zero; the corresponding eigenvector gives the steady-state density matrix, for which the energy takes on its value in thermal equilibrium.   Moreover, in the low temperature limit of interest here, $\mathbf{B}$ is a lower-triangular matrix.    In this case, we have 
\begin{equation} \label{Eq:lambdaa}
-\lambda_a =\mathbf{B}_{aa} =  \sum_k \Gamma_{k,-} 
 n_{\text{p}}(a, E_a - k \omega) 
  ,
\end{equation}
 where $n_{\text{p}}(a, E_a-k \omega)$ is the total number of states of energy $E_a - k \omega$ that can be reached from state $a$ using single-particle processes.  In other words, $- \lambda_a$ is given by the total number of relaxation paths from $a$ to any state of lower energy,  weighted by the corresponding decay rates $\Gamma_{k,-}$.

We can use Eq. (\ref{Psolution}) to solve for the instantateous relaxation rate $\gamma(t)$ as follows. Defining $\vec{\epsilon}$ as a row vector of energies in the basis used to construct $\vec{\rho}$, we have 
\begin{align}
\langle E (t) \rangle = \vec{\epsilon} \cdot \vec{\rho} (t) = \vec{\epsilon} \cdot  \sum_a c_a e^{- \lambda_a t} \vec{x}_a \equiv 
\sum_a  c_a  E_a(t)
\end{align}
where $E_a(t) =  \vec{\epsilon}\cdot\vec{x}_a e^{-  \lambda_a t}$
 is a single dissipation mode associated with eigenvector $\vec{x}_a$.
The instantaneous relaxation rate is $\gamma(t) \equiv -d \log(\langle E \rangle)/ dt$ thus given by:
\begin{align}\label{gamma_N}
\gamma(t) =  \frac{\sum_a c_a E_a(t) \lambda_a}{\sum_a c_a E_a(t)} .
\end{align}
Equations (\ref{Eq:lambdaa}) and (\ref{gamma_N}) quantify how the  instantaneous relaxation rate is determined by the number of relaxation paths into and out of of each state $a$.

For a given initial configuration  in the low temperature limit,
the total energy $E = E(t)$ decays monotonically with $t$,  such that we can express $t$ in terms of the average energy $E$ to obtain $\gamma = \gamma(E) $.   In Appendix (\ref{Sec:AppPath}), we use this fact to estimate how relaxation rates of channels $k=1$, $k=2,$ and $k=3$ depend on energy at low temperatures for very high-energy initial states.

\subsection{Universal Lindblad dynamics of HES} \label{Sec:UniLin}

Having argued that differences in relaxation rates stem from differences in the number of available relaxation paths in the presence of constraints, we now focus on understanding how the number of relaxation paths available to a particular system depends on the HES parameter $g$.  We will show that for all rational $g \neq 0,1$ the number of relaxation paths is universal.  In other words, there is a universal relaxation rate common to all HES systems.  We further show that with a single interaction channel $k=1$ this relaxation rate is identical to that of spinless fermions, while for $k>1$ HES systems generically have slower relaxation rates than spinless fermions.  This explains the striking dynamics observed in Fig. (\ref{fig28}).

\begin{figure}[tb]
\includegraphics[height=7.5cm,width=8cm]{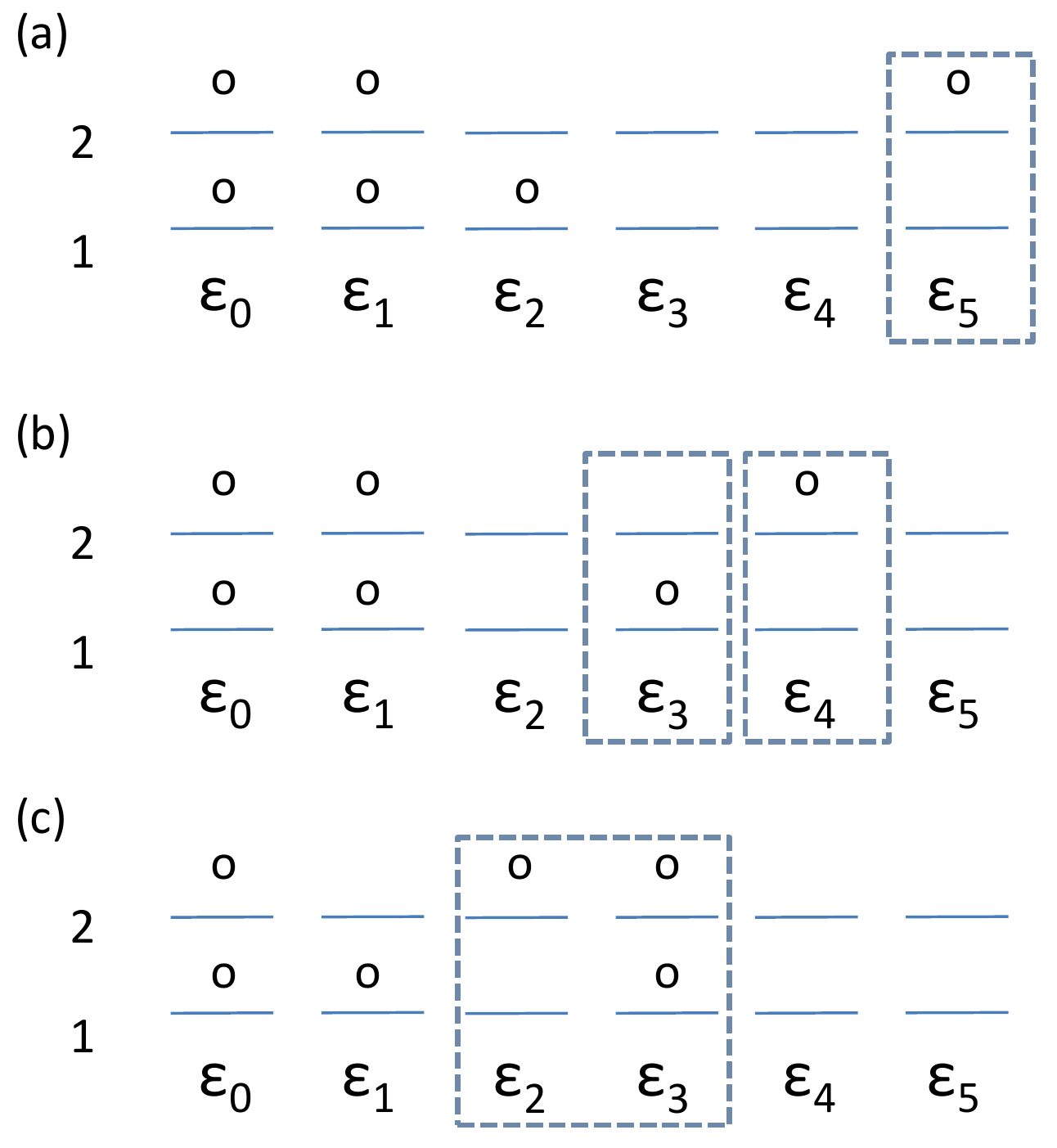}
\caption{ The three  many-body states of $g=1/2$ HES particles with energy $3\omega$ [(a)-(c)] correspond to the three energy assignments $[3]_e$, $[21]_e$ and $[111]_e$ respectively.  All excited particles are boxed.  }
\label{fig15}
\end{figure}

To establish this universality, we first show a one-to-one correspondence between ordered integer partitions $[x_1, x_2, ... x_n]_e$ ($x_1 \geq x_2 \geq ... \geq x_n$) and many-body eigenstates of both HES particles and spinless fermions.  (Note that this representation of the excited states is distinct from that of Murthy and Shankar\cite{Murthy} described in Sec. \ref{Sec:Constraints}.)  
In both cases, we set the energy of the many-body ground state to zero, and take the single-particle spectrum to consist of discrete energy levels $\epsilon_i$, with $\epsilon_{i+1}- \epsilon_i = \omega$.    We assign an index $i$  to each particle in the many-body ground state as follows.  First, particles in cells farther from the  Fermi surface have higher index: for $g=1/p$, $i=1,...,p$ for particles in the first cell below the Fermi surface, and so on.  Second, for particles in the same cell, particles with higher pseudospin have lower index.  Then a many-body excited state can be described by a sequence of integers $[x_1, x_2, ... x_n]_e$, where $x_i$ indicates how much energy the $i^{th}$ particle has gained relative to the many-body ground state, in units of $\omega$-- i.e. particle $1$ moves up by $x_1\omega$, particle $2$ moves up by $x_2 \omega$, etc.   The energy $E$ of the corresponding excited state is given by $E=\omega\sum_i x_i $ -- i.e. $[x_1, x_2, ... x_n]_e$ is an ordered integer  partition of $E/\omega$. Figure (\ref{fig15}) shows three states $[3]_e$, $[21]_e$ and $[111]_e$ with total energy $3\omega$ for $g=1/2$.

Clearly, all ordered integer partitions of $m$ give excited states with energy $m\omega$.  Moreover, this correspondence is one-to-one: every excited state with energy $m\omega$ corresponds to an ordered integer partition of $m$. To obtain excited states with energy $m  \omega$ above the ground state, the energy $m \omega$ must be distributed among some number of excited particles, i.e. the excited state is necessarily described by some partition of $m$.  
For $g \in (0,1)$, however, the HES occupancy constraints ensure that this partition must be ordered, since every excited state must respect the pseudospin ordering, and hence is equivalent to a configuration that can be obtained by exciting particles from the ground state such that the ordering of their energies is preserved, i.e., any particles of  index $i$ gains at least as much energy as every particle of index $j>i$.\footnote{In more detail, this ordering can be violated only by permuting entire cells' worth of particles.  However, as for spinless fermions, any such permutation leads to an occupancy that is identical to that obtained from an ordered partition.}
For $g=1$, or spinless fermions, there is no pseudospin, and hence no ordering constraint.  However, since only states corresponding to distinct occupancy patterns of excited levels are distinct, not every partition of $m$ describes a distinct excited state.  Indeed, since every occupancy pattern can be obtained by exciting  particles from the ground state while preserving the ordering of their energies, each distinct excited state can be identified with an {\it ordered} partition of $m$. 
Thus for any $g \in (0,1]$, we have a one-to-one corresponce between the ordered integer partitions of $m$ and excited states with energy $m \omega$.

To illustrate how this works, consider the many-body eigenstates of $E - E_0 =3\omega$.  Regardless of the choice of $g$ we find three such states, corresponding to the three integer partitions of 3 , i.e. $[3]_e$, $[21]_e$ and $[111]_e$ (see Fig. (\ref{fig15})).      For semions, for example, where the ground state is $\ket{...2222000...}$,  $[3]_e$, $[21]_e$ and $[111]_e$ respectively refer to the states $\ket{...222100100...}$,  $\ket{...22201100...}$ and $\ket{...2211200...}$. For spinless fermions, where the ground state is $\ket{...11111000...}$,   $[3]_e$, $[21]_e$ and $[111]_e$ respectively correspond to the states $\ket{...1111000100...}$, $\ket{...111010100...}$ and $\ket{...11011100...}$.

Thus, we see that any excited state of energy $E$ in a HES system with any rational $0 < g \leq 1$ can be specified by an ordered partition of $E/\omega$.  What about the allowed transitions between such states?  For spinless fermions ($g=1)$, a transition between states $[x_1, x_2, ... x_j, ... x_n]_e$ and $[x_1, x_2, ... x'_j, ... x_n]_e$ exists if $x_j' \neq x_i + i-j$ for any $i$, and $|x'_j-x_j| \leq k$, where $k$ is the largest energy transfer allowed with the bath.   The second partition need not be ordered; in general the equivalent ordered partition  differs from the original one at more than one point in the sequence.  For HES particles ($0<g<1$), however, we have seen above that preserving the pseudospin ordering also requires that transitions cannot move HES particles past each other.  (More precisely, such transitions would require moving an entire cell's worth of particles, i.e. it would require multi-particle transitions, which we do not include here).  For a single interaction channel with  $k=1$ in Eq.(\ref{Hdynamics}), these two constraints amount to the same thing, and the number of relaxation paths out of a given eigenstate is the same for HES particles as for spinless fermions.  For $k >1$, generically we find that some transitions that are allowed for spinless fermions are generically forbidden for HES particles, reducing the number of available relaxation paths and slowing the overall relaxation rate.

To see how this leads to universal Lindblad dynamics, observe that the 1-1 correspondence between the excited states of our HES system and those of a system of spinless fermions always allows us to write the relevant Lindblad operators in a common basis, given by the integer partitions $[x_1, x_2, ...x_n]_e, x_{i+1} \leq x_i$.  For $k=1$, we also find that in this basis the allowed transitions are the same for any rational $g \in (0, 1]$.  The non-vanishing matrix elements of single-particle-hopping operators are necessarily $+1$ since no fermion exchange is allowed.   It follows that the $k=1$ Lindblad matrices of  all species of HES particles are identical.  Thus, we have shown the surprising result that for $k=1$ the relaxation dynamics of any HES system is identical to that of a system with spinless fermions, as seen in   Figs. \ref{fig28}(a) and \ref{fig28}(c). 
For $k>1$, we have shown that the dynamics of different species of HES particles are identical for rational $g \in (0,1)$, though in this case they are not equivalent to spinless fermions, as seen in Figs. \ref{fig28}(b) and \ref{fig28}(d).

Our proof of universal dynamics also explains why different species of HES particles have the same thermal equilibirum energy, shown in Fig.(\ref{fig21}), assuming the many-body ground state energy is 0. In the basis of ordered integer partitions, different species of HES particles have the same Lindblad transition matrices for all interaction channels. Therefore, the Lindblad dynamics of different species of HES particles will relax to the same steady-state Gibbs density matrix in the basis of ordered integer partitions, which leads to identical thermal energy.

\section{Discussion and summary} \label{Sec:Discussion}

This work details the Lindblad dynamics of a system obeying Haldane Exclusion statistics (HES), coupled to a bosonic thermal bath.  We have shown how the constraints on the occupancy of different energy levels that define HES generically lead to slower energy relaxation, and in some situations can even lead to blockaded relaxation and stable excited states.  Moreover, we have shown that for rational $g \in (0,1)$, HES systems coupled quadratically to a bosonic thermal bath exhibit {\it universal} energy relaxation dynamics, in the sense that the relaxation is independent of the parameter $g$.  An interesting corolary of this result is that the energy of an HES system in thermal equilibrium is also independent of $g$.  

One key takeaway from our study is that, unlike in the context of thermodynamics, where neglecting inter-level occupancy constraints leads at worst to small quantitative corrections in most measurable quantities, in the context of dynamics, the constraints between different energy levels have distinctive, physically observable consequences.  For example, the slower relaxation dynamics of semions relative to spinless fermions is a direct consequence of the inter-level occupancy constraints, which reduces the number of relaxation paths for semions.   Indeed both the universal relaxation dynamics, and the presence of relaxation blockade, are possible only when such constraints exist.

It is worth pointing out that we do not expect these qualitative results to be sensitive to the specific method used to model the bath, and can thus reasonably be expected to apply to energy relaxation even in situations where the approximations leading to the Lindblad master equation fail to apply.  This is because the mapping between HES and spinless fermions is a feature only of the system, and hence any model of energy relaxation for HES particles can be mapped onto a model of energy relaxation in spinless fermions.  

In establishing these results, we have also described a second-quantized treatment appropriate for HES particles, which is based on a comprehensive microscopic picture of the corresponding allowed occupancy patterns, which generalizes the results of Ref. \onlinecite{Murthy}.   This second-quantized description gives an appealingly simple picture of the constraints, in terms of multiple flavors of fermions that are restricted to respect a particular pseudospin ordering in energy space.  Though other frameworks for second quantization of particles with novel exclusion statistics exist\cite{karabali,Ilinski,Speliotopoulos97,Meljanac_1999}, ours has the advantage of allowing a straightforward computation of matrix elements, with projections that are easily implemented numerically.  As such, this formalism may be useful for investigating the impact of inter-occupancy constraints on other aspects of HES systems.

For example, Wu's approximation, which ignores inter-level occupancy constraints, has been employed to study many response functions in HES systems, including electrical and thermal conductance\cite{Rego} and shot noise\cite{Gomila}.  The premise of these studies is that the main role of statistics is to alter the underlying distribution in occupancies of various states, and that this distribution is well described by Wu's approximation.  If the number of states at the relevant energy density is large, we expect this to be correct; however, it would be interesting to search for transport regimes where this approximation may break down.  

It is also interesting to compare our results, which describe a universal dynamics independent of $g$, to the dynamics that emerges from the Boltzmann equation. The latter has been studied  by several authors \cite{Bhaduri,Kaniadakis96,Isakov98,Arkeryd}, using techniques in which occupancy constraints between different energy levels are neglected.   In general, the associated dynamics -- including effective relaxation rates as calculated by a modified Fermi's golden rule\cite{Bhaduri}-- depends on $g$, although some transport quantities are found to be universal.\cite{Isakov98,Rego}   Though our analysis does not immediately lend itself to studying transport, it does suggest that the occupancy constraints could be qualitatively important for Boltzmann relaxation dynamics as well, and it would be interesting to understand how incorporating the exact constraints changes these results.

{\it Acknowledgments}	
This work has been supported by the national science foundation under award number NSF-DMR 1928166.  FJB is grateful for the support of the Carnegie Corporation of New York, and the Institute for Advanced Study.

\titleformat{\section}{\centering\bfseries}{\appendixname~\thesection :}{0.5em}{}
\begin{appendices}

\section{Dimension of many-body Hilbert space at arbitrary particle number} \label{Sec:CountingApp}

Eq.(\ref{D_N}), which appears elsewhere in the literature\cite{Wu, Nayak1, Murthy}, does not in general produce an integral many-body Hilbert space dimension.  
This is because for $g=q/p$, the expression $d(N+\Delta N)= d(N)-\Delta N q/p$ gives an integer single-particle Hilbert space dimension only if $\Delta N = n p$.   Thus we may take $d(N_i) = d_i(1) - g (N_i-1)$ only if $N_i=n p + 1$ with $n$ an integer.

For a generic particle number $N = np + m$ ($1 \leq m \leq p$),
we can instead use our understanding of the HES constraints to count the single-particle and many-body Hilbert space dimensions $d(N)$ and $D_N(g)$.  $d(N)$ can be calculated using Eq.(\ref{d}): $d(N) = d(m) - nq$, where $d(m) = d(1) - \text{floor}((m-1)q/p)$. Here, we use the fact that if the particle number is enough to fully occupy one cell, the number of accessible cells to each particle decreases by one.

In the context of the CSM, as discussed in the main text, the dimension of the many-body Hilbert space is:
\begin{align}\label{appA_D}
D_N(g) = \binom{k_{\text{max}}-k_{\text{min}}+1}{N} .
\end{align}
where $k_{\text{max}}$ and $k_{\text{min}}$ are determined by the maximum and minimum shifted single-particle energies $\epsilon_s^{\text{max}}$ and $\epsilon_s^{\text{min}}$, respectively.   We wish to show that this is the correct Hilbert space dimension for our HES particles, which will be true if $k_{\text{max}}-k_{\text{min}}+1 = d(N)+ N -1$.  Setting $\epsilon_s^{\text{min}} = k_{\text{min}} =0$ for simplicity, this will be true provided that
\begin{align}
d(N) = k_{\text{max}} +2 -N.
\end{align}
With the shifted energy levels divided into cells of size $\omega$, i.e. $[0,\omega), [\omega, 2\omega), [2\omega, 3\omega) ...$, we find $d(1) = \text{floor}(\epsilon_s^{\text{max}}/\omega)+1$. Substituting $\epsilon_s^{\text{max}}/\omega = k_{\text{max}} - (1-q/p) (N-1)$, we obtain $d(1) = k_{\text{max}}+2-N+nq+\text{floor}((m-1)q/p)$. Thus, 
\begin{align}
d(N) &= d(m) -nq \nonumber \\
&= d(1) - \text{floor}((m-1)q/p) -nq \nonumber \\ 
&= k_{\text{max}} +2-N.
\end{align}

\begin{widetext}

\section{Matrix elements of generic single-particle operators for $g= 1/2$}\label{Sec:Appe}

For $g=1/2$, the matrix elements for hopping a particle to a neighboring cell are:
\begin{align}
P f_{i+1}^{(1) \dagger} f_{i}^{(1)} P \ket{..., n_i^{(1)}, n_i^{(2)}, n_{i+1}^{(1)}, n_{i+1}^{(2)}, ... }= n_i^{(1)}  (1-n_{i+1}^{(1)}) c_{1}^{(1/2)}(i) P \ket{..., n_i^{(1)}-1, n_i^{(2)}, n_{i+1}^{(1)}+1, n_{i+1}^{(2)}, ... }, \\
 P f_{i-1}^{(1) \dagger} f_{i}^{(1)} P\ket{..., n_{i-1}^{(1)}, n_{i-1}^{(2)}, n_{i}^{(1)}, n_{i}^{(2)}, ... }= n_i^{(1)} (1-  n_{i-1}^{(1)}) c_{1}^{(1/2)}(i-1) P \ket{..., n_{i-1}^{(1)}+1, n_{i-1}^{(2)}, n_{i}^{(1)}-1, n_{i}^{(2)}, ... }, \nonumber  \\
 P f_{i+1}^{(2) \dagger} f_{i}^{(2)} P \ket{..., n_i^{(1)}, n_i^{(2)}, n_{i+1}^{(1)}, n_{i+1}^{(2)}, ... }= n_i^{(2)} (1-n_{i+1}^{(2)}) c_{2}^{(1/2)}(i) P \ket{..., n_i^{(1)}, n_i^{(2)}-1, n_{i+1}^{(1)}, n_{i+1}^{(2)}+1, ... }, \nonumber \\
 P f_{i-1}^{(2) \dagger} f_{i}^{(2)} P\ket{..., n_{i-1}^{(1)}, n_{i-1}^{(2)}, n_{i}^{(1)}, n_{i}^{(2)}, ... }= n_i^{(2)} (1-n_{i-1}^{(2)}) c_{2}^{(1/2)}(i-1) P\ket{..., n_{i-1}^{(1)}, n_{i-1}^{(2)}+1, n_{i}^{(1)}, n_{i}^{(2)}-1, ... }, \nonumber 
\end{align} 
where $c_{1}^{(1/2)}(i)=1-n_i^{(2)}$, 
$c_{2}^{(1/2)}(i)=(1-n_{i+1}^{(1)})$ describe the effect of the left-most projector.  
For hopping processes to more distant cells, we again have the appropriate fermionic matrix element, multiplied by a projector that ensures that our particle cannot hop past any particle of the opposite pseudo-spin (which, in configurations allowed by the constraint, is sufficient to also ensure it cannot hop past any particles of the same type) : 
\begin{align}
P f_{i+k}^{(1) \dagger} f_{i}^{(1)} P \ket{.., n_i^{(1)}, n_i^{(2)},.., n_{i+k}^{(1)}, n_{i+k}^{(2)},.. }=& n_i^{(1)} (1-n_{i+k}^{(1)}) \prod_{j=1}^{k}  c_{1}^{(1/2)}(i+j-1) 
P\ket{.., n_i^{(1)}-1, n_i^{(2)},.., n_{i+k}^{(1)}+1, n_{i+k}^{(2)},.. }, \nonumber  \\
P f_{i-k}^{(1) \dagger} f_{i}^{(1)} P \ket{.., n_{i-k}^{(1)}, n_{i-k}^{(2)},.., n_{i}^{(1)}, n_{i}^{(2)},.. }=&n_i^{(1)} (1-n_{i-k}^{(1)})  \prod_{j=1}^{k}  c_{1}^{(1/2)}(i-j) 
P \ket{.., n_{i-k}^{(1)}+1, n_{i-k}^{(2)},.., n_{i}^{(1)}-1, n_{i}^{(2)},.. },\nonumber  \\
P f_{i+k}^{(2) \dagger} f_{i}^{(2)} P\ket{.., n_i^{(1)}, n_i^{(2)},.., n_{i+k}^{(1)}, n_{i+k}^{(2)},.. }=  &n_i^{(2)} (1-n_{i+k}^{(2)}) \prod_{j=1}^{k}  c_{2}^{(1/2)}(i+j-1)
P \ket{.., n_i^{(1)}, n_i^{(2)}-1,..,n_{i+k}^{(1)}, n_{i+k}^{(2)}+1,.. },\nonumber \\
P f_{i-k}^{(2) \dagger} f_{i}^{(2)} P\ket{.., n_{i-k}^{(1)}, n_{i-k}^{(2)},.., n_{i}^{(1)}, n_{i}^{(2)},.. }= &n_i^{(2)} (1-n_{i-k}^{(2)}) \prod_{j=1}^{k}  c_{2}^{(1/2)}(i-j)
P \ket{.., n_{i-k}^{(1)}, n_{i-k}^{(2)}+1,.., n_{i}^{(1)}, n_{i}^{(2)}-1,.. }, \nonumber 
\end{align}
Again, we find that the non-zero matrix elements of single-particle-hopping processes can only be 1. 

Note that within a given cell, only one of the two species of particles can have a non-zero hopping matrix element in either direction: if $n_i^{(2)} = 0$ then there is no particle on the top ladder to hop; on the other hand, if $n_i^{(2)} = 1$, then particles of type $1$ cannot hop to cells of higher energy.  Similarly, if $n_i^{(1)} \neq 0$, particles of type $2$ cannot hop to cells of lower energy.  
Hence  for an allowed occupancy $| ... N_{i}, N_{i+1}, ... N_{i+k}, ...  \rangle $ of the energy cells, 
\begin{equation}
P h^\dag_{i+k} h_{i} P| ... N_{i}, N_{i+1}, ... N_{i+k}, ...  \rangle = | ... N_{i}-1, N_{i+1}, ... N_{i+k}+1, ... \rangle \ .
\end{equation}

Next, we consider a similar calculation for $g=1/3$.  Substituting the expressions of $h_{p+1}^{\dagger}$ and $h_{p}$ into  $P h_{p+1}^{\dagger} h_p P \ket{..., N_p, N_{p+1}, ...}$, we get
\begin{align}
 P h_{p+1}^{\dagger} h_p P\ket{..., N_p, N_{p+1}, ...} = &   P \left[ f_ {p+1}^{(1)\dagger} f_{p}^{(1)} + f_ {p+1}^{(2)\dagger} f_{p}^{(2)} + f_ {p+1}^{(3) P\dagger} f_{p}^{(3)} \right] P \ket{..., n_{p}^{(1)}, n_{p}^{(2)}, n_{p}^{(3)}, n_{p+1}^{(1)}, n_{p+1}^{(2)}, n_{p+1}^{(3)},...}    \\
= &  n_p^{(1)} (1-n_{p+1}^{(1)}) (1-n_{p}^{(2)}) (1-n_{p}^{(3)}) P \ket{..., n_{p}^{(1)}-1, n_{p}^{(2)}, n_{p}^{(3)}, n_{p+1}^{(1)}+1, n_{p+1}^{(2)}, n_{p+1}^{(3)},...} \nonumber \\
 &+   n_p^{(2)} (1-n_{p+1}^{(2)}) (1-n_{p}^{(3)}) (1-n_{p+1}^{(1)}) P \ket{..., n_{p}^{(1)}, n_{p}^{(2)}-1, n_{p}^{(3)}, n_{p+1}^{(1)}, n_{p+1}^{(2)}+1, n_{p+1}^{(3)},...}\nonumber   \\
 &+   n_p^{(3)} (1-n_{p+1}^{(3)}) (1-n_{p+1}^{(1)}) (1-n_{p+1}^{(2)}) P \ket{..., n_{p}^{(1)}, n_{p}^{(2)}, n_{p}^{(3)}-1, n_{p+1}^{(1)}, n_{p+1}^{(2)}, n_{p+1}^{(3)}+1,...}.\nonumber 
\end{align}
Evidently, all the three of the final microscopic states are identified with the same HES configuration $\ket{..., N_p -1, N_{p+1} +1, ...}$.

As for $g=1/2$, the constraints impose considerable structure on the non-vanishing matrix elements.  
First, the signs $(-1)^{n_{p}^{(2)}+n_{p}^{(3)}}$, $(-1)^{n_{p}^{(3)}+n_{p+1}^{(1)}}$ and $(-1)^{n_{p+1}^{(1)}+n_{p+1}^{(2)}}$ have been dropped, since particles cannot hop past each other.  Second, only one of the three matrix elements can take the value $+1$; the other two are zero.  
For example, $n_i^{(1)} = 1$, then matrix elements hopping particles of type $2$ or $3$ to a cell of lower energy vanish, with similar constraints for other configurations.   Indeed, both of these conditions are a generic feature of single-particle-hopping operators ($P h_k^{\dagger} h_l P$) for  any $g=q/p$.  This structure simply reflects the fact that pseudospin-energy ordering prevents particles from hopping past each other:
a fermion of pseudospin $i_1$ cannot move up in energy past a fermion of pseudospin $j_1 (j_1 > i_1)$, and   a fermion of pseudospin $i_2$ cannot move down past a fermion of pseudospin $j_2 (j_2 < i_2)$.   Thus at most one of the possible fermionic hoppings is allowed.

\end{widetext}

\section{Estimating relaxation rates at high energies}\label{Sec:AppPath}
Since the instantaneous relaxation rate depends on the weighted average of  relaxation paths,  we can estimate the energy dependence of typical relaxation rates in high energy states  by examining the distribution $P(N(E))$ of available transition paths at each energy $E$, for a given choice of interaction channels.

\begin{figure}[tb]
\includegraphics[height=6.5cm,width=8cm]{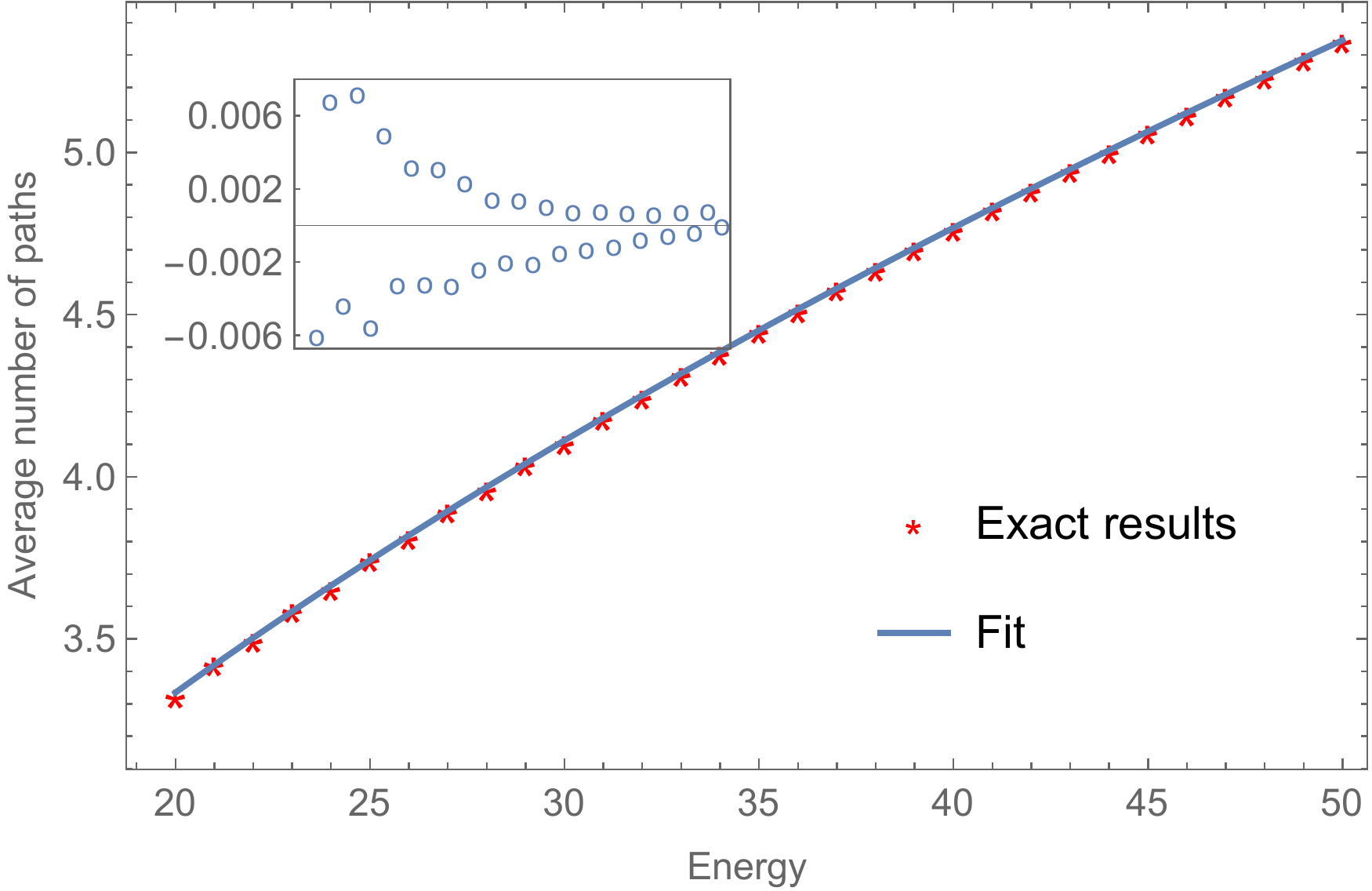}
\caption{The average number of relaxation paths $\bar{N}(E)$ of all states with energy $E$ for the interaction channel $k=1$  vs energy.  The exact result is fitted by a trial solution $\bar{N}(E) = a + \sqrt{b + d E}$, where $a=-0.1449$, $b=0.0795$ and $d=0.6014/\omega$. The inset shows the differences between exact results and the fit.  }
\label{fig31}
\end{figure} 

\begin{figure}[tb]
\includegraphics[height=6.5cm,width=8cm]{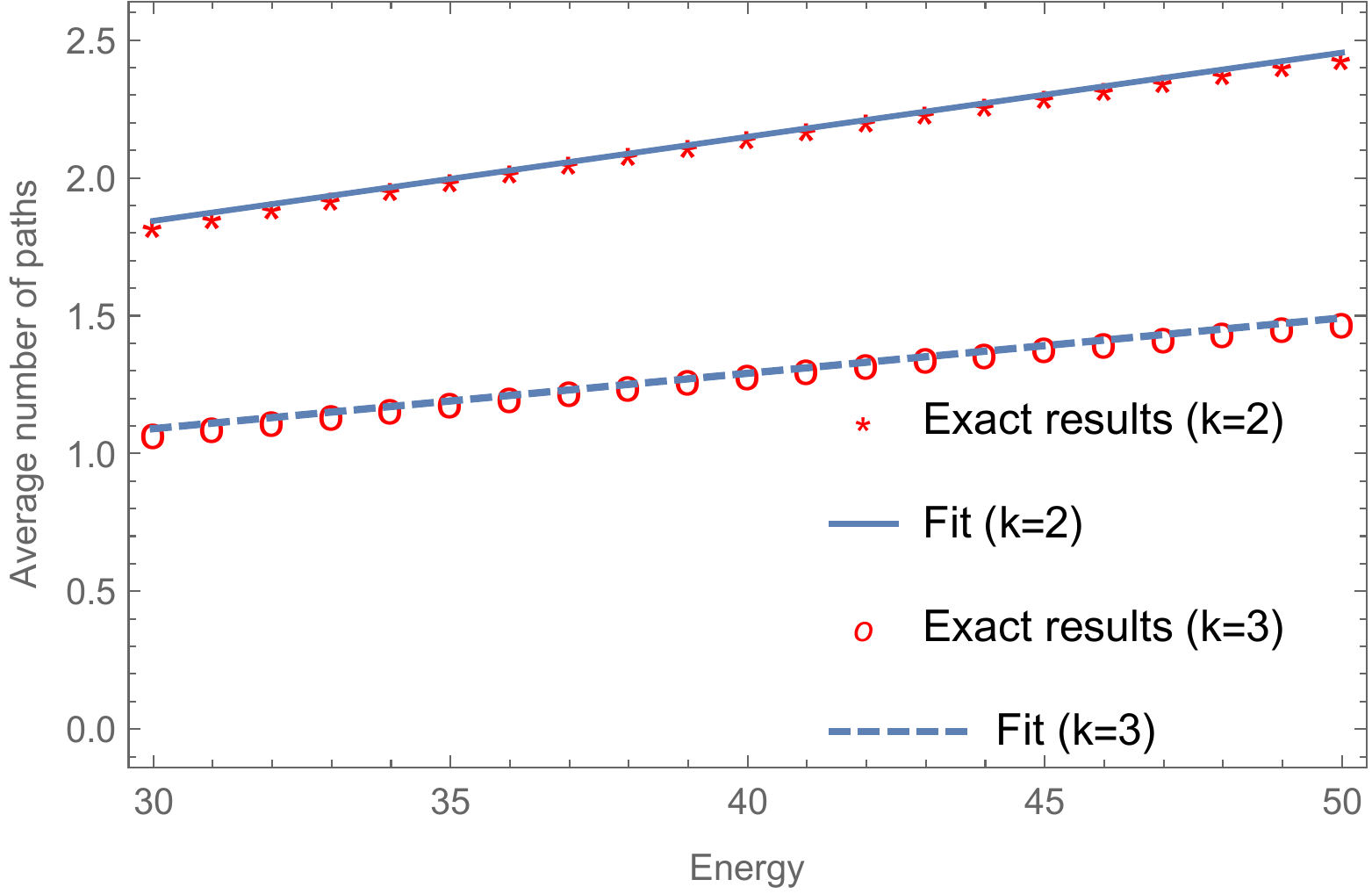}
\caption{The average number of relaxation paths $\bar{N}(E)$ of all states with energy $E$ for the interaction channel $k=2$ and $k=3$  vs energy.  The data fit well to the line $\bar{N}(E) = a E + b$, with $a=0.0305/\omega$ and $b=0.9282$ ($a=0.0201/\omega$ and $b=0.4875$) for $k=2$ ($k=3)$. $R^2$ of the two fits are $0.999$.}
\label{fig32}
\end{figure}

We illustrate this for the case of a single interaction channel, $k=1$.    A  many-body excited state of energy $E$ is described by an ordered integer partition of $E$, $[x_1, x_2, ... x_n]| \omega \sum_i x_i = E$, with $x_{j=1} \geq x_j$.  For $k=1$, transitions to a state represented by $[x_1, x_2, ... x_j -1, ... x_n]$ are allowed only if $x_j  \geq x_{j-1}+1$.   We call a consecutive set of particles with the same $x_j$ a cluster; when $x_{j+1} \geq x_{j-1} + 1$, we say there is a gap between two adjacent clusters.  For example, in Fig. \ref{fig15},  the configurations shown in (a) and (c) have only one cluster, while the configuration in (b) has two. Then the number of relaxation paths available to a given configuration is equal to the number of clusters.  

To understand the typical relaxation dynamics, we wish to describe the average number of relaxation paths $\bar{N}(E)$, and hence the average number of clusters, in a configuration of total energy $E$.  
Suppose that we have a typical configuration $\bar{N}(E)$, and let us consider the possible configurations with total energy $E+1$ that can be obtained by exciting a single particle in this configuration.  
If we excite a particle that is in a cluster of size greater than one, then a new cluster will be created if and only if the gap to the next highest energy cluster has size greater than $1$ -- i.e we take $x_j \rightarrow x_j +1$, with $x_{j+1} - x_j > 1$.  If the particle that we excite is in a cluster of size one, then we can either decrease the number of clusters (if $x_{j+1} - x_j =1$), or leave it unchanged.  
Since there are a total of $\bar{N}(E)+1$ clusters (including the lowest energy cluster, i.e. the Fermi sea), and we must average over all possible moves that raise $E$ by $\omega$, we conclude that:
\begin{align}
\bar{N}(E+1)  = \bar{N}(E) + \frac{c(E)}{\bar{N}(E) + 1},
\end{align}
where $c(E) = (p_g(E) + p_c(E) -1)\bar{N}(E) +1$, with $p_c(E)$ the probability of a cluster containing more than one particle, and $p_g(E)$ being the probability that the gap between two clusters is greater than one.  At high energies, we can approximate this by a differential equation $d N/dE = c/N$, where for $k=1$ we find a good fit to our numerical data (shown in Fig.(\ref{fig31})) by taking $c(E) =c$ to be independent of energy.  This gives an approximate functional form $\bar{N}(E) = a + \sqrt{b + d E}$ with $a$, $b$ and $d$ constant.    This suggests that for highly excited states that have been relaxing for some time (such that they have attained a typical distribution of clusters), we will find $\bar{\gamma}(E) \sim \sqrt{ E}$ for $k=1$.We note that the exact simulations presented in the main text are performed on relatively small energies, and hence need not exhibit this regime.

For $k=m > 1$ (but still with a single relaxation channel) we may follow a similar logic, by defining a cluster as a contiguous set of $x_j$ such that $x_{j_1} - x_j < m$.  In other words, our new clusters contain energy gaps up to size  $m \omega$, since HES particles on the interior of such clusters cannot relax.  Following the same logic as for $k=1$, we can obtain a similar difference equation relating $\bar{N}(E+1)$ and $\bar{N}(E)$, with $c(E)$ replaced by $c'(E) = (p'_g(E) + p_c(E) -1)\bar{N}(E) +1$. Here, $p'_g(E)$ is the probability of an energy gap  greater than $m$ energy units.  Numerically, we find in this case that  $c(E) \sim c \bar{N}(E) +1$ (see Fig. \ref{fig32}), such that at high energies, we have $d N/dE = c$ and solutions depend linearly on $E$. Thus, the best estimation of $\gamma(E)$ from our result for $k=2$ and $k=3$  is $\gamma(E) \sim E$.

\end{appendices}

\bibliography{HES}

\begin{thebibliography}{65}%
\makeatletter
\providecommand \@ifxundefined [1]{%
 \@ifx{#1\undefined}
}%
\providecommand \@ifnum [1]{%
 \ifnum #1\expandafter \@firstoftwo
 \else \expandafter \@secondoftwo
 \fi
}%
\providecommand \@ifx [1]{%
 \ifx #1\expandafter \@firstoftwo
 \else \expandafter \@secondoftwo
 \fi
}%
\providecommand \natexlab [1]{#1}%
\providecommand \enquote  [1]{``#1''}%
\providecommand \bibnamefont  [1]{#1}%
\providecommand \bibfnamefont [1]{#1}%
\providecommand \citenamefont [1]{#1}%
\providecommand \href@noop [0]{\@secondoftwo}%
\providecommand \href [0]{\begingroup \@sanitize@url \@href}%
\providecommand \@href[1]{\@@startlink{#1}\@@href}%
\providecommand \@@href[1]{\endgroup#1\@@endlink}%
\providecommand \@sanitize@url [0]{\catcode `\\12\catcode `\$12\catcode
  `\&12\catcode `\#12\catcode `\^12\catcode `\_12\catcode `\%12\relax}%
\providecommand \@@startlink[1]{}%
\providecommand \@@endlink[0]{}%
\providecommand \url  [0]{\begingroup\@sanitize@url \@url }%
\providecommand \@url [1]{\endgroup\@href {#1}{\urlprefix }}%
\providecommand \urlprefix  [0]{URL }%
\providecommand \Eprint [0]{\href }%
\providecommand \doibase [0]{https://doi.org/}%
\providecommand \selectlanguage [0]{\@gobble}%
\providecommand \bibinfo  [0]{\@secondoftwo}%
\providecommand \bibfield  [0]{\@secondoftwo}%
\providecommand \translation [1]{[#1]}%
\providecommand \BibitemOpen [0]{}%
\providecommand \bibitemStop [0]{}%
\providecommand \bibitemNoStop [0]{.\EOS\space}%
\providecommand \EOS [0]{\spacefactor3000\relax}%
\providecommand \BibitemShut  [1]{\csname bibitem#1\endcsname}%
\let\auto@bib@innerbib\@empty
\bibitem [{\citenamefont {Turner}\ \emph {et~al.}(2018)\citenamefont {Turner},
  \citenamefont {Michailidis}, \citenamefont {Abanin}, \citenamefont {Serbyn},\
  and\ \citenamefont {Papi'c}}]{PXP-Scar}%
  \BibitemOpen
  \bibfield  {author} {\bibinfo {author} {\bibfnamefont {C.~J.}\ \bibnamefont
  {Turner}}, \bibinfo {author} {\bibfnamefont {A.~A.}\ \bibnamefont
  {Michailidis}}, \bibinfo {author} {\bibfnamefont {D.~A.}\ \bibnamefont
  {Abanin}}, \bibinfo {author} {\bibfnamefont {M.}~\bibnamefont {Serbyn}},\
  and\ \bibinfo {author} {\bibfnamefont {Z.}~\bibnamefont {Papi'c}},\
  }\bibfield  {title} {\bibinfo {title} {Weak ergodicity breaking from quantum
  many-body scars},\ }\href {https://doi.org/10.1038/s41567-018-0137-5}
  {\bibfield  {journal} {\bibinfo  {journal} {Nature Physics}\ }\textbf
  {\bibinfo {volume} {14}},\ \bibinfo {pages} {745} (\bibinfo {year}
  {2018})}\BibitemShut {NoStop}%
\bibitem [{\citenamefont {Prem}\ \emph {et~al.}(2017)\citenamefont {Prem},
  \citenamefont {Haah},\ and\ \citenamefont {Nandkishore}}]{Prem17}%
  \BibitemOpen
  \bibfield  {author} {\bibinfo {author} {\bibfnamefont {A.}~\bibnamefont
  {Prem}}, \bibinfo {author} {\bibfnamefont {J.}~\bibnamefont {Haah}},\ and\
  \bibinfo {author} {\bibfnamefont {R.}~\bibnamefont {Nandkishore}},\
  }\bibfield  {title} {\bibinfo {title} {Glassy quantum dynamics in translation
  invariant fracton models},\ }\href
  {https://doi.org/10.1103/PhysRevB.95.155133} {\bibfield  {journal} {\bibinfo
  {journal} {Phys. Rev. B}\ }\textbf {\bibinfo {volume} {95}},\ \bibinfo
  {pages} {155133} (\bibinfo {year} {2017})}\BibitemShut {NoStop}%
\bibitem [{\citenamefont {Chen}\ \emph {et~al.}(2018)\citenamefont {Chen},
  \citenamefont {Burnell},\ and\ \citenamefont {Chandran}}]{Constrained-MBL}%
  \BibitemOpen
  \bibfield  {author} {\bibinfo {author} {\bibfnamefont {C.}~\bibnamefont
  {Chen}}, \bibinfo {author} {\bibfnamefont {F.}~\bibnamefont {Burnell}},\ and\
  \bibinfo {author} {\bibfnamefont {A.}~\bibnamefont {Chandran}},\ }\bibfield
  {title} {\bibinfo {title} {How does a locally constrained quantum system
  localize?},\ }\href {https://doi.org/10.1103/PhysRevLett.121.085701}
  {\bibfield  {journal} {\bibinfo  {journal} {Phys. Rev. Lett.}\ }\textbf
  {\bibinfo {volume} {121}},\ \bibinfo {pages} {085701} (\bibinfo {year}
  {2018})}\BibitemShut {NoStop}%
\bibitem [{\citenamefont {Chandran}\ \emph {et~al.}(2016)\citenamefont
  {Chandran}, \citenamefont {Schulz},\ and\ \citenamefont
  {Burnell}}]{FibonacciETH}%
  \BibitemOpen
  \bibfield  {author} {\bibinfo {author} {\bibfnamefont {A.}~\bibnamefont
  {Chandran}}, \bibinfo {author} {\bibfnamefont {M.~D.}\ \bibnamefont
  {Schulz}},\ and\ \bibinfo {author} {\bibfnamefont {F.~J.}\ \bibnamefont
  {Burnell}},\ }\bibfield  {title} {\bibinfo {title} {The eigenstate
  thermalization hypothesis in constrained hilbert spaces: A case study in
  non-abelian anyon chains},\ }\href
  {https://doi.org/10.1103/PhysRevB.94.235122} {\bibfield  {journal} {\bibinfo
  {journal} {Phys. Rev. B}\ }\textbf {\bibinfo {volume} {94}},\ \bibinfo
  {pages} {235122} (\bibinfo {year} {2016})}\BibitemShut {NoStop}%
\bibitem [{\citenamefont {Timpanaro}\ \emph {et~al.}(2019)\citenamefont
  {Timpanaro}, \citenamefont {Wald}, \citenamefont {Semi{\~a}o},\ and\
  \citenamefont {Landi}}]{timpanaro}%
  \BibitemOpen
  \bibfield  {author} {\bibinfo {author} {\bibfnamefont {A.~M.}\ \bibnamefont
  {Timpanaro}}, \bibinfo {author} {\bibfnamefont {S.}~\bibnamefont {Wald}},
  \bibinfo {author} {\bibfnamefont {F.}~\bibnamefont {Semi{\~a}o}},\ and\
  \bibinfo {author} {\bibfnamefont {G.~T.}\ \bibnamefont {Landi}},\ }\bibfield
  {title} {\bibinfo {title} {Dynamical chaotic phases and constrained quantum
  dynamics},\ }\href@noop {} {\bibfield  {journal} {\bibinfo  {journal}
  {Physical Review A}\ }\textbf {\bibinfo {volume} {100}},\ \bibinfo {pages}
  {012117} (\bibinfo {year} {2019})}\BibitemShut {NoStop}%
\bibitem [{\citenamefont {Ho}\ \emph {et~al.}(2019)\citenamefont {Ho},
  \citenamefont {Choi}, \citenamefont {Pichler},\ and\ \citenamefont
  {Lukin}}]{Ho2019}%
  \BibitemOpen
  \bibfield  {author} {\bibinfo {author} {\bibfnamefont {W.~W.}\ \bibnamefont
  {Ho}}, \bibinfo {author} {\bibfnamefont {S.}~\bibnamefont {Choi}}, \bibinfo
  {author} {\bibfnamefont {H.}~\bibnamefont {Pichler}},\ and\ \bibinfo {author}
  {\bibfnamefont {M.~D.}\ \bibnamefont {Lukin}},\ }\bibfield  {title} {\bibinfo
  {title} {Periodic orbits, entanglement, and quantum many-body scars in
  constrained models: Matrix product state approach},\ }\href
  {https://doi.org/10.1103/PhysRevLett.122.040603} {\bibfield  {journal}
  {\bibinfo  {journal} {Phys. Rev. Lett.}\ }\textbf {\bibinfo {volume} {122}},\
  \bibinfo {pages} {040603} (\bibinfo {year} {2019})}\BibitemShut {NoStop}%
\bibitem [{\citenamefont {Khemani}\ \emph {et~al.}(2020)\citenamefont
  {Khemani}, \citenamefont {Hermele},\ and\ \citenamefont
  {Nandkishore}}]{KhemaniShatter}%
  \BibitemOpen
  \bibfield  {author} {\bibinfo {author} {\bibfnamefont {V.}~\bibnamefont
  {Khemani}}, \bibinfo {author} {\bibfnamefont {M.}~\bibnamefont {Hermele}},\
  and\ \bibinfo {author} {\bibfnamefont {R.}~\bibnamefont {Nandkishore}},\
  }\bibfield  {title} {\bibinfo {title} {Localization from hilbert space
  shattering: From theory to physical realizations},\ }\href
  {https://doi.org/10.1103/PhysRevB.101.174204} {\bibfield  {journal} {\bibinfo
   {journal} {Phys. Rev. B}\ }\textbf {\bibinfo {volume} {101}},\ \bibinfo
  {pages} {174204} (\bibinfo {year} {2020})}\BibitemShut {NoStop}%
\bibitem [{\citenamefont {Iadecola}\ and\ \citenamefont
  {Schecter}(2020)}]{IadecolaSchecter}%
  \BibitemOpen
  \bibfield  {author} {\bibinfo {author} {\bibfnamefont {T.}~\bibnamefont
  {Iadecola}}\ and\ \bibinfo {author} {\bibfnamefont {M.}~\bibnamefont
  {Schecter}},\ }\bibfield  {title} {\bibinfo {title} {Quantum many-body scar
  states with emergent kinetic constraints and finite-entanglement revivals},\
  }\href {https://doi.org/10.1103/PhysRevB.101.024306} {\bibfield  {journal}
  {\bibinfo  {journal} {Phys. Rev. B}\ }\textbf {\bibinfo {volume} {101}},\
  \bibinfo {pages} {024306} (\bibinfo {year} {2020})}\BibitemShut {NoStop}%
\bibitem [{\citenamefont {Moudgalya}\ \emph {et~al.}(2019)\citenamefont
  {Moudgalya}, \citenamefont {Prem}, \citenamefont {Nandkishore},\ and\
  \citenamefont {Regnault}}]{MoudgalyaConstrained}%
  \BibitemOpen
  \bibfield  {author} {\bibinfo {author} {\bibfnamefont {S.}~\bibnamefont
  {Moudgalya}}, \bibinfo {author} {\bibfnamefont {A.}~\bibnamefont {Prem}},
  \bibinfo {author} {\bibfnamefont {R.}~\bibnamefont {Nandkishore}},\ and\
  \bibinfo {author} {\bibfnamefont {B.~A.}\ \bibnamefont {Regnault},
  \bibfnamefont {Nicolas~andBernevig}},\ }\bibfield  {title} {\bibinfo {title}
  {Thermalization and its absence within krylov subspaces of a constrained
  hamiltonian}} (\bibinfo {year} {2019}),\ \bibinfo {note}
  {arXiv:1910.14048}\BibitemShut {NoStop}%
\bibitem [{\citenamefont {Gromov}\ \emph {et~al.}(2020)\citenamefont {Gromov},
  \citenamefont {Lucas},\ and\ \citenamefont {Nandkishore}}]{Gromov20}%
  \BibitemOpen
  \bibfield  {author} {\bibinfo {author} {\bibfnamefont {A.}~\bibnamefont
  {Gromov}}, \bibinfo {author} {\bibfnamefont {A.}~\bibnamefont {Lucas}},\ and\
  \bibinfo {author} {\bibfnamefont {R.~M.}\ \bibnamefont {Nandkishore}},\
  }\bibfield  {title} {\bibinfo {title} {Fracton hydrodynamics},\ }\href
  {https://doi.org/10.1103/PhysRevResearch.2.033124} {\bibfield  {journal}
  {\bibinfo  {journal} {Phys. Rev. Research}\ }\textbf {\bibinfo {volume}
  {2}},\ \bibinfo {pages} {033124} (\bibinfo {year} {2020})}\BibitemShut
  {NoStop}%
\bibitem [{\citenamefont {Magnifico}\ \emph {et~al.}(2020)\citenamefont
  {Magnifico}, \citenamefont {Dalmonte}, \citenamefont {Facchi}, \citenamefont
  {Pascazio}, \citenamefont {Pepe},\ and\ \citenamefont
  {Ercolessi}}]{Magnifico2020}%
  \BibitemOpen
  \bibfield  {author} {\bibinfo {author} {\bibfnamefont {G.}~\bibnamefont
  {Magnifico}}, \bibinfo {author} {\bibfnamefont {M.}~\bibnamefont {Dalmonte}},
  \bibinfo {author} {\bibfnamefont {P.}~\bibnamefont {Facchi}}, \bibinfo
  {author} {\bibfnamefont {S.}~\bibnamefont {Pascazio}}, \bibinfo {author}
  {\bibfnamefont {F.~V.}\ \bibnamefont {Pepe}},\ and\ \bibinfo {author}
  {\bibfnamefont {E.}~\bibnamefont {Ercolessi}},\ }\bibfield  {title} {\bibinfo
  {title} {Real {T}ime {D}ynamics and {C}onfinement in the {$\mathbb{Z}_{n}$}
  {S}chwinger-{W}eyl lattice model for 1+1 {QED}},\ }\href
  {https://doi.org/10.22331/q-2020-06-15-281} {\bibfield  {journal} {\bibinfo
  {journal} {{Quantum}}\ }\textbf {\bibinfo {volume} {4}},\ \bibinfo {pages}
  {281} (\bibinfo {year} {2020})}\BibitemShut {NoStop}%
\bibitem [{\citenamefont {Deutsch}(1991)}]{Deutsch}%
  \BibitemOpen
  \bibfield  {author} {\bibinfo {author} {\bibfnamefont {J.~M.}\ \bibnamefont
  {Deutsch}},\ }\bibfield  {title} {\bibinfo {title} {Quantum statistical
  mechanics in a closed system},\ }\href
  {https://doi.org/10.1103/PhysRevA.43.2046} {\bibfield  {journal} {\bibinfo
  {journal} {Phys. Rev. A}\ }\textbf {\bibinfo {volume} {43}},\ \bibinfo
  {pages} {2046} (\bibinfo {year} {1991})}\BibitemShut {NoStop}%
\bibitem [{\citenamefont {Srednicki}(1994)}]{Srednicki}%
  \BibitemOpen
  \bibfield  {author} {\bibinfo {author} {\bibfnamefont {M.}~\bibnamefont
  {Srednicki}},\ }\bibfield  {title} {\bibinfo {title} {Chaos and quantum
  thermalization},\ }\href {https://doi.org/10.1103/PhysRevE.50.888} {\bibfield
   {journal} {\bibinfo  {journal} {Phys. Rev. E}\ }\textbf {\bibinfo {volume}
  {50}},\ \bibinfo {pages} {888} (\bibinfo {year} {1994})}\BibitemShut
  {NoStop}%
\bibitem [{\citenamefont {Mesterh\'azy}\ \emph {et~al.}(2015)\citenamefont
  {Mesterh\'azy}, \citenamefont {Stockemer},\ and\ \citenamefont
  {Tanizaki}}]{Mesterhazy2015}%
  \BibitemOpen
  \bibfield  {author} {\bibinfo {author} {\bibfnamefont {D.}~\bibnamefont
  {Mesterh\'azy}}, \bibinfo {author} {\bibfnamefont {J.~H.}\ \bibnamefont
  {Stockemer}},\ and\ \bibinfo {author} {\bibfnamefont {Y.}~\bibnamefont
  {Tanizaki}},\ }\bibfield  {title} {\bibinfo {title} {From quantum to
  classical dynamics: The relativistic $o(n)$ model in the framework of the
  real-time functional renormalization group},\ }\href
  {https://doi.org/10.1103/PhysRevD.92.076001} {\bibfield  {journal} {\bibinfo
  {journal} {Phys. Rev. D}\ }\textbf {\bibinfo {volume} {92}},\ \bibinfo
  {pages} {076001} (\bibinfo {year} {2015})}\BibitemShut {NoStop}%
\bibitem [{\citenamefont {Wald}\ \emph {et~al.}(2021)\citenamefont {Wald},
  \citenamefont {Henkel},\ and\ \citenamefont {Gambassi}}]{Wald2021}%
  \BibitemOpen
  \bibfield  {author} {\bibinfo {author} {\bibfnamefont {S.}~\bibnamefont
  {Wald}}, \bibinfo {author} {\bibfnamefont {M.}~\bibnamefont {Henkel}},\ and\
  \bibinfo {author} {\bibfnamefont {A.}~\bibnamefont {Gambassi}},\ }\bibfield
  {title} {\bibinfo {title} {Non-equilibrium dynamics of the open quantum
  o(n)-model with non-markovian noise: exact results}} (\bibinfo {year}
  {2021}),\ \bibinfo {note} {arXiv:2106.08237}\BibitemShut {NoStop}%
\bibitem [{\citenamefont {Haldane}(1991)}]{Haldane}%
  \BibitemOpen
  \bibfield  {author} {\bibinfo {author} {\bibfnamefont {F.~D.~M.}\
  \bibnamefont {Haldane}},\ }\bibfield  {title} {\bibinfo {title} {``fractional
  statistics'' in arbitrary dimensions: A generalization of the pauli
  principle},\ }\href {https://doi.org/10.1103/PhysRevLett.67.937} {\bibfield
  {journal} {\bibinfo  {journal} {Phys. Rev. Lett.}\ }\textbf {\bibinfo
  {volume} {67}},\ \bibinfo {pages} {937} (\bibinfo {year} {1991})}\BibitemShut
  {NoStop}%
\bibitem [{\citenamefont {Leinaas}\ and\ \citenamefont
  {Myrheim}(1977)}]{Leinaas}%
  \BibitemOpen
  \bibfield  {author} {\bibinfo {author} {\bibfnamefont {J.~M.}\ \bibnamefont
  {Leinaas}}\ and\ \bibinfo {author} {\bibfnamefont {J.}~\bibnamefont
  {Myrheim}},\ }\bibfield  {title} {\bibinfo {title} {On the theory of
  identical particles},\ }\href {https://doi.org/10.1007/BF02727953} {\bibfield
   {journal} {\bibinfo  {journal} {Il Nuovo Cimento B (1971-1996)}\ }\textbf
  {\bibinfo {volume} {37}},\ \bibinfo {pages} {1} (\bibinfo {year}
  {1977})}\BibitemShut {NoStop}%
\bibitem [{\citenamefont {Wilczek}(1982)}]{Wilczek1}%
  \BibitemOpen
  \bibfield  {author} {\bibinfo {author} {\bibfnamefont {F.}~\bibnamefont
  {Wilczek}},\ }\bibfield  {title} {\bibinfo {title} {Quantum mechanics of
  fractional-spin particles},\ }\href
  {https://doi.org/10.1103/PhysRevLett.49.957} {\bibfield  {journal} {\bibinfo
  {journal} {Phys. Rev. Lett.}\ }\textbf {\bibinfo {volume} {49}},\ \bibinfo
  {pages} {957} (\bibinfo {year} {1982})}\BibitemShut {NoStop}%
\bibitem [{\citenamefont {Laughlin}(1983)}]{Laughlin}%
  \BibitemOpen
  \bibfield  {author} {\bibinfo {author} {\bibfnamefont {R.~B.}\ \bibnamefont
  {Laughlin}},\ }\bibfield  {title} {\bibinfo {title} {Anomalous quantum hall
  effect: An incompressible quantum fluid with fractionally charged
  excitations},\ }\href {https://doi.org/10.1103/PhysRevLett.50.1395}
  {\bibfield  {journal} {\bibinfo  {journal} {Phys. Rev. Lett.}\ }\textbf
  {\bibinfo {volume} {50}},\ \bibinfo {pages} {1395} (\bibinfo {year}
  {1983})}\BibitemShut {NoStop}%
\bibitem [{\citenamefont {Camino}\ \emph {et~al.}(2005)\citenamefont {Camino},
  \citenamefont {Zhou},\ and\ \citenamefont {Goldman}}]{Camino}%
  \BibitemOpen
  \bibfield  {author} {\bibinfo {author} {\bibfnamefont {F.}~\bibnamefont
  {Camino}}, \bibinfo {author} {\bibfnamefont {W.}~\bibnamefont {Zhou}},\ and\
  \bibinfo {author} {\bibfnamefont {V.}~\bibnamefont {Goldman}},\ }\bibfield
  {title} {\bibinfo {title} {Realization of a laughlin quasiparticle
  interferometer: Observation of fractional statistics},\ }\href@noop {}
  {\bibfield  {journal} {\bibinfo  {journal} {Physical Review B}\ }\textbf
  {\bibinfo {volume} {72}},\ \bibinfo {pages} {075342} (\bibinfo {year}
  {2005})}\BibitemShut {NoStop}%
\bibitem [{\citenamefont {Murthy}\ and\ \citenamefont
  {Shankar}(1994{\natexlab{a}})}]{Murthy2}%
  \BibitemOpen
  \bibfield  {author} {\bibinfo {author} {\bibfnamefont {M.~V.~N.}\
  \bibnamefont {Murthy}}\ and\ \bibinfo {author} {\bibfnamefont
  {R.}~\bibnamefont {Shankar}},\ }\bibfield  {title} {\bibinfo {title} {Haldane
  exclusion statistics and second virial coefficient},\ }\href
  {https://doi.org/10.1103/PhysRevLett.72.3629} {\bibfield  {journal} {\bibinfo
   {journal} {Phys. Rev. Lett.}\ }\textbf {\bibinfo {volume} {72}},\ \bibinfo
  {pages} {3629} (\bibinfo {year} {1994}{\natexlab{a}})}\BibitemShut {NoStop}%
\bibitem [{\citenamefont {Bernard}\ and\ \citenamefont {shi
  Wu}(1994)}]{bernard1994note}%
  \BibitemOpen
  \bibfield  {author} {\bibinfo {author} {\bibfnamefont {D.}~\bibnamefont
  {Bernard}}\ and\ \bibinfo {author} {\bibfnamefont {Y.}~\bibnamefont {shi
  Wu}},\ }\bibfield  {title} {\bibinfo {title} {A note on statistical
  interactions and the thermodynamic bethe ansatz},\ }in\ \href@noop {} {\emph
  {\bibinfo {booktitle} {Nankai Lecture Notes on Mathematical Physics, (World
  Scientific)}}}\ (\bibinfo {year} {1994})\BibitemShut {NoStop}%
\bibitem [{\citenamefont {Isakov}(1994)}]{isakov1994statistical}%
  \BibitemOpen
  \bibfield  {author} {\bibinfo {author} {\bibfnamefont {S.~B.}\ \bibnamefont
  {Isakov}},\ }\bibfield  {title} {\bibinfo {title} {Statistical mechanics for
  a class of quantum statistics},\ }\href
  {https://doi.org/10.1103/PhysRevLett.73.2150} {\bibfield  {journal} {\bibinfo
   {journal} {Phys. Rev. Lett.}\ }\textbf {\bibinfo {volume} {73}},\ \bibinfo
  {pages} {2150} (\bibinfo {year} {1994})}\BibitemShut {NoStop}%
\bibitem [{\citenamefont {Sutherland}(1997)}]{SutherlandHES}%
  \BibitemOpen
  \bibfield  {author} {\bibinfo {author} {\bibfnamefont {B.}~\bibnamefont
  {Sutherland}},\ }\bibfield  {title} {\bibinfo {title} {Microscopic theory of
  exclusion statistics},\ }\href {https://doi.org/10.1103/PhysRevB.56.4422}
  {\bibfield  {journal} {\bibinfo  {journal} {Phys. Rev. B}\ }\textbf {\bibinfo
  {volume} {56}},\ \bibinfo {pages} {4422} (\bibinfo {year}
  {1997})}\BibitemShut {NoStop}%
\bibitem [{\citenamefont {Hatsugai}\ \emph {et~al.}(1996)\citenamefont
  {Hatsugai}, \citenamefont {Kohmoto}, \citenamefont {Koma},\ and\
  \citenamefont {Wu}}]{Hatsugai}%
  \BibitemOpen
  \bibfield  {author} {\bibinfo {author} {\bibfnamefont {Y.}~\bibnamefont
  {Hatsugai}}, \bibinfo {author} {\bibfnamefont {M.}~\bibnamefont {Kohmoto}},
  \bibinfo {author} {\bibfnamefont {T.}~\bibnamefont {Koma}},\ and\ \bibinfo
  {author} {\bibfnamefont {Y.-S.}\ \bibnamefont {Wu}},\ }\bibfield  {title}
  {\bibinfo {title} {Mutual-exclusion statistics in exactly solvable models in
  one and higher dimensions at low temperatures},\ }\href
  {https://doi.org/10.1103/PhysRevB.54.5358} {\bibfield  {journal} {\bibinfo
  {journal} {Phys. Rev. B}\ }\textbf {\bibinfo {volume} {54}},\ \bibinfo
  {pages} {5358} (\bibinfo {year} {1996})}\BibitemShut {NoStop}%
\bibitem [{\citenamefont {Bhaduri}\ \emph
  {et~al.}(1996{\natexlab{a}})\citenamefont {Bhaduri}, \citenamefont {Murthy},\
  and\ \citenamefont {Srivastava}}]{Bhaduri1}%
  \BibitemOpen
  \bibfield  {author} {\bibinfo {author} {\bibfnamefont {R.~K.}\ \bibnamefont
  {Bhaduri}}, \bibinfo {author} {\bibfnamefont {M.~V.~N.}\ \bibnamefont
  {Murthy}},\ and\ \bibinfo {author} {\bibfnamefont {M.~K.}\ \bibnamefont
  {Srivastava}},\ }\bibfield  {title} {\bibinfo {title} {Fractional exclusion
  statistics and two dimensional electron systems},\ }\href
  {https://doi.org/10.1103/PhysRevLett.76.165} {\bibfield  {journal} {\bibinfo
  {journal} {Phys. Rev. Lett.}\ }\textbf {\bibinfo {volume} {76}},\ \bibinfo
  {pages} {165} (\bibinfo {year} {1996}{\natexlab{a}})}\BibitemShut {NoStop}%
\bibitem [{\citenamefont {Wu}(1994)}]{Wu}%
  \BibitemOpen
  \bibfield  {author} {\bibinfo {author} {\bibfnamefont {Y.-S.}\ \bibnamefont
  {Wu}},\ }\bibfield  {title} {\bibinfo {title} {Statistical distribution for
  generalized ideal gas of fractional-statistics particles},\ }\href
  {https://doi.org/10.1103/PhysRevLett.73.922} {\bibfield  {journal} {\bibinfo
  {journal} {Phys. Rev. Lett.}\ }\textbf {\bibinfo {volume} {73}},\ \bibinfo
  {pages} {922} (\bibinfo {year} {1994})}\BibitemShut {NoStop}%
\bibitem [{\citenamefont {Nayak}\ and\ \citenamefont {Wilczek}(1994)}]{Nayak1}%
  \BibitemOpen
  \bibfield  {author} {\bibinfo {author} {\bibfnamefont {C.}~\bibnamefont
  {Nayak}}\ and\ \bibinfo {author} {\bibfnamefont {F.}~\bibnamefont
  {Wilczek}},\ }\bibfield  {title} {\bibinfo {title} {Exclusion statistics:
  Low-temperature properties, fluctuations, duality, and applications},\ }\href
  {https://doi.org/10.1103/PhysRevLett.73.2740} {\bibfield  {journal} {\bibinfo
   {journal} {Phys. Rev. Lett.}\ }\textbf {\bibinfo {volume} {73}},\ \bibinfo
  {pages} {2740} (\bibinfo {year} {1994})}\BibitemShut {NoStop}%
\bibitem [{\citenamefont {Polychronakos}(1996)}]{Polychronakos}%
  \BibitemOpen
  \bibfield  {author} {\bibinfo {author} {\bibfnamefont {A.~P.}\ \bibnamefont
  {Polychronakos}},\ }\bibfield  {title} {\bibinfo {title} {Probabilities and
  path-integral realization of exclusion statistics},\ }\href@noop {}
  {\bibfield  {journal} {\bibinfo  {journal} {Physics Letters B}\ }\textbf
  {\bibinfo {volume} {365}},\ \bibinfo {pages} {202} (\bibinfo {year}
  {1996})}\BibitemShut {NoStop}%
\bibitem [{\citenamefont {Chaturvedi}\ and\ \citenamefont
  {Srinivasan}(1997)}]{chaturvedi1997microscopic}%
  \BibitemOpen
  \bibfield  {author} {\bibinfo {author} {\bibfnamefont {S.}~\bibnamefont
  {Chaturvedi}}\ and\ \bibinfo {author} {\bibfnamefont {V.}~\bibnamefont
  {Srinivasan}},\ }\bibfield  {title} {\bibinfo {title} {Microscopic
  interpretation of haldane's semion statistics},\ }\href
  {https://doi.org/10.1103/PhysRevLett.78.4316} {\bibfield  {journal} {\bibinfo
   {journal} {Phys. Rev. Lett.}\ }\textbf {\bibinfo {volume} {78}},\ \bibinfo
  {pages} {4316} (\bibinfo {year} {1997})}\BibitemShut {NoStop}%
\bibitem [{\citenamefont {Calogero}(1969{\natexlab{a}})}]{calogero1}%
  \BibitemOpen
  \bibfield  {author} {\bibinfo {author} {\bibfnamefont {F.}~\bibnamefont
  {Calogero}},\ }\bibfield  {title} {\bibinfo {title} {Solution of a three-body
  problem in one dimension},\ }\href {https://doi.org/10.1063/1.1664820}
  {\bibfield  {journal} {\bibinfo  {journal} {Journal of Mathematical Physics}\
  }\textbf {\bibinfo {volume} {10}},\ \bibinfo {pages} {2191} (\bibinfo {year}
  {1969}{\natexlab{a}})}\BibitemShut {NoStop}%
\bibitem [{\citenamefont {Calogero}(1969{\natexlab{b}})}]{calogero2}%
  \BibitemOpen
  \bibfield  {author} {\bibinfo {author} {\bibfnamefont {F.}~\bibnamefont
  {Calogero}},\ }\bibfield  {title} {\bibinfo {title} {Ground state of a
  one-dimensional n-body system},\ }\href {https://doi.org/10.1063/1.1664821}
  {\bibfield  {journal} {\bibinfo  {journal} {Journal of Mathematical Physics}\
  }\textbf {\bibinfo {volume} {10}},\ \bibinfo {pages} {2197} (\bibinfo {year}
  {1969}{\natexlab{b}})}\BibitemShut {NoStop}%
\bibitem [{\citenamefont {Sutherland}(1971{\natexlab{a}})}]{sutherland1}%
  \BibitemOpen
  \bibfield  {author} {\bibinfo {author} {\bibfnamefont {B.}~\bibnamefont
  {Sutherland}},\ }\bibfield  {title} {\bibinfo {title} {Quantum many-body
  problem in one dimension: Ground state},\ }\href
  {https://doi.org/10.1063/1.1665584} {\bibfield  {journal} {\bibinfo
  {journal} {Journal of Mathematical Physics}\ }\textbf {\bibinfo {volume}
  {12}},\ \bibinfo {pages} {246} (\bibinfo {year}
  {1971}{\natexlab{a}})}\BibitemShut {NoStop}%
\bibitem [{\citenamefont {Sutherland}(1971{\natexlab{b}})}]{sutherland2}%
  \BibitemOpen
  \bibfield  {author} {\bibinfo {author} {\bibfnamefont {B.}~\bibnamefont
  {Sutherland}},\ }\bibfield  {title} {\bibinfo {title} {Quantum many-body
  problem in one dimension: Thermodynamics},\ }\href
  {https://doi.org/10.1063/1.1665585} {\bibfield  {journal} {\bibinfo
  {journal} {Journal of Mathematical Physics}\ }\textbf {\bibinfo {volume}
  {12}},\ \bibinfo {pages} {251} (\bibinfo {year}
  {1971}{\natexlab{b}})}\BibitemShut {NoStop}%
\bibitem [{\citenamefont {Sutherland}(1971{\natexlab{c}})}]{sutherland3}%
  \BibitemOpen
  \bibfield  {author} {\bibinfo {author} {\bibfnamefont {B.}~\bibnamefont
  {Sutherland}},\ }\bibfield  {title} {\bibinfo {title} {Exact results for a
  quantum many-body problem in one dimension},\ }\href
  {https://doi.org/10.1103/PhysRevA.4.2019} {\bibfield  {journal} {\bibinfo
  {journal} {Phys. Rev. A}\ }\textbf {\bibinfo {volume} {4}},\ \bibinfo {pages}
  {2019} (\bibinfo {year} {1971}{\natexlab{c}})}\BibitemShut {NoStop}%
\bibitem [{\citenamefont {Sutherland}(1972)}]{sutherland4}%
  \BibitemOpen
  \bibfield  {author} {\bibinfo {author} {\bibfnamefont {B.}~\bibnamefont
  {Sutherland}},\ }\bibfield  {title} {\bibinfo {title} {Exact results for a
  quantum many-body problem in one dimension. ii},\ }\href
  {https://doi.org/10.1103/PhysRevA.5.1372} {\bibfield  {journal} {\bibinfo
  {journal} {Phys. Rev. A}\ }\textbf {\bibinfo {volume} {5}},\ \bibinfo {pages}
  {1372} (\bibinfo {year} {1972})}\BibitemShut {NoStop}%
\bibitem [{\citenamefont {Ha}(1995)}]{Ha-HES}%
  \BibitemOpen
  \bibfield  {author} {\bibinfo {author} {\bibfnamefont {Z.}~\bibnamefont
  {Ha}},\ }\bibfield  {title} {\bibinfo {title} {Fractional statistics in one
  dimension: view from an exactly solvable model},\ }\href
  {https://doi.org/https://doi.org/10.1016/0550-3213(94)00537-O} {\bibfield
  {journal} {\bibinfo  {journal} {Nuclear Physics B}\ }\textbf {\bibinfo
  {volume} {435}},\ \bibinfo {pages} {604} (\bibinfo {year}
  {1995})}\BibitemShut {NoStop}%
\bibitem [{\citenamefont {Murthy}\ and\ \citenamefont
  {Shankar}(1999)}]{Murthy}%
  \BibitemOpen
  \bibfield  {author} {\bibinfo {author} {\bibfnamefont {M.~V.~N.}\
  \bibnamefont {Murthy}}\ and\ \bibinfo {author} {\bibfnamefont
  {R.}~\bibnamefont {Shankar}},\ }\bibfield  {title} {\bibinfo {title}
  {Exclusion statistics: A resolution of the problem of negative weights},\
  }\href {https://doi.org/10.1103/PhysRevB.60.6517} {\bibfield  {journal}
  {\bibinfo  {journal} {Phys. Rev. B}\ }\textbf {\bibinfo {volume} {60}},\
  \bibinfo {pages} {6517} (\bibinfo {year} {1999})}\BibitemShut {NoStop}%
\bibitem [{\citenamefont {Isakov}(1996)}]{Isakov96a}%
  \BibitemOpen
  \bibfield  {author} {\bibinfo {author} {\bibfnamefont {S.~B.}\ \bibnamefont
  {Isakov}},\ }\bibfield  {title} {\bibinfo {title} {Bosonic and fermionic
  single-particle states in the haldane approach to statistics for identical
  particles},\ }\href {https://doi.org/10.1103/PhysRevB.53.6585} {\bibfield
  {journal} {\bibinfo  {journal} {Phys. Rev. B}\ }\textbf {\bibinfo {volume}
  {53}},\ \bibinfo {pages} {6585} (\bibinfo {year} {1996})}\BibitemShut
  {NoStop}%
\bibitem [{\citenamefont {Ha}(1994)}]{Ha}%
  \BibitemOpen
  \bibfield  {author} {\bibinfo {author} {\bibfnamefont {Z.~N.~C.}\
  \bibnamefont {Ha}},\ }\bibfield  {title} {\bibinfo {title} {Exact dynamical
  correlation functions of calogero-sutherland model and one-dimensional
  fractional statistics},\ }\href {https://doi.org/10.1103/PhysRevLett.73.1574}
  {\bibfield  {journal} {\bibinfo  {journal} {Phys. Rev. Lett.}\ }\textbf
  {\bibinfo {volume} {73}},\ \bibinfo {pages} {1574} (\bibinfo {year}
  {1994})}\BibitemShut {NoStop}%
\bibitem [{\citenamefont {Bhaduri}\ \emph
  {et~al.}(1996{\natexlab{b}})\citenamefont {Bhaduri}, \citenamefont
  {Bhalerao},\ and\ \citenamefont {Murthy}}]{Bhaduri}%
  \BibitemOpen
  \bibfield  {author} {\bibinfo {author} {\bibfnamefont {R.~K.}\ \bibnamefont
  {Bhaduri}}, \bibinfo {author} {\bibfnamefont {R.~S.}\ \bibnamefont
  {Bhalerao}},\ and\ \bibinfo {author} {\bibfnamefont {M.~V.~N.}\ \bibnamefont
  {Murthy}},\ }\bibfield  {title} {\bibinfo {title} {Haldane exclusion
  statistics and the boltzmann equation},\ }\href
  {https://doi.org/10.1007/BF02183398} {\bibfield  {journal} {\bibinfo
  {journal} {Journal of Statistical Physics}\ }\textbf {\bibinfo {volume}
  {82}},\ \bibinfo {pages} {1659} (\bibinfo {year}
  {1996}{\natexlab{b}})}\BibitemShut {NoStop}%
\bibitem [{\citenamefont {Kaniadakis}\ \emph {et~al.}(1996)\citenamefont
  {Kaniadakis}, \citenamefont {Lavagno},\ and\ \citenamefont
  {Quarati}}]{Kaniadakis96}%
  \BibitemOpen
  \bibfield  {author} {\bibinfo {author} {\bibfnamefont {G.}~\bibnamefont
  {Kaniadakis}}, \bibinfo {author} {\bibfnamefont {A.}~\bibnamefont
  {Lavagno}},\ and\ \bibinfo {author} {\bibfnamefont {P.}~\bibnamefont
  {Quarati}},\ }\bibfield  {title} {\bibinfo {title} {Kinetic approach to
  fractional exclusion statistics},\ }\href
  {https://doi.org/https://doi.org/10.1016/0550-3213(96)00040-5} {\bibfield
  {journal} {\bibinfo  {journal} {Nuclear Physics B}\ }\textbf {\bibinfo
  {volume} {466}},\ \bibinfo {pages} {527 } (\bibinfo {year}
  {1996})}\BibitemShut {NoStop}%
\bibitem [{\citenamefont {Isakov}(1998)}]{Isakov98}%
  \BibitemOpen
  \bibfield  {author} {\bibinfo {author} {\bibfnamefont {S.~B.}\ \bibnamefont
  {Isakov}},\ }\bibfield  {title} {\bibinfo {title} {Quantum liquids of
  particles with generalized statistics},\ }\href
  {https://doi.org/https://doi.org/10.1016/S0375-9601(98)00220-5} {\bibfield
  {journal} {\bibinfo  {journal} {Physics Letters A}\ }\textbf {\bibinfo
  {volume} {242}},\ \bibinfo {pages} {130 } (\bibinfo {year}
  {1998})}\BibitemShut {NoStop}%
\bibitem [{\citenamefont {Arkeryd}(2010)}]{Arkeryd}%
  \BibitemOpen
  \bibfield  {author} {\bibinfo {author} {\bibfnamefont {L.}~\bibnamefont
  {Arkeryd}},\ }\bibfield  {title} {\bibinfo {title} {A quantum boltzmann
  equation for haldane statistics and hard forces; the space-homogeneous
  initial value problem},\ }\href {https://doi.org/10.1007/s00220-010-1046-3}
  {\bibfield  {journal} {\bibinfo  {journal} {Communications in Mathematical
  Physics}\ }\textbf {\bibinfo {volume} {298}},\ \bibinfo {pages} {573}
  (\bibinfo {year} {2010})}\BibitemShut {NoStop}%
\bibitem [{\citenamefont {Rego}\ and\ \citenamefont {Kirczenow}(1999)}]{Rego}%
  \BibitemOpen
  \bibfield  {author} {\bibinfo {author} {\bibfnamefont {L.~G.~C.}\
  \bibnamefont {Rego}}\ and\ \bibinfo {author} {\bibfnamefont {G.}~\bibnamefont
  {Kirczenow}},\ }\bibfield  {title} {\bibinfo {title} {Fractional exclusion
  statistics and the universal quantum of thermal conductance: A unifying
  approach},\ }\href {https://doi.org/10.1103/PhysRevB.59.13080} {\bibfield
  {journal} {\bibinfo  {journal} {Phys. Rev. B}\ }\textbf {\bibinfo {volume}
  {59}},\ \bibinfo {pages} {13080} (\bibinfo {year} {1999})}\BibitemShut
  {NoStop}%
\bibitem [{\citenamefont {Gomila}\ and\ \citenamefont
  {Reggiani}(2001)}]{Gomila}%
  \BibitemOpen
  \bibfield  {author} {\bibinfo {author} {\bibfnamefont {G.}~\bibnamefont
  {Gomila}}\ and\ \bibinfo {author} {\bibfnamefont {L.}~\bibnamefont
  {Reggiani}},\ }\bibfield  {title} {\bibinfo {title} {Fractional exclusion
  statistics and shot noise in ballistic conductors},\ }\href
  {https://doi.org/10.1103/PhysRevB.63.165404} {\bibfield  {journal} {\bibinfo
  {journal} {Phys. Rev. B}\ }\textbf {\bibinfo {volume} {63}},\ \bibinfo
  {pages} {165404} (\bibinfo {year} {2001})}\BibitemShut {NoStop}%
\bibitem [{\citenamefont {Murthy}\ and\ \citenamefont
  {Shankar}(1994{\natexlab{b}})}]{Murthy1}%
  \BibitemOpen
  \bibfield  {author} {\bibinfo {author} {\bibfnamefont {M.~V.~N.}\
  \bibnamefont {Murthy}}\ and\ \bibinfo {author} {\bibfnamefont
  {R.}~\bibnamefont {Shankar}},\ }\bibfield  {title} {\bibinfo {title}
  {Thermodynamics of a one-dimensional ideal gas with fractional exclusion
  statistics},\ }\href {https://doi.org/10.1103/PhysRevLett.73.3331} {\bibfield
   {journal} {\bibinfo  {journal} {Phys. Rev. Lett.}\ }\textbf {\bibinfo
  {volume} {73}},\ \bibinfo {pages} {3331} (\bibinfo {year}
  {1994}{\natexlab{b}})}\BibitemShut {NoStop}%
\bibitem [{\citenamefont {Rajagopal}(1995)}]{Rajagopal}%
  \BibitemOpen
  \bibfield  {author} {\bibinfo {author} {\bibfnamefont {A.~K.}\ \bibnamefont
  {Rajagopal}},\ }\bibfield  {title} {\bibinfo {title} {von neumann entropy
  associated with the haldane exclusion statistics},\ }\href
  {https://doi.org/10.1103/PhysRevLett.74.1048} {\bibfield  {journal} {\bibinfo
   {journal} {Phys. Rev. Lett.}\ }\textbf {\bibinfo {volume} {74}},\ \bibinfo
  {pages} {1048} (\bibinfo {year} {1995})}\BibitemShut {NoStop}%
\bibitem [{\citenamefont {Sen}\ and\ \citenamefont {Bhaduri}(1995)}]{Sen}%
  \BibitemOpen
  \bibfield  {author} {\bibinfo {author} {\bibfnamefont {D.}~\bibnamefont
  {Sen}}\ and\ \bibinfo {author} {\bibfnamefont {R.~K.}\ \bibnamefont
  {Bhaduri}},\ }\bibfield  {title} {\bibinfo {title} {Thomas-fermi method for
  particles obeying generalized exclusion statistics},\ }\href
  {https://doi.org/10.1103/PhysRevLett.74.3912} {\bibfield  {journal} {\bibinfo
   {journal} {Phys. Rev. Lett.}\ }\textbf {\bibinfo {volume} {74}},\ \bibinfo
  {pages} {3912} (\bibinfo {year} {1995})}\BibitemShut {NoStop}%
\bibitem [{\citenamefont {Fukui}\ and\ \citenamefont {Kawakami}(1995)}]{Fukui}%
  \BibitemOpen
  \bibfield  {author} {\bibinfo {author} {\bibfnamefont {T.}~\bibnamefont
  {Fukui}}\ and\ \bibinfo {author} {\bibfnamefont {N.}~\bibnamefont
  {Kawakami}},\ }\bibfield  {title} {\bibinfo {title} {Haldane's fractional
  exclusion statistics for multicomponent systems},\ }\href
  {https://doi.org/10.1103/PhysRevB.51.5239} {\bibfield  {journal} {\bibinfo
  {journal} {Phys. Rev. B}\ }\textbf {\bibinfo {volume} {51}},\ \bibinfo
  {pages} {5239} (\bibinfo {year} {1995})}\BibitemShut {NoStop}%
\bibitem [{\citenamefont {Joyce}\ \emph {et~al.}(1996)\citenamefont {Joyce},
  \citenamefont {Sarkar}, \citenamefont {Spal/ek},\ and\ \citenamefont
  {Byczuk}}]{Joyce}%
  \BibitemOpen
  \bibfield  {author} {\bibinfo {author} {\bibfnamefont {G.~S.}\ \bibnamefont
  {Joyce}}, \bibinfo {author} {\bibfnamefont {S.}~\bibnamefont {Sarkar}},
  \bibinfo {author} {\bibfnamefont {J.}~\bibnamefont {Spal/ek}},\ and\ \bibinfo
  {author} {\bibfnamefont {K.}~\bibnamefont {Byczuk}},\ }\bibfield  {title}
  {\bibinfo {title} {Thermodynamic properties of particles with intermediate
  statistics},\ }\href {https://doi.org/10.1103/PhysRevB.53.990} {\bibfield
  {journal} {\bibinfo  {journal} {Phys. Rev. B}\ }\textbf {\bibinfo {volume}
  {53}},\ \bibinfo {pages} {990} (\bibinfo {year} {1996})}\BibitemShut
  {NoStop}%
\bibitem [{\citenamefont {Isakov}\ \emph {et~al.}(1996)\citenamefont {Isakov},
  \citenamefont {Arovas}, \citenamefont {Myrheim},\ and\ \citenamefont
  {Polychronakos}}]{Isakov96b}%
  \BibitemOpen
  \bibfield  {author} {\bibinfo {author} {\bibfnamefont {S.~B.}\ \bibnamefont
  {Isakov}}, \bibinfo {author} {\bibfnamefont {D.~P.}\ \bibnamefont {Arovas}},
  \bibinfo {author} {\bibfnamefont {J.}~\bibnamefont {Myrheim}},\ and\ \bibinfo
  {author} {\bibfnamefont {A.~P.}\ \bibnamefont {Polychronakos}},\ }\bibfield
  {title} {\bibinfo {title} {Thermodynamics for fractional exclusion
  statistics},\ }\href
  {https://doi.org/https://doi.org/10.1016/0375-9601(96)00157-0} {\bibfield
  {journal} {\bibinfo  {journal} {Physics Letters A}\ }\textbf {\bibinfo
  {volume} {212}},\ \bibinfo {pages} {299} (\bibinfo {year}
  {1996})}\BibitemShut {NoStop}%
\bibitem [{\citenamefont {Potter}\ \emph {et~al.}(2007)\citenamefont {Potter},
  \citenamefont {M\"uller},\ and\ \citenamefont {Karbach}}]{Potter}%
  \BibitemOpen
  \bibfield  {author} {\bibinfo {author} {\bibfnamefont {G.~G.}\ \bibnamefont
  {Potter}}, \bibinfo {author} {\bibfnamefont {G.}~\bibnamefont {M\"uller}},\
  and\ \bibinfo {author} {\bibfnamefont {M.}~\bibnamefont {Karbach}},\
  }\bibfield  {title} {\bibinfo {title} {Thermodynamics of ideal quantum gas
  with fractional statistics in $\mathcal{D}$ dimensions},\ }\href
  {https://doi.org/10.1103/PhysRevE.75.061120} {\bibfield  {journal} {\bibinfo
  {journal} {Phys. Rev. E}\ }\textbf {\bibinfo {volume} {75}},\ \bibinfo
  {pages} {061120} (\bibinfo {year} {2007})}\BibitemShut {NoStop}%
\bibitem [{\citenamefont {del Campo}(2016)}]{Campo2016}%
  \BibitemOpen
  \bibfield  {author} {\bibinfo {author} {\bibfnamefont {A.}~\bibnamefont {del
  Campo}},\ }\bibfield  {title} {\bibinfo {title} {Exact quantum decay of an
  interacting many-particle system: the calogero--sutherland model},\
  }\href@noop {} {\bibfield  {journal} {\bibinfo  {journal} {New Journal of
  Physics}\ }\textbf {\bibinfo {volume} {18}},\ \bibinfo {pages} {015014}
  (\bibinfo {year} {2016})}\BibitemShut {NoStop}%
\bibitem [{\citenamefont {Ilinski}\ \emph {et~al.}(1996)\citenamefont
  {Ilinski}, \citenamefont {Gunn},\ and\ \citenamefont {Ilinskaia}}]{Ilinski}%
  \BibitemOpen
  \bibfield  {author} {\bibinfo {author} {\bibfnamefont {K.~N.}\ \bibnamefont
  {Ilinski}}, \bibinfo {author} {\bibfnamefont {J.~M.~F.}\ \bibnamefont
  {Gunn}},\ and\ \bibinfo {author} {\bibfnamefont {A.~V.}\ \bibnamefont
  {Ilinskaia}},\ }\bibfield  {title} {\bibinfo {title} {Fractional-dimensional
  fock spaces, second quantization, and dynamical interaction of particles with
  haldane's exclusion statistics},\ }\href
  {https://doi.org/10.1103/PhysRevB.53.2615} {\bibfield  {journal} {\bibinfo
  {journal} {Phys. Rev. B}\ }\textbf {\bibinfo {volume} {53}},\ \bibinfo
  {pages} {2615} (\bibinfo {year} {1996})}\BibitemShut {NoStop}%
\bibitem [{\citenamefont {Iguchi}(1997)}]{Iguchi}%
  \BibitemOpen
  \bibfield  {author} {\bibinfo {author} {\bibfnamefont {K.}~\bibnamefont
  {Iguchi}},\ }\bibfield  {title} {\bibinfo {title} {Quantum statistical
  mechanics of an ideal gas with fractional exclusion statistics in arbitrary
  dimensions},\ }\href {https://doi.org/10.1103/PhysRevLett.78.3233} {\bibfield
   {journal} {\bibinfo  {journal} {Phys. Rev. Lett.}\ }\textbf {\bibinfo
  {volume} {78}},\ \bibinfo {pages} {3233} (\bibinfo {year}
  {1997})}\BibitemShut {NoStop}%
\bibitem [{\citenamefont {Iguchi}(1998)}]{Iguchi1}%
  \BibitemOpen
  \bibfield  {author} {\bibinfo {author} {\bibfnamefont {K.}~\bibnamefont
  {Iguchi}},\ }\bibfield  {title} {\bibinfo {title} {Generalization of a fermi
  liquid to a liquid with fractional exclusion statistics in arbitrary
  dimensions: Theory of a haldane liquid},\ }\href
  {https://doi.org/10.1103/PhysRevLett.80.1698} {\bibfield  {journal} {\bibinfo
   {journal} {Phys. Rev. Lett.}\ }\textbf {\bibinfo {volume} {80}},\ \bibinfo
  {pages} {1698} (\bibinfo {year} {1998})}\BibitemShut {NoStop}%
\bibitem [{\citenamefont {Iguchi}(2000)}]{Iguchi2}%
  \BibitemOpen
  \bibfield  {author} {\bibinfo {author} {\bibfnamefont {K.}~\bibnamefont
  {Iguchi}},\ }\bibfield  {title} {\bibinfo {title} {Haldane liquid with mutual
  exclusion statistics},\ }\href {https://doi.org/10.1103/PhysRevB.61.12757}
  {\bibfield  {journal} {\bibinfo  {journal} {Phys. Rev. B}\ }\textbf {\bibinfo
  {volume} {61}},\ \bibinfo {pages} {12757} (\bibinfo {year}
  {2000})}\BibitemShut {NoStop}%
\bibitem [{\citenamefont {Karabali}\ and\ \citenamefont
  {Nair}(1995)}]{karabali}%
  \BibitemOpen
  \bibfield  {author} {\bibinfo {author} {\bibfnamefont {D.}~\bibnamefont
  {Karabali}}\ and\ \bibinfo {author} {\bibfnamefont {V.}~\bibnamefont
  {Nair}},\ }\bibfield  {title} {\bibinfo {title} {Many-body states and
  operator algebra for exclusion statistics},\ }\href
  {https://doi.org/https://doi.org/10.1016/0550-3213(95)00003-B} {\bibfield
  {journal} {\bibinfo  {journal} {Nuclear Physics B}\ }\textbf {\bibinfo
  {volume} {438}},\ \bibinfo {pages} {551 } (\bibinfo {year}
  {1995})}\BibitemShut {NoStop}%
\bibitem [{\citenamefont {Speliotopoulos}(1997)}]{Speliotopoulos97}%
  \BibitemOpen
  \bibfield  {author} {\bibinfo {author} {\bibfnamefont {A.~D.}\ \bibnamefont
  {Speliotopoulos}},\ }\bibfield  {title} {\bibinfo {title} {Turning bosons
  into fermions: exclusion statistics, fractional statistics and the simple
  harmonic oscillator},\ }\href {https://doi.org/10.1088/0305-4470/30/17/024}
  {\bibfield  {journal} {\bibinfo  {journal} {Journal of Physics A:
  Mathematical and General}\ }\textbf {\bibinfo {volume} {30}},\ \bibinfo
  {pages} {6177} (\bibinfo {year} {1997})}\BibitemShut {NoStop}%
\bibitem [{\citenamefont {Meljanac}\ \emph {et~al.}(1999)\citenamefont
  {Meljanac}, \citenamefont {Milekovic},\ and\ \citenamefont
  {Stojic}}]{Meljanac_1999}%
  \BibitemOpen
  \bibfield  {author} {\bibinfo {author} {\bibfnamefont {S.}~\bibnamefont
  {Meljanac}}, \bibinfo {author} {\bibfnamefont {M.}~\bibnamefont
  {Milekovic}},\ and\ \bibinfo {author} {\bibfnamefont {M.}~\bibnamefont
  {Stojic}},\ }\bibfield  {title} {\bibinfo {title} {Exclusion statistics,
  operator algebras and fock space representations},\ }\href
  {https://doi.org/10.1088/0305-4470/32/7/004} {\bibfield  {journal} {\bibinfo
  {journal} {Journal of Physics A: Mathematical and General}\ }\textbf
  {\bibinfo {volume} {32}},\ \bibinfo {pages} {1115} (\bibinfo {year}
  {1999})}\BibitemShut {NoStop}%
\bibitem [{\citenamefont {Raitzsch}\ \emph {et~al.}(2009)\citenamefont
  {Raitzsch}, \citenamefont {Heidemann}, \citenamefont {Weimer}, \citenamefont
  {Butscher}, \citenamefont {Kollmann}, \citenamefont {Löw}, \citenamefont
  {Büchler},\ and\ \citenamefont {Pfau}}]{Raitzsch2009}%
  \BibitemOpen
  \bibfield  {author} {\bibinfo {author} {\bibfnamefont {U.}~\bibnamefont
  {Raitzsch}}, \bibinfo {author} {\bibfnamefont {R.}~\bibnamefont {Heidemann}},
  \bibinfo {author} {\bibfnamefont {H.}~\bibnamefont {Weimer}}, \bibinfo
  {author} {\bibfnamefont {B.}~\bibnamefont {Butscher}}, \bibinfo {author}
  {\bibfnamefont {P.}~\bibnamefont {Kollmann}}, \bibinfo {author}
  {\bibfnamefont {R.}~\bibnamefont {Löw}}, \bibinfo {author} {\bibfnamefont
  {H.~P.}\ \bibnamefont {Büchler}},\ and\ \bibinfo {author} {\bibfnamefont
  {T.}~\bibnamefont {Pfau}},\ }\bibfield  {title} {\bibinfo {title}
  {Investigation of dephasing rates in an interacting rydberg gas},\ }\href
  {https://doi.org/10.1088/1367-2630/11/5/055014} {\bibfield  {journal}
  {\bibinfo  {journal} {New Journal of Physics}\ }\textbf {\bibinfo {volume}
  {11}},\ \bibinfo {pages} {055014} (\bibinfo {year} {2009})}\BibitemShut
  {NoStop}%
\bibitem [{\citenamefont {Barreiro}\ \emph {et~al.}(2011)\citenamefont
  {Barreiro}, \citenamefont {Müller}, \citenamefont {Schindler}, \citenamefont
  {Nigg}, \citenamefont {Monz}, \citenamefont {Chwalla}, \citenamefont
  {Hennrich}, \citenamefont {Roos}, \citenamefont {Zoller},\ and\ \citenamefont
  {Blatt}}]{Barreiro2011}%
  \BibitemOpen
  \bibfield  {author} {\bibinfo {author} {\bibfnamefont {J.~T.}\ \bibnamefont
  {Barreiro}}, \bibinfo {author} {\bibfnamefont {M.}~\bibnamefont {Müller}},
  \bibinfo {author} {\bibfnamefont {P.}~\bibnamefont {Schindler}}, \bibinfo
  {author} {\bibfnamefont {D.}~\bibnamefont {Nigg}}, \bibinfo {author}
  {\bibfnamefont {T.}~\bibnamefont {Monz}}, \bibinfo {author} {\bibfnamefont
  {M.}~\bibnamefont {Chwalla}}, \bibinfo {author} {\bibfnamefont
  {M.}~\bibnamefont {Hennrich}}, \bibinfo {author} {\bibfnamefont {C.~F.}\
  \bibnamefont {Roos}}, \bibinfo {author} {\bibfnamefont {P.}~\bibnamefont
  {Zoller}},\ and\ \bibinfo {author} {\bibfnamefont {R.}~\bibnamefont
  {Blatt}},\ }\bibfield  {title} {\bibinfo {title} {An open-system quantum
  simulator with trapped ions},\ }\href {https://doi.org/10.1038/nature09801}
  {\bibfield  {journal} {\bibinfo  {journal} {Nature}\ }\textbf {\bibinfo
  {volume} {470}},\ \bibinfo {pages} {486} (\bibinfo {year}
  {2011})}\BibitemShut {NoStop}%
\bibitem [{\citenamefont {Manzano}(2020)}]{Manzano2020}%
  \BibitemOpen
  \bibfield  {author} {\bibinfo {author} {\bibfnamefont {D.}~\bibnamefont
  {Manzano}},\ }\bibfield  {title} {\bibinfo {title} {A short introduction to
  the lindblad master equation},\ }\href
  {https://doi.org/https://doi.org/10.1063/1.5115323} {\bibfield  {journal}
  {\bibinfo  {journal} {AIP Advances}\ }\textbf {\bibinfo {volume} {10}},\
  \bibinfo {pages} {025106} (\bibinfo {year} {2020})}\BibitemShut {NoStop}%
\bibitem [{\citenamefont {Breuer}\ \emph {et~al.}(2002)\citenamefont {Breuer},
  \citenamefont {Petruccione} \emph {et~al.}}]{Breuer}%
  \BibitemOpen
  \bibfield  {author} {\bibinfo {author} {\bibfnamefont {H.-P.}\ \bibnamefont
  {Breuer}}, \bibinfo {author} {\bibfnamefont {F.}~\bibnamefont {Petruccione}},
  \emph {et~al.},\ }\href@noop {} {\emph {\bibinfo {title} {The theory of open
  quantum systems}}}\ (\bibinfo  {publisher} {Oxford University Press on
  Demand},\ \bibinfo {year} {2002})\BibitemShut {NoStop}%
\end{thebibliography}%

\end{document}